\documentstyle [12pt,epsf,rotate] {article}
\newcommand{\nwc}{\newcommand}
\nwc{\be}  {\begin{equation}}
\nwc{\ee}  {\end{equation}}
\nwc{\ba}  {\begin{array}}
\nwc{\ea}  {\end{array}}
\nwc{\bdm} {\begin{displaymath}}
\nwc{\edm} {\end{displaymath}}
\nwc{\bea} {\be\ba{rcl}}
\nwc{\eea} {\ear\ee}
\nwc{\bear} {\begin{eqnarray}}
\nwc{\ear} {\end{eqnarray}}
\nwc{\Tr} {\rm Tr}
\nwc{\LL} {\cal L}
\renewcommand{\theequation}{\arabic{section}.\arabic{equation}}
\begin{document}
\begin{titlepage}
\begin{flushright}
HD--THEP--00--35
\end{flushright}
\quad\\
\vspace{1.8cm}

\begin{center}
{\Large Spontaneously broken color}\\
\vspace{2cm}
Christof Wetterich\footnote{e-mail: C.Wetterich@thphys.uni-heidelberg.de}\\
\bigskip
Institut  f\"ur Theoretische Physik\\
Universit\"at Heidelberg\\
Philosophenweg 16, D-69120 Heidelberg\\
\vspace{3cm}

\end{center}

\begin{abstract}
The vacuum of QCD is characterized by the Higgs mechanism.
Color is ``spontaneously broken'' by a quark-antiquark
condensate in the octet representation. The massive gluons
carry integer electric charges and are identified
with the vector mesons.  The fermionic excitations consist
of the low mass baryon octet and a singlet. The interactions
between these particles and the light pseudoscalar octet are
largely determined by chiral symmetry and a nonlinear local symmetry.
A consistent phenomenological picture of strong interactions
at long distances arises from a simple effective action.

\end{abstract}
\end{titlepage}

\section{Introduction}

An understanding of the properties of the vacuum is a central
goal in the description of strong interactions by quantum chromodynamics
(QCD). Recently, it has been proposed that the local color symmetry
is broken spontaneously by a dynamical quark-antiquark condensate
\cite{W1}\cite{W2}. The Higgs mechanism provides a mass for all gluons,
thereby squeezing the gauge fields between color charges
into flux tubes. The separation  of two color charges
leads therefore to an increasing potential until the string breaks due to
particle production. This results in a simple picture
of confinement. The Higgs mechanism also gives
integer electric charge to all physical particles. Furthermore,
the expectation value of the quark-antiquark color octet breaks the
global chiral symmetry. In consequence the fermions become massive
and the spectrum contains light pions and kaons as pseudo-Goldstone
bosons. In the limit of equal masses for the three light quarks
the global vector-like  SU(3)-symmetry of the ``eightfold way''
remains unbroken and can be used for a classification of the particle
spectrum. The nine quarks transform as an octet and a singlet.
We identify the octet with the low mass baryons -- this is quark-baryon
duality. Similarly, the eight gluons carry the quantum numbers of the light
vector mesons $\rho,K^*$ etc..  The identification of
massive gluons with the vector mesons is called gluon-meson duality.

Strictly speaking, local symmetries cannot be broken spontaneously
in the vacuum. This has led to the realisation that the
confinement and the Higgs description are not necessarily associated to
mutually exclusive phases. They may only be different
facets of one and same
physical state \cite{Banks}. We stress that this observation
is not only of formal importance. For example, the high temperature
phase transition of electroweak interactions with a small Higgs scalar
mass ends for a larger scalar mass in a critical endpoint. Beyond
this endpoint the phase transition is replaced by an analytical crossover
\cite{ReWe}. In this region -- which is relevant for a realistic Higgs
mass in the standard model -- a Higgs and a confinement description
can be used simultaneously. Our picture of the QCD vacuum ressembles in many
aspects the ``strongly coupled electroweak theory'' at high
temperature\footnote{Without a direct connection to the
standard model the $SU(2)$-Yang-Mills theory with stong
gauge coupling and fundamental scalar has been first simulated
on the lattice in \cite{SU2L}.}.

For strong interactions, the complementarity between the Higgs- and
confinement description of the vacuum has been mainly discussed
in toy models with additional fundamental colored scalar
fields\cite{H}\footnote{See also ref. \cite{M} for early discussions of strong 
interactions with additional fundamental colored scalar fields.}.
In contrast, our approach concentrates on standard QCD with the
gauge coupling and the current quark masses as the only free parameters.
Similarly to chiral symmetry breaking the relevant dynamical
condensate is provided by a quark-antiquark pair. The main
difference to the usual treatment of chiral symmetry breaking
is the assumption that both color singlet and octet composite
fields acquire a vacuum expectation value.

We propose that besides the effective running of the gauge
coupling the main ingredient of low-momentum QCD consists of
effective  scalar fields representing  quark-antiquark bound states.
Once these composite operators are treated on the
same footing as the quark and gluon fields, the description of
propagators and vertices in terms of an effective action becomes again
very simple. The long-sought  dual description of long-distance
strong interactions can be realized by the addition of fields for composites.
This is very similar to the asymptotically free nonlinear sigma model
in two dimensions where the addition of the composite ``radial excitation''
provides for  a simple dual description of the low momentum behavior
\cite{1}. We will call this particular version of duality where
a simple effective action for the low momentum degrees of freedom can be
achieved by the addition of fields for composite operators by the
German word\footnote{To be translated roughly by
``completion''.} ``Vervollst\"andigung''.

Spontaneous breaking of color has also been proposed \cite{4} for
situations with a very high baryon density, as perhaps in the interior
of neutron stars. In this proposal a condensation of diquark operators
is responsible for color superconductivity and spontaneous breaking
of baryon number. In particular, the suggestion of
color-flavor locking \cite{CF} offers analogies to our description
of the vacuum, even though different physical situations are
described (va\-cuum vs. high density state) and the pattern of spontaneous
color-symmetry breaking is distinct (quark-antiquark vs. quark-quark
condensate; conserved vs. broken baryon number). This analogy
may be an important key for the understanding of possible phase
transitions to a high density phase of QCD.

After spontaneous color symmetry breaking by an octet quark-antiquark
condensate all quantum numbers of the excitations above the vacuum
match with the observed spectrum of low mass particles. (These
quantum numbers include baryon number, strangeness, isospin and spin
as well as parity  and C-parity.) We have therefore little doubt that
such a description of the QCD vacuum is, in principle, possible.
In the present paper we investigate the more specific hypothesis of
``scalar Vervollst\"andigung''. This hypothesis states that the
effective action becomes simple once one adds to the quark and
gluon fields the scalar fields with the quantum numbers of
quark-antiquark pairs. More precisely, we assume that in leading
order only invariants with mass dimension smaller or equal to
four have to be included, where the scalar fields are counted
according to their canonical dimension. This
``renormalizable\footnote{Renormalizability is here no fundamental
property since the validity of the effective action needs not to
cover a large range of momenta.} effective action'' is assumed
to be valid for low momenta. It exhibits a number of new couplings
which describe the scalar potential and the Yukawa couplings
between scalars and quarks. In principle, these couplings
are calculable in QCD as a function of the gauge coupling or
$\Lambda_{\rm QCD}$. In this work we make no attempt of
such a computation. (See ref. \cite{W2} for first steps in this
direction.) We rather perform  a phenomenological analysis
treating the new couplings as free parameters. We concentrate
on the limit of equal current quark masses. Besides the explicit
chiral symmetry breaking by the current quark mass seven effective
couplings are relevant for our discussion. Four of them appear only
in the mass formulae for the light baryon octet and singlet and the
$\eta'$-meson. The masses and interactions (including interactions
with baryons) of the pseudoscalar and vector mesons carrying isospin
or strangeness are governed by only three  effective
parameters: the expectation values of the scalar octet $(\chi_0)$
and singlet $(\sigma_0)$and the effective gauge coupling $(g)$.
We fix $\chi_0,\sigma_0$ and $g$ by the observed values of the vector
meson mass, the pseudoscalar decay constant and the electromagnetic
decays of the vector mesons. ``Predictions'' for other quantities
in this sector like the decay width for $\rho\to 2\pi$ involve
then no further free parameters. We will argue that the hypothesis of scalar
Vervollst\"andigung gives indeed a satisfactory description for
long-distance strong interactions in leading order.

In sect. 2 we describe our setting in more detail and specify the
transformation properties of the fields under the various symmetry  
transformations. In particular, the emergence of an integer
electric charge for the excitations above the vacuum (physical particles) is
discussed in sect. 3. Sect. 4 introduces a nonlinear description
for the fields corresponding to physical particles. This makes
the complementarity between the ``Higgs picture''
and the ``confinement picture'' manifest. In the nonlinear
description the local color symmetry remains unbroken and all
physical particles transform as color singlets. The remnant
of the color symmetry for the interactions of the physical
particles is a nonlinear local reparametrization  symmetry. This is
investigated in sect. 5. After a gauge fixing consistent with
a simple realization of parity we arrive at the effective action
for the nonlinear ``physcial fields''. In sect. 6 we turn to the
issue of baryon number $B$. We address the puzzle how quarks with $B=1/3$ and  
baryons with $B=1$ can be described by the same field. In the Higgs
language this issue is obscured by the fact that the gauge fixing
is not compatible with the global transformation associated to $B$.
A careful treatment reveals that baryons  carry indeed three times
the baryon number of quarks.

In sect. 7 we turn to the electromagnetic interactions for the nonlinear
physical fields. They are largely governed by the nonlinear local
reparametrization symmetry. We recover the well established concept
of vector dominance as well as successful relations between hadronic
and electromagnetic decays of the $\rho$-meson. Some shortcomings
of the scalar Vervollst\"andigung concerning the physics of
the vector mesons are listed in sect. 8. Here we also propose an
extension to ``scalar-vector Vervollst\"andigung'' by adding fields
for quark-antiquark bilinears in the vector and axial-vector channels.
This completes the spectrum of light mesons by
the ninth vector meson (the $SU(3)$-singlet)
and the axial-vector mesons. In
sect. 9 we discuss the effective interactions between the pseudoscalar
mesons which follow from the hypothesis of scalar Vervollst\"andigung.
Their generic form is determined by chiral symmetry and we compute
some of the effective couplings $L_i$ which appear in next-to-leading
order in chiral perturbation theory. The agreement with observation
is very satisfactory.
The weak interactions are introduced
in sect. 10. There we show that the $\Delta I=1/2$ rule for the hadronic
kaon decays arises naturally in our setting. In sect. 11 our discussion
is extended to diquark fields which presumably play an important role
for QCD in a medium with high baryon density. This opens new
perspectives on the qualitative features of the QCD-phase diagram
for large temperature and chemical potential. The short sect. 12
sketches the inclusion of the heavy quarks charm, beauty and top
in our approach. We finally present a summary and conclusions in
sect. 13.

\section{Quark-antiquark condensates}
\noindent {\bf a) Effective action}

In this section we describe the effective action
for long-distance strong interactions according to the hypothesis
of scalar Vervoll\-st\"an\-digung. By de\-finition the effective action
generates the one-particle-irreducible (1PI) correlation functions
and includes all quantum fluctuations. It therefore contains
the direct information about the propagators and proper vertices.

In addition to the quark and gluon fields we consider scalar
fields with the transformation properties of
quark-antiquark pairs. With respect to the color and chiral flavor
rotations $SU(3)_C\times SU(3)_L\times SU(3)_R$ the three
light left-handed and right-handed quarks $\psi_L,\psi_R$
transform as (3,3,1) and (3,1,3), respectively. Quark-antiquark
bilinears therefore contain a color singlet $\Phi(1,\bar 3, 3)$ and
a color octet $\chi(8,\bar 3, 3)$
\bear\label{2.1}
\gamma_{ij}&=&\chi_{ij}+\frac{1}{\sqrt3}\phi\ \delta_{ij},\nonumber\\
\phi&=&\frac{1}{\sqrt3}\gamma_{ii}\quad,\quad \chi_{ii}=0\ear
Here we use a matrix notation for the flavor indices and write
the color indices $ i,j$
explicitly, e.g. $\chi
_{ij,ab}\equiv\chi_{ij},\ \phi_{ab}
\equiv\phi$. Then $\gamma_{ij}$ contains 81 complex scalar fields.
Similarly, the quark fields are represented as three flavor vectors
$\psi_{ai}\equiv\psi_i$, $\bar\psi_{ia}\equiv\bar\psi_i$. The infinitesimal
$SU(3)_C\times SU(3)_L\times SU(3)_R$ transformations are given
by
\bear\label{2.2}
&&(\delta\psi_L)_i=i\psi_{L,j}(\Theta^T_C)_{ji}+i\Theta_L\psi_{L,i}
\nonumber\\
&&(\delta\psi_R)_i=i\psi_{R,j}(\Theta^T_C)_{ji}+i\Theta_R\psi_{R,i}
\ear
with
\be\label{2.3}
(\Theta_C)_{ij}=\frac{1}{2}\theta^z_C(x)(\lambda_z)_{ij},\ \Theta_{L,R}=
\frac{1}{2}\theta^z_{L,R}\lambda_z\ee
Here $\lambda_z$ are the eight Gell-Mann matrices normalized
according to $Tr(\lambda_y\lambda_z)=2\delta_{yz}$ and
$\Theta_C$ is a scalar in flavor space. Correspondingly, the scalar
fields $\phi$ and $\gamma_{ij}$ transform as
\bear\label{2.4}
&&\delta\phi=i\Theta_R\phi-i\phi\Theta_L,\nonumber\\
&&\delta\chi_{ij}=i\Theta_R\chi_{ij}-i\chi_{ij}\Theta_L+i(\Theta_C)
_{ik}\chi_{kj}-i\chi_{ik}(\Theta_C)_{kj}\ear
As usual we represent the eight $SU(3)_C$-gauge fields by
\bear\label{2.5}
&&A_{ij,\mu}=\frac{1}{2}A^z_\mu(\lambda_z)_{ij},\nonumber\\
&&(\delta A)_{ij,\mu}=i(\Theta_C)_{ik}A_{kj,\mu}-iA_{ik,\mu}
(\Theta_C)_{kj}+\frac{1}{g}\partial_\mu(\Theta_C)_{ij}\ear

We consider a very simple
effective Lagrangian containing only terms with dimension
up to  four\footnote{Four our conventions for fermions see
\cite{Conv}.}
\bear\label{2.6}
{\cal L}&=&iZ_\psi\bar\psi_i\gamma^\mu\partial_\mu\psi_i+g
Z_\psi\bar\psi_i
\gamma^\mu A_{ij,\mu}\psi_j+\frac{1}{2}G^{\mu\nu}_{ij}
G_{ji,\mu\nu}\nonumber\\
&&+Tr\{(D^\mu\gamma_{ij})^\dagger D_\mu \gamma_{ij}\}+U(\gamma)
\nonumber\\
&&+Z_\psi\bar\psi_i[(h\phi\delta_{ij}+\tilde h\chi_{ij})
\frac{1+\gamma_5}{2}-(h\phi^\dagger\delta_{ij}+\tilde h
\chi^\dagger_{ji})\frac{1-\gamma_5}{2}]\psi_j\ear
Here $G_{ij,\mu\nu}=\partial_\mu A_{ij,\nu}-\partial_\nu A_{ij,\mu}-ig
A_{ik,\mu}A_{kj,\nu}+ig A_{ik,\nu}A_{kj,\mu}$ and
the interaction between gluons and $\chi$ arise
from the covariant derivative
\be\label{2.6a}
D_\mu\gamma_{ij}=\partial_\mu\gamma_{ij}-
ig A_{ik,\mu}\gamma_{kj}+ig\gamma_{ik}A_{kj,\mu}\ee
In our notation the
transposition acts only on flavor indices, e.g. $(\gamma^\dagger
_{ij})_{ab}=\gamma^*_{ij,ba}$.
The effective potential
\bear\label{2.7}
&&U(\gamma)=U_0(\chi,\phi)-\frac{1}{2}\nu (\det\ \phi+\det \phi^\dagger)
-\frac{1}{2}\nu'
(E(\phi,\chi)+E^*(\phi,\chi))\nonumber\\
&&E(\phi,\chi)=\frac{1}{6}\epsilon_{a_1a_2a_3}\epsilon_{b_1b_2b_3}
\phi_{a_1b_1}\chi_{ij,a_2b_2}\chi_{ji,a_3b_3}\ear
conserves axial $U(1)$ symmetry except for the 't Hooft
terms \cite{Anom} $\sim\nu,
\nu'$.
For the purpose of this section the only information
we need from $U_0$ concerns the expectation values $\sigma_0$ and
$\chi_0$ for the singlet and octet in the limit of equal quark
masses. The hypothesis of scalar Vervollst\"andigung states
that the leading behavior of low energy QCD can be directly
extracted from the propagators and vertices of the effective
action (\ref{2.6}).

Finally, explicit chiral symmetry breaking is induced
by a linear term \cite{6}
\bear\label{2.8}
{\cal L}_j&=&-\frac{1}{2}Z^{-1/2}_\phi\ Tr\ (j^\dagger\phi+\phi^\dagger
j)\nonumber\\
j=j^\dagger&=&a_q\bar m=a_q\  diag(\bar m_u,\bar m_d,\bar m_s)\ear
with $\bar m_q$ the current quark masses
normalized at some appropriate scale, say $\mu=2$ GeV. We note that
the quark wave function renormalization $Z_\psi$ can be absorbed
by a rescaling of $\psi$. We keep it here in order to discuss later
a possible simple bridge between our picture and the nonrelativistic
quark model.

Besides the explicit chiral symmetry breaking in (\ref{2.8}) the model
contains the seven real parameters $g,h,\tilde h,\sigma_0,\chi_0, \nu$
and $\nu'$. In fact, the effective action (\ref{2.6}) is the most
general\footnote{The only exception is the omission of a possible
$U(1)_A$-violating term cubic in $\chi$ which is not present in the 't Hooft
interaction \cite{Anom} and also not generated by one-loop fluctuations.}
one which contains only ``renormalizable'' interactions and is
consistent with
$SU(3)_C\times SU(3)_L\times SU(3)_R$ symmetry, as well as with the
discrete transformations parity and charge conjugation ($c$ is
the charge conjugation matrix)
\bear\label{2.9}
P&:&\psi_L\to-\psi_R\ ,\ \psi_R\to \psi_L\ ,\ A_{ij,\mu}\to A_{ij,\mu},
\nonumber\\
&\ &\phi\to\phi^\dagger\ ,\ \chi_{ij,ab}\to\chi^*_{ji,ba}\nonumber\\
&&\nonumber\\
C&:&\psi_L\to c\bar\psi_R\ ,\ \psi_R\to-c\bar\psi_L\ ,\
A_{ij,\mu}\to-A_{ji,\mu},\nonumber\\
&\ & \phi\to\phi^T\ ,\ \chi_{ij,ab}\to\chi_{ji,ba}\ear
We will show that the masses and interactions of the lightest
octets of baryons as well as pseudoscalar and vector mesons are well
described by the effective action (\ref{2.6}), (\ref{2.8}).
A few shortcomings in the vector meson sector  and a sketch of a possible
extension are discussed  in sect. 8.

Actually, there is no reason why only interactions between
scalar and pseudoscalar quark-antiquark bilinears should be
present as in (\ref{2.6}). Effective four-quark interactions
(1PI) in the vector or axial-vector channel are certainly
induced by fluctuations, and we will discuss them in appendix A.
There is also no need that the effective action has to be
of the ``renormalizable form'' (\ref{2.6}). In fact, we have at present no  
strong argument why higher-order operators have to be small.
In the spirit that a useful dual description should not be too
complicated, we simply investigate in this paper to what
extent the hypothesis of
scalar Vervollst\"andigung (\ref{2.6}) is compatible with observation.
If successful, the neglected subleading terms may be considered
later for increased quantitative accuracy. Furthermore, we know
that QCD contains higher resonances like the $\Delta$ or the axial-vector
mesons. They are not
described by  the  effective action (\ref{2.6}). One may include them
by the introduction of additional fields for bound states (see appendix A).
This is, however, not the purpose of the present paper. We only mention
here that ``integrating out'' the missing resonances with
lowest mass presumably gives the leading contribution to the neglected
higher-order operators. Those will typically contain ``nonlocal
behavior'' in the momentum range characteristic for the resonances.

\medskip
\noindent{\bf b) Connection to perturbative QCD}

In a renormalization group framework the parameters appearing in (\ref{2.6})
can be considered as running coupling constants. For instance,
we may associate  $\Gamma_k=\int d^4x{\cal L}$ with the effective
average action \cite{5} which exhibits an infrared cutoff $k$.
Then only quantum fluctuations with momenta larger than $k$
are included. The vacuum
properties and the physical particle spectrum should be extracted
for $k=0$. For short-distance scattering processes, however, one
may account for the momentum dependence of the vertices by
using for $k$ an appropriate momentum scale. Our description should
coincide with perturbative QCD for large enough $k$.

Let us look first at the range of validity of perturbative QCD. For large
$k$ (say $k\stackrel {>}{\scriptstyle \sim}$ 2 GeV)
we expand $U_0=m^2_\phi\ Tr\ \phi^\dagger\phi+m^2_\chi\ Tr\ \chi
^\dagger_{ij}\chi_{ij}+...$ and assume that
both $m_\phi^2$ and $m^2_\chi$ are large and positive. Then
the additional scalar excitations will effectively decouple.
They can be eliminated by solving  the scalar field equations
in terms of $\psi,\bar\psi$ and $A$ and reinserting this
solution into the effective action. For $\psi=\bar\psi=0,\ A_\mu
=0$ the expectation value $<\chi>$ vanishes. On the other hand one finds
$<\phi>=\frac{1}{2}m^{-2}_\phi Z^{-1/2}_\Phi j$ and therefore
effective quark masses
\be\label{2.10}
\bar m_q=\frac{1}{2}hm_\Phi^{-2}Z_\Phi^{-1/2}j_q.\ee
This is the only relevant coupling induced by the additional degrees
of freedom and exactly what is needed for perturbative
QCD! We have explicitly checked that in leading order the
running of the quark masses only arises from gluon diagrams.
(The sources $j$ are constant and the effects of the
running scalar wave function renormalization $Z_\Phi$ drop out.)
Comparing with (\ref{2.8}) we find for $k=\mu=2$ GeV the relation
\be\label{2.11}
a_q(\mu)=2m^2_\phi(\mu) Z_\phi^{1/2}(\mu) h^{-1}(\mu)\ee

For $A_\mu=0$ the exchange of $\phi$ and $\chi$ produces a correction
$\Delta{\cal L}$
\bear\label{2.12}
&&\Delta{\cal L}=h^2\left(\bar\psi_{ia}\left(\frac{1-\gamma_5}
{2}\right)\psi_{bi}\right)(m^2_\phi-\partial^\mu
\partial_\mu)^{-1}\left(\bar\psi_{jb}\frac{1+\gamma_5}{2}\psi_{aj}
\right)\nonumber\\
&&+\tilde h^2\Bigl\{\left(\bar\psi_{aj}\frac{1-\gamma_5}{2}\psi_{bi}\right
)(m^2_\chi
-\partial^\mu\partial_\mu)^{-1}\left(\bar\psi_{bi}\frac{1+\gamma_5}{2}
\psi_{aj}\right)\nonumber\\
&&-\frac{1}{3}\left(\bar\psi_{ia}\frac{1-\gamma_5}{2}\psi_{bi}\right)
(m^2_\chi-\partial
^\mu\partial_\mu)^{-1}\left(\bar\psi_{jb}\frac{1+\gamma_5}{2}\psi_{aj}\right)
\Bigr\}\ear
This corresponds to a one-particle irreducible effective four-quark
interaction which can be computed perturbatively by evaluating the
relevant box diagrams with two internal gluon and quark lines. One
finds
\be\label{2.13}
\frac{h^2(k)}{m^2_\phi(k)}=\frac{23l^4_3}{144\pi^2}\frac{g^4(k)}
{k^2}\qquad ,\qquad \frac{\tilde h(k)}{m_\chi^2(k)}=\frac{13 l^4_3}{384\pi^2}
\frac{g^4(k)}{k^2}\ee
with $l^4_3$ a constant of order one which depends on the precise choice
of the infrared cutoff $k$. We emphasize the composite character
of the scalar fields which implies that the $k$-dependence
of $m_{\phi^2}$ or $h$ is essentially determined by the QCD-box diagrams
if $k$ is large. Since the source
$j$ is independent of the renormalization scale $\mu$, one infers from
(\ref{2.8}), (\ref{2.11}) and (\ref{2.13}) that the $k$-dependence
of $Z_{\phi}$  obeys for the perturbative range
\be\label{2.14}
Z_\phi^{1/2}(k)h(k)\sim\frac{g^4(k)}{k^2\bar m_q(k)}\ee
A small value of the scalar wave function renormalization $Z_\phi$ for
large $k$ is consistent with a composite scalar field.

>From the exchange of $\chi$ we also obtain for $A_\mu\not=0$
higher-order gluon
interactions by expanding $-Tr\ \ln(m^2_\chi-D^2[A])$ with $D^2$
the covariant Laplacian.
All these corrections are
nonrenormalizable interactions which
are ``irrelevant'' by standard universality arguments.
This means precisely that they are computable within
the range of validity of perturbation theory. In fact, the
scalar fields in $\Gamma_k$ may be viewed as a shorthand for the
presence of these nonrenormalizable quark and gluon interactions
in the effective action.

Due to the presence of quark fluctuations the mass terms $m^2_\Phi$
and $m^2_\chi$ decrease as $k$ is lowered. Perturbation
theory typically breaks down once
$m^2_\Phi$ or $m^2_\chi$ are of the same size as $k^2$.
Furthermore, the relative importance of the anomaly
terms increases. These cubic interactions tend to destabilize
the minimum of the potential at $\phi=\chi=0$ (for $j=0$). It
is our central postulate that  in the
nonperturbative region of small $k$ the minimum of the effective
potential $U$ is located at  nonzero expectation
values\footnote{The $SU(3)_C$ color breaking by nonzero $<\chi>$
occurs only in a gauge-fixed version, whereas an explicitly gauge-invariant
formulation is used below.} for both $<\phi>$ and  $<\chi>$
even for $j=0$. A first dynamical analysis based on a mean field
approximation indeed suggests \cite{W2} that this postulate
is reasonable.

\medskip
\noindent{\bf c) Spontaneous symmetry breaking}

For small enough and equal quark masses the usual color singlet
chiral quark condensate $<\bar q q>$ can
be related to the expectaton value of $\phi$
\be\label{2.15}
<\phi_{ab}>=\sigma_0\delta_{ab}\ee
by saturating the expectation value of the explicit chiral symmetry
breaking term. From ${\cal L}_j=\sum_q\bar m_q(\mu)<\bar qq>(\mu)$ and
(\ref{2.8}), (\ref{2.10}) one finds
\be\label{2.16}
<\bar qq>(\mu)=-Z_\phi^{-1/2}(0)j_q\sigma_0\bar m_q(\mu)^{-1}
=-\frac{2m^2_\phi(\mu)}{h(\mu)}\left(\frac{Z_\phi(\mu)}
{Z_\phi(0)}\right)^{1/2}\sigma_0\ee
The expectation value (\ref{2.15}) breaks chiral symmetry. It preserves
a vector-like flavor symmetry $SU(3)_{\hat V}$ with transformations
given by $\theta^z_L=\theta_R^z$ and
color symmetry.

We suggest that the essential features of confinement are described by
nonzero $<\chi>$. For equal quark masses we propose
\be\label{2.17}
<\chi_{ij,ab}>=\frac{1}{\sqrt{24}}\chi_0\lambda^z_{ji}\lambda^z_{ab}
=\frac{1}{\sqrt6}\chi_0(\delta_{ia}\delta_{jb}-\frac{1}{3}
\delta_{ij}\delta_{ab})\ee
with real $\chi_0>0$. This expectation value is invariant under vectorlike
$SU(3)_V$ transformations by which identical
left and right flavor rotations and transposed color
rotations are performed with opposite angles.
More explicitly, they are given by $\theta^z_L=\theta^z_R=\epsilon
^{(z)}\theta^z_C\equiv\theta^z_V$
with $\epsilon^{(z)}=-1$ for $z=1,3,4,6,8$ and $\epsilon^{(z)}=1$
for $z=2,5,7$. The $SU(3)_V$-transformation properties
of the various fields follow by inserting in eqs. (\ref{2.2}), (\ref{2.4})
and (\ref{2.5}) $\Theta_L=\Theta_R=-(\Theta_C)^T\equiv\Theta_V$
\bear\label{2.18}
\delta_V\psi_{ai}&=&i(\Theta_V)_{ab}\psi_{bi}-i\psi_{aj}(\Theta_V)_{ji}
\nonumber\\
\delta_V A^T_{ij,\mu}&=&i(\Theta_V)_{ik}A^T_{kj,\mu}-iA^T_{ik,\mu}
(\Theta_V)_{kj}\nonumber\\
\delta_V\Phi_{ab}&=&i(\Theta_V)_{ac}\phi_{cb}-i\Phi_{ac}(\Theta_V)_{cb}
\nonumber\\
\delta_V\chi_{ij,ab}&=&i(\Theta_V)_{ac}\chi_{ij,cb}-i\chi_{ij,ac}
(\Theta_V)_{cb}\nonumber\\
&&-i(\Theta_V)_{ki}\chi_{kj,ab}+i\chi_{ik,ab}(\Theta_V)_{jk}\ear
The invariance of the expectation value
$\delta_V<\chi_{ij,ab}>=0$ is easily checked.
With respect to
the transposed color rotations the quarks behave as antitriplets.
In consequence, under the combined
transformation they  transform as an octet plus a singlet,
with all quantum numbers identical to the baryon octet/singlet.
This is most obvious if we also use a
matrix notation for the fermions $\psi\equiv\psi_{ai}$ such
that $\delta_V\psi=i[\Theta_V,\psi]$.  In
particular, the electric charges, as given by the generator $Q=
\frac{1}{2}\lambda_3+\frac{1}{2\sqrt3}\lambda_8$ of $SU(3)_V$, are integer.
We therefore identify the quark field with the lowest baryon
octet and singlet -- this is quark-baryon duality. Similarly,
the gluons transform as an octet of vector-mesons, again with
the standard charges for the $\rho, K^*$ and $\omega/\phi$ mesons.
We therefore describe these vector mesons by the gluon field --
this is gluon-meson duality.

Due to the Higgs mechanism all gluons acquire a mass. For equal
quark masses conserved $SU(3)_V$ symmetry implies that all masses are
equal. They should be identified with the average mass of the vector
meson multiplet
$\bar M_\rho^2=\frac{2}{3} M^2_{K^*}+\frac{1}{3}M^2_\rho=(850\
{\rm MeV})^2$. One finds
\be\label{2.18a}
\bar M_\rho=g\chi_0=850\ {\rm MeV}\ee
The breaking of chiral symmetry by $\sigma_0$ and
$\chi_0$ also induces masses for the baryon octet and singlet.
For $\chi_0\not=0$ the singlet mass is larger than the octet mass
(see below).

We note that gluon-meson duality is a necessary consequence
of the Higgs picture of QCD. It persists if additional vector-meson
fields are introduced as color singlet quark-antiquark composite fields.
There will simply be a mixing between the ``gluons'' and the
``composite vector fields'' (see sect. 8). On the other hand, quark
baryon duality strongly depends on the proper assignment of the baryon
number. We will discuss this issue in detail in sect. 6.

\section{Integer electric charges}
\setcounter{equation}{0}

In order to get familiar with the Higgs picture of QCD, it seems
useful to understand in more detail the consequences of
spontaneous color symmetry breaking for the electromagnetic
interactions of hadrons. We start by adding to the effective
Lagrangian (\ref{2.6}) a coupling to a $U(1)$-gauge field $\tilde B_\mu$
by making derivatives covariant
\bear\label{3.1}
D_\mu\psi&=&(\partial_\mu-i\tilde e\tilde Q\tilde B_\mu)\psi\nonumber\\
D_\mu\gamma_{ij}&=&\partial_\mu\gamma_{ij}-i\tilde e[
\tilde Q,\gamma_{ij}]\tilde B_\mu-...\ear
with $\tilde Q=\frac{1}{2}\lambda_3+\frac{1}{2\sqrt3}\lambda_8
=diag(\frac{2}{3},-\frac{1}{3}, -\frac{1}{3})$
acting on the flavor indices. Furthermore we supplement the Maxwell
term
\be\label{3.2}
{\cal L}_{\tilde B}=\frac{1}{4}\tilde B^{\mu\nu}\tilde B_{\mu\nu}
\ , \ \tilde B_{\mu\nu}=\partial_\mu\tilde B_\nu-\partial_\nu
\tilde B_\mu\ee
Whereas the quarks carry fractional $\tilde Q$, the abelian
charges of the scalars are integer. In particular, the expectation value
$<\phi>$ is neutral, $[\tilde Q,<\phi>]=0$, whereas some components
of $<\chi_{ij,ab}>$ carry charge, namely for $k=2,3$
\bear\label{3.3}
{}[\tilde Q,<\chi_{1k,1k}>]&=&<\chi_{1k,1k}>\nonumber\\
{}[\tilde Q,<\chi_{k1,k1}>]&=&-<\chi_{k1,k1}>\ear
The expectation value (\ref{2.17}) therefore also breaks
the local $U(1)$ symmetry associated with $\tilde Q$. The abelian color
charge $(Q_C)_{ij}=\frac{1}{2}
(\lambda_3)_{ij}+\frac{1}{2\sqrt3}(\lambda_8)_{ij}$ of these fields
is, however, equal to $\tilde Q$. In consequence, a local abelian
symmetry with generator $\tilde Q-Q_C$ remains unbroken
\be\label{3.4}
\tilde Q_{ac}<\chi_{ij,cb}>-<\chi_{ij,ac}>\tilde Q_{cb}-
(Q_C)_{il}<\chi_{lj,ab}>+<\chi_{il,ab}>(Q_C)_{lj}=0\ee
The corresponding gauge field corresponds to the photon.

The situation encountered here is completely analogous to
the Higgs mechanism in electroweak symmetry breaking. The mixing
between the hypercharge boson and the $W_3$ boson in the electroweak
theory appears here as a mixing between $\tilde B_\mu$ and a particular
gluon field $\tilde G_\mu$ which corresponds to $A_{ij,\mu}=\frac{\sqrt
3}{2}(Q_C)_{ij}\tilde G_\mu$. Let us restrict the discussion
to the gauge bosons $\tilde B_\mu$ and $\tilde G_\mu$. Then the
covariant derivative for fields with a fixed value of $\tilde Q$ and
$Q_C$ is given by
\be\label{3.5}
D_\mu=\partial_\mu-i\tilde e\tilde B_\mu \tilde Q-i\tilde g\tilde
G_\mu Q_C\ee
with $\tilde g=\frac{\sqrt3}{2}g$. Due to the ``charged''
expectation values (\ref{3.3}) a linear combination of
$\tilde B_\mu$ and $\tilde G_\mu$ gets massive, as can be
seen from the quadratic Lagrangian
\be\label{3.6}
{\cal L}^{(2)}_{em}=\frac{1}{4}\tilde B^{\mu\nu}\tilde B_{\mu\nu}
+\frac{1}{4}\tilde G^{\mu\nu}\tilde G_{\mu\nu}+
\frac{2}{3}\chi^2_0(\tilde g\tilde G^\mu+\tilde e\tilde B^\mu)
(\tilde g\tilde G_\mu+\tilde e\tilde B_\mu)\ee
The massive neutral vector meson $R_\mu$ and the massless photon
$B_\mu$ are related to $\tilde G_\mu$ and $\tilde B_\mu$
by a mixing angle
\bear\label{3.7}
R_\mu&=&\cos\theta_{em}\tilde G_\mu+\sin\theta_{em}\tilde B_\mu\nonumber\\
B_\mu&=&\cos\theta_{em}\tilde B_\mu-\sin\theta_{em}\tilde G_\mu\nonumber\\
tg\theta_{em}&=&\frac{\tilde e}{\tilde g}\ear
and we note that the mass of the neutral vector meson is somewhat
enhanced by the mixing
\be\label{3.8}
M_{V_0}=g\chi_0/\cos \theta_{em}\ee
The mixing is, however, tiny for the large value $\bar\alpha
_s=g^2/4\pi\approx 3$ that we will find below. In terms of
the mass eigenstates the covariant derivative (\ref{3.5})
reads now
\be\label{3.9}
D_\mu=\partial_\mu-ieQB_\mu-i\tilde g\cos\theta_{em}(Q_C
+tg^2\theta_{em}\tilde Q)R_\mu\ee
and we observe the universal electromagnetic coupling
\be\label{3.10}
e=\tilde e\cos\theta_{em}\ee
of all particles with electric charge $Q=\tilde Q-Q_C$.
This coupling is exactly the same for the colored quarks and the
colorless leptons as it should be for the neutrality of atoms.
For an illustration we show the charges $Q_C,\tilde Q$ and $Q$
for the nine light quarks in table 1.
\begin{table}
\begin{center}
\begin{tabular}{l|ccc|c|}
&$\tilde Q$&$Q_c$&$Q$&\\
\hline
$u_1$&$2/3$&$2/3$&0&$\Sigma^0,\Lambda^0,S^0$\\
$u_2$&$2/3$&$-1/3$&1&$\Sigma^+$\\
$u_3$&$2/3$&$-1/3$&1&$p$\\
\hline
$d_1$&$-1/3$&$2/3$&-1&$\Sigma^-$\\
$d_2$&$-1/3$&$-1/3$&0&$\Sigma^0,\Lambda^0,S^0$\\
$d_3$&$-1/3$&$-1/3$&0&$n$\\
\hline
$s_1$&$-1/3$&$2/3$&-1&$\Xi^-$\\
$s_2$&$-1/3$&$-1/3$&0&$\Xi^0$\\
$s_3$&$-1/3$&$-1/3$&0&$\Lambda^0,S^0$
\end{tabular}\\
\end{center}

\medskip
{\em Table 1}: Abelian charges of quarks and association
with baryons. The baryon
singlet is denoted by $S^0$.
\end{table}

 We note the difference between
the coupling $\tilde e$ which would be computed in a grand
unified theory and the true electromagnetic coupling $e$. They are related
by
\be\label{3.11}
\frac{1}{\tilde e^2}=\frac{1}{e^2}-\frac{4}{3g^2}\ee
For $g=6$ the relative correction is only of the order of $10^{-3}$.

\section{Nonlinear meson interactions}
\setcounter{equation}{0}

The interactions of the light mesons are most easily described
in an equivalent gauge-invariant picture, using nonlinear
fields\footnote{On the level of the effective action the quantum
fluctuations are already included. Therefore Jacobians for nonlinear
field transformations play no role. All ``coordinate choices''
in the space of fields are equivalent.}. For the ``Goldstone directions''
we introduce unitary matrices $W_L, W_R, v$ and define
\bear\label{4.1}
\psi_L&=&Z_\psi^{-1/2}W_LN_Lv\ ,\ \psi_R=Z_\psi^{-1/2}W_RN_Rv,\nonumber\\
\bar\psi_L&=&Z_\psi^{-1/2}v^\dagger\bar N_LW^\dagger_L\ , \ \bar\psi_R
=Z_\psi^{-1/2}v^\dagger\bar N_RW_R^\dagger\nonumber\\
A_\mu&=&-v^TV^T_\mu v^*-\frac{i}{g}\partial_\mu v^Tv^*\ ,\
U=W_RW_L^\dagger,\nonumber\\
\phi&=&W_RSW^\dagger_L\ , \ \chi_{ij,ab}=(W_R)_{ac}v^T_{ik}
X_{kl,cd}v^*_{lj}(W^\dagger_L)_{db}
\ear
Here we have again
extended the matrix notation to the color indices
with the quark/baryon nonet represented by a complex $3\times 3$ matrix
$\psi\equiv\psi_{ai},\bar\psi\equiv\bar\psi_{ia}$. The decomposition
of $\phi$ is such that $S$ is a hermitean matrix. The
restrictions on $X$, which are necessary to avoid double counting,
play no role in the present work.
Besides $v$ all nonlinear
fields are color singlets. The field $v^\dagger$ transforms
as a color anti-triplet similar to a quark-quark pair or diquark.
In consequence, the baryons $N\sim \psi v^\dagger$ transform as a color
singlet.
More precisely, the $SU(3)_C\times SU(3)_L
\times SU(3)_R$ transformations of the nonlinear fields are given by
\bear\label{4.2}
\delta W_L&=&i\Theta_LW_L-iW_L\Theta_P\ ,\ \delta W_R=i\Theta_R
W_R-iW_R\Theta_P\ ,\nonumber\\
\delta v&=&i\Theta_P v+iv\Theta_C^T\ ,\ \delta U=i\Theta_R U-iU\Theta_L\nonumber\\
\delta N_L&=&i[\Theta_P,N_L]\ ,\ \delta N_R=i[\Theta_P,N_R]\ ,\
\delta V_\mu=i[\Theta_P,V_\mu]+\frac{1}{g}\partial_\mu\Theta_P,
\nonumber\\
\delta S&=&i[\Theta_P,S]\ ,\ (\delta X)_{ij,ab}=i(\Theta_P)_{ac}X_{ij,cb}-iX
_{ij,ac}(\Theta_P)_{cb}\nonumber\\
&&\qquad\qquad\qquad\qquad\ \ -i(\Theta_P^T)_{ik}X_{kj,ab}+iX_{ik,ab}
(\Theta_P^T)_{kj}
\ear

We observe the appearance of a new local symmetry $U(3)_P=
SU(3)_P\times U(1)_P$ under
which the nucleons $N$ and the vector meson fields  $V_\mu$ transform
as octets and singlets
\be\label{4.2AA}
\Theta_P=\frac{1}{2}(\theta_P^z(x)\lambda_z+\theta^0_P(x)\lambda_0\ ,\
\lambda_0\equiv\frac{2}{\sqrt6}.\ee
This symmetry acts only on the nonlinear fields whereas
$\psi, A_\mu, \phi$ and $\chi$ are invariant. It
reflects the possibility of local reparametrizations of the
nonlinear fields. It will play the role of the hidden gauge symmetry
underlying vector dominance. Furthermore, there is an additional local
abelian symmetry $U(1)_N$ (not shown in eq. (\ref{4.2})
\be\label{ZA9}
v\to e^{-i\alpha(x)}v\ ,\ N\to e^{i\alpha(x)}N\ ,\
V_\mu\to V_\mu-\frac{1}{g}\partial_\mu \alpha(x)\ee
and a global symmetry (with charge denoted by $B_V$)
\be\label{4.3a}
\psi\to e^{i\frac{\gamma}{3}}\psi\ ,\ W_{L,R}\to e^{i\frac{\gamma}{3}}
W_{L,R}\ee

For the present investigation we omit the scalar
excitations except for the ``Goldstone directions'' contained in
$W_L, W_R, v$. We can therefore replace $X_{kl,ab}$ by the expectation
value (\ref{2.17}) and use $<S>=\sigma_0$ such that
\be\label{4.3}
\Phi=\sigma_0U\quad,\quad
\chi_{ij,ab}=\frac{1}{\sqrt6}\chi_0\{(W_Rv)_{ai}(v^\dagger
W^\dagger_L)_{jb}
-\frac{1}{3}U_{ab}\delta_{ij}\}\ee
In terms of the nonlinear field coordinates the Lagrangian
(\ref{2.6}) reads
\bear\label{4.4}
{\cal L}&=&Tr\{i\bar N\gamma^\mu(\partial_\mu
+iv_\mu-i\gamma^5 a_\mu)N-g\bar N\gamma^\mu NV_\mu
\}\nonumber\\
&&+\frac{1}{2}Tr\{V^{\mu\nu} V_{\mu\nu}\}
+g^2\chi^2_0\ Tr\ \{\tilde V^\mu \tilde V_\mu\}
+(\frac{2}{9}\nu'\chi_0^2\sigma_0-\nu\sigma^3_0)\cos\theta\nonumber\\
&&+(
h\sigma_0-\frac{\tilde h}{3\sqrt6}\chi_0)
\ Tr\{\bar N \gamma_5N\}+\frac{\tilde h}{\sqrt6}\ Tr\ \bar N\gamma^5\ Tr\
N\nonumber\\
&&+(\sigma^2_0+\frac{7}{36}\chi_0^2)\ Tr\ \{\partial^\mu U^\dagger
\partial_\mu
U\}+\frac{1}{12}\chi^2_0
\ \partial^\mu\theta\ \partial_\mu\theta\nonumber\\
&&+\chi_0^2\ Tr\{\tilde v^\mu\tilde v_\mu\}+2g\chi^2_0\ Tr\{\tilde V
^\mu\tilde v_\mu\}\ear
with
\bear\label{4.4a}
&&v_\mu=-\frac{i}{2}(W^\dagger_L\partial_\mu W_L+W_R^\dagger
\partial_\mu W_R),\nonumber\\
&&a_\mu=\frac{i}{2}(W_L^\dagger\partial_\mu W_L-W_R^\dagger\partial_\mu W_R)
=-\frac{i}{2}W_R^\dagger \partial_\mu UW_L,\nonumber\\
&&Tr\ v_\mu=-\frac{i}{2}\partial_\mu(\ln \det\ W_L+\ln\det W_R),
\nonumber\\
&&Tr\ a_\mu=-\frac{i}{2}Tr\ U^\dagger\partial_\mu  
U=-\frac{1}{2}\partial_\mu\theta,\nonumber\\
&&V_{\mu\nu}=\partial_\mu V_\nu-\partial_\nu V_\mu \ -ig[V_\mu,V
_\nu]=\partial_\mu\tilde V_\nu-\partial_\nu\tilde V_\mu-
ig[\tilde V_\mu,\tilde V_\nu],\nonumber\\
&&\tilde V_\mu=V_\mu-\frac{1}{3}\ Tr\ V_\mu\ ,\ \tilde v_\mu=v_\mu-\frac{1}
{3}\ Tr\ v_\mu\ear
As it
should be, ${\cal L}$ does not depend on $v$ and is therefore invariant
under $SU(3)_C$ transformations. For a check of local $U(3)_P$
invariance we note
\be\label{4.4b}
\delta v_\mu=i[\Theta_P,v_\mu]-\partial_\mu\Theta_P\ ,\
\delta a_\mu=i[\Theta_P,a_\mu]\ee
such that $v_\mu$ transforms the same as $-gV_\mu$ and
$v_\mu+gV_\mu$ transforms homogeneously. One observes that the
mass term for $\tilde V_\mu$ appears in the
$U(3)_P$-invariant combination $\chi_0^2\ Tr\{(\tilde v^\mu+g\tilde
V^\mu)(\tilde v_\mu+g\tilde V_\mu)\}$.
The nonlinear field  $U$ only appears
in derivative terms except for the phase $\theta$ associated
to the $\eta'$-meson. We associate $U$ in the standard
way with the pseudoscalar octet of Goldstone bosons $\Pi^z$
\be\label{4.5}
U=\exp(-\frac{i}{3}\theta)\exp\left(i\frac{\Pi^z\lambda_z}{f}\right
)={\exp}\left(-\frac{i}{3}\theta\right)\tilde U\ee
where the decay constant $f$ will be specified below.

>From eq. (\ref{4.4}) one can directly read the average
mass of the baryon octet and singlet\footnote{The lowest state
with the appropriate quantum numbers for the singlet is $\Lambda$(1600).
Since we expect a very broad decay width of the singlet this
assignment is very uncertain and we add in brackets a somewhat higher
value.}
\be\label{4.6}
M_8=h\sigma_0-\frac{\tilde h}{3\sqrt6}\chi_0=
1.15\ {\rm GeV}\ ,\ M_1=h\sigma_0+\frac{8}{3}\frac{\tilde h}{\sqrt6}\chi_0
=1.6(1.8)\ {\rm GeV}
\ee
The singlet-octet mass splitting is proportional to the octet
condensate $\chi_0$
\be\label{4.7}
M_1-M_8=\frac{3}{\sqrt6}\tilde h\chi_0=450(650)\ {\rm MeV}\ee
With eq. (\ref{2.18a}) this determines the ratio
\be\label{4.7a}
\frac{\tilde h}{g}=0.43(0.62)\ee
Similarly, one finds
\be\label{4.7b}
h\sigma_0=\frac{1}{9}(8M_8+M_1)=1.2(1.22)\ {\rm GeV}\ee

\section{Local reparametrization symmetry}
\setcounter{equation}{0}
Before extracting the couplings between the physical mesons and
the baryons we have to understand the role of the local
reparametrization transformations $U(3)_P=SU(3)_P\times U(1)_P$.
First of all, we note that $Tr\  v_\mu$ can be eliminated by $U(1)_P$-gauge
transformations and is therefore a pure gauge degree of freedom.
Similarly, $Tr\ V_\mu=-\frac{i}{g}\partial_\mu\ln\ \det\ v$
is the gauge degree of freedom corresponding to $U(1)_N$ transformations.
This explains why $Tr\  v_\mu$ and $Tr\  V_\mu$ only appear
in the term bilinear in the nucleon fields. The vector fields
$\tilde V_\mu$ are the gauge bosons of $SU(3)_P$. The nucleons
transform as an octet and a singlet under $SU(3)_P$ and the same
holds for the bilinear $a_\mu$. The fields contained
in $W_L,W_R$ are antitriplets and $U$ is a singlet.

In a gauge-fixed version the vacuum corresponds to $W_L=W_R=v=1$.
The remaining global symmetry\footnote{The abelian part of the symmetry
will be discussed in detail in sect. 6.} $SU(3)_{\tilde V}$
is a linear combination of $SU(3)_L, SU(3)_R, SU(3)_C$ and
$SU(3)_P$ given by transformations obeying
\be\label{5.1}
\Theta_{\tilde V} =\Theta_L=\Theta_R=-(\Theta_C)^T=\Theta_V
=\Theta_P\ee
With respect to $SU(3)_{\tilde V}$ the nonlinear fields and
bilinears $W_L, W_R, v$,\\ $ U, N, V_\mu,\\ a_\mu, v_\mu$
transform all as octets or singlets. On the level of the nonlinear
fields the global $SU(3)_{\tilde V}$ symmetry plays
the same role as $SU(3)_V$ for the linear fields. One may identify
$SU(3)_V$ and $SU(3)_{\tilde V}$ by specifying the
$SU(3)_V$-transformations of the nonlinear fields by the choice
$\Theta_P=\Theta_V$.

For a discussion of the mesons we have to eliminate the
redundance of the local $U(3)_P$-transformations.
We first argue that the appropriate choice of a gauge fixing
requires some care. As an example, one may explore the possible
gauge fixing $W_R=v=1$. Then $U=W_L^\dagger,
\ v_\mu=-a_\mu=-\frac{i}{2}U\partial_\mu U^\dagger=\frac{i}{2}
\partial_\mu UU^\dagger$ would yield
\be\label{5.2}
\frac{7}{36}Tr\{\partial^\mu U^\dagger \partial_\mu U\}+\frac{1}{12}
\partial^\mu\theta\ \partial_\mu\theta+ Tr\{\tilde v^\mu\tilde v_\mu\}
=\frac{4}{9}Tr\{\partial^\mu U^\dagger\partial_\mu U\}\ee
\be\label {5.3}
2Tr\{\tilde V^\mu\tilde v_\mu\}=-iTr\{U\partial_\mu U^\dagger V^\mu\}\ee
This is the language used\footnote{The
field $V_\mu$ in ref. \cite{W1} corresponds to
$-V_\mu ^T$ in the present notation.}
in ref \cite{W1}. If one would associate
the physical pions with $\Pi_z$, one would obtain for the decay constant
$f=2\sqrt{\sigma_0^2+4\chi_0^2/9}$. We note, however, that in this
language the vector bilinear $v_\mu$ has a contribution
linear in the pion fields, $v_\mu=-(1/2f)\lambda_z\partial_\mu
\Pi^z+\partial_\mu\theta/6+...$. The term (\ref{5.3})
leads therefore to a field mixing in quadratic oder $\sim  
\partial^\mu\Pi^zV^z_\mu$. We conclude that the propagators for
$\Pi^z$ and $V^z_\mu$ are not in standard diagonal form
in this gauge. The corresponding fields can therefore not be associated
with physical particles in this language. This point was overlooked
in ref. \cite{W1}.

On the other hand, the gauge choice
\be\label{5.3a}
W_L^\dagger=W_R=\xi\ ,\ v=1\ee
implies
\bear\label{5.4}
 &&U=\xi^2\ ,\ v_\mu=-\frac{i}{2}(\xi^\dagger \partial_\mu
\xi+\xi\partial_\mu\xi^\dagger),\nonumber\\
&&a_\mu=-\frac{i}{2}(\xi^\dagger\partial_\mu\xi-\xi\partial_\mu
\xi^\dagger)=-\frac{i}{2}\xi^\dagger\partial_\mu U\xi^\dagger,\nonumber\\
&&Tr v_\mu=0\ ,\quad Tr V_\mu=0\nonumber\\
&&\phi=\sigma_0U\ ,\ \xi_{ij,ab}=\frac{1}{\sqrt6}\chi_0(\xi_{ai}\xi_{jb}-
\frac{1}{3}U_{ab}\delta_{ij}),\nonumber\\
&&(D_\mu\chi)_{ij,ab}=\frac{1}{\sqrt6}\chi_0.\\
&&\qquad [\partial_\mu(\xi_{ai}\xi_{jb}
-\frac{1}{3}U_{ab}\delta_{ij})
+ig((\xi\tilde V_\mu)_{ai}\xi_{jb}-\xi_{ai}(\tilde V_\mu\xi)_{jb})]\nonumber\ear
This gauge is singled out by the fact that $v_\mu$ contains
no term linear in $\Pi^z$ or $\theta$
\be\label{5.5}
v_\mu=\tilde v_\mu=-\frac{i}{2f^2}[\Pi,\partial_\mu\Pi]=
-\frac{i}{8f^2}[\lambda_y,\lambda_z]\Pi^y\partial_\mu
\Pi^z+...\ee
where we have chosen the parametrization
\bear\label{X12}
&&\xi=\tilde\xi\exp\left(-\frac{i}{6}\theta\right)\ ,
\ det \tilde\xi=1\nonumber\\
&&\tilde\xi=\exp\left(\frac{i}{2f}\Pi^z\lambda_z\right)
=\exp\left(\frac{i}{f}\Pi\right)\ear
As for any gauge where $v_\mu$ contains no term linear in $\Pi$
or $\theta$, the term $\sim Tr\tilde v^\mu\tilde V_\mu$ contains
only interactions and does not affect the propagators. We can
therefore associate $\Pi^z$ with the physical pions. Also the
term $Tr\tilde v^\mu\tilde v_\mu$ involves at least
four meson fields and is irrelevant for the propagator of the pseudoscalar
mesons.

The bilinear $a_\mu$ reads in the gauge (\ref{5.3a})
\be\label{5.13}
a_\mu=\frac{1}{2f}\lambda_z\partial_\mu\Pi^z-\frac{1}{6}\partial_\mu\theta
+O(\Pi^3,..)\ee
The effective action (\ref{4.4}) contains therefore a cubic coupling
between two baryons and a pion. As it should be for pseudo-Goldstone
bosons, this is a derivative coupling. Actually, the advantage
of the gauge (\ref{5.3a}) can also be understood from the point
of view of the discrete symmetries $P$ and $C$. A simple realisation
of the discrete transformations on the level of
nonlinear fields is\footnote{Note that in our matrix notation
$C$ acts as $\psi_L\to c\bar\psi_R^T,\ \psi_R\to-c\bar\psi^T_L$.}
\bear\label{5.14}
P&:&N_L\to -N_R,\ N_R\to N_L,\nonumber\\
&&W_L\to W_R,\ W_R\to W_L,\ v\to v,\ U\to U^\dagger\ ,\
\xi\to\xi^\dagger\nonumber\\
&&v_\mu\to v_\mu,\ a_\mu\to -a_\mu,\ V_\mu\to V_\mu\nonumber\\
&&\nonumber\\
C&:& N_L\to c\bar N^T_R,\ N_R\to-c\bar N^T_L,\nonumber\\
&&W_L\to W_R^*,\ W_R\to W_L^*,\ v\to v^*,\ U\to U^T
\ ,\ \xi\to\xi^T\nonumber\\
&&v_\mu\to-v_\mu^T,\ a_\mu\to a_\mu^T,\ V_\mu\to -V_\mu^T\ear
The gauge condition (\ref{5.3a}) can also be
written as $W_LW_R=W_RW_L=1$ and is manifestly invariant under $P$ and $C$.
The coupling ${\rm Tr}\bar N\gamma^\mu\gamma^5a_\mu N$ is a standard
coupling of baryons to the axial vector current. In contrast,
the gauge $W_R=1$ is not compatible with
the transformation (\ref{5.14}). (For this gauge
there exists an alternative
realization of $P$ and $C$ on the level of nonlinear fields.
Under these modified $P$ and $C$ transformations $a_\mu$ and
$v_\mu$ do not transform, however, as axial vectors and vectors,
respectively. This can be seen from the identification $v_\mu=-a_\mu$
in this gauge.) Having found an appropriate gauge fixing for the
reparametrization symmetry, we may now directly proceed to the
analysis of the physical content of the effective action
(\ref{4.4}).

The kinetic term for the pseudoscalar mesons can be inferred
from
\bear\label{5.6}
{\cal L}^{(P)}_{kin}&=&(\sigma_0^2+\frac{7}{36}\chi_0^2)\ Tr\{\partial
^\mu U^\dagger\partial_\mu U\}+\frac{1}{12}\chi^2_0\ \partial^\mu
\theta\ \partial_\mu\theta\nonumber\\
&=&\frac{2}{f^2}(\sigma_0^2+\frac{7}{36}\chi^2_0)\ \partial^\mu\Pi^z
\partial_\mu\Pi_z+\frac{1}{3}(\sigma_0^2+\frac{4}{9}\chi_0^2)\partial^\mu
\theta\partial_\mu\theta
\ear
Its standard normalization fixes a combination of $\sigma_0$ and $\chi_0$
in terms of
the pseudoscalar decay constant
$f$ which corresponds to an average in the octet \cite{8}
$f=\frac{2}{3}f_K+\frac{1}{3}f_\pi=106$ MeV. For an optimal
quantitative estimate of the expectation values $\sigma_0, \chi_0$
the partial Higgs effect should be included.
As we show in
appendix A, the
mixing between the pseudoscalars contained in $U\ (\sim i\bar\psi\psi)$ and
those in the divergence of the axial-vector current  
$(\sim\partial_\mu(\bar\psi\gamma^\mu\gamma^5\psi)$)
introduces an additional negative contribution (partial Higgs effects)
\be\label{5.8a}
\Delta{\cal L}_{kin}^{(P)}=-\frac{\Delta^2_f}{4}
{\rm Tr}\{\partial^\mu \tilde U^\dagger \partial_\mu \tilde U\}
-\Delta^2_\theta \partial
^\mu\theta \partial_\mu\theta\ee
One infers a standard renormalization of the kinetic term for
\be\label{5.8b}
f^2=4\sigma^2_0+\frac{7}{9}\chi^2_0-\Delta^2_f=\kappa^2_f
(4\sigma_0^2+\frac{7}{9}\chi^2)\ee
with
\be\label{5.8c}
\kappa_f=(1-\frac{\Delta_f^2}{4}(\sigma_0^2+\frac{7}
{36}\chi^2_0)^{-1})^{1/2}=(1+\frac{\Delta_f^2}{f^2})^{-1/2}\leq1\ee
The deviation of $\kappa_f$ from one is one of the most
important effects of effective interactions not included in eq. (\ref{2.6}).
The effective interactions responsible for $\Delta^2_f$ also induce
$SU(3)$-violating contributions to the pseudoscalar wave function
renormalizations related to the strange quark mass\footnote{In the
notation of ref. \cite{8} one has  
$\Delta^2_f/f^2=-X_\phi^-\bar\sigma_0^2Z_m^{-1}$.}. This has led
to a phenomenological estimate \cite{8}
\be\label{5.8d}
\Delta^2_f\approx 0.45 f^2,\ \kappa_f\approx 0.83\ee
The pseudoscalar decay constants provide now for a quantitative
estimate of the expectation values $\sigma_0$ and $\chi_0$
\be\label{5.8}
f_0=2\sqrt{\sigma^2_0+\frac{7}{36}\chi^2_0}=f/\kappa_f=(\frac{2}{3}
f_K+\frac{1}{3}f_\pi)/\kappa_f=128\ {\rm MeV}\ee
In the approximation (\ref{2.6}) of scalar Vervollst\"andigung
one should use the leading order relation $f=f_0$, keeping in mind
that nonleading effects lower $f$ to its physical value.

We may also use this relation for a first estimate of
the effective gauge coupling $g$. Let
us denote the relative size of the octet and singlet contributions
to the squared meson decay constant by
\be\label{5.9}
x=\frac{7}{36}\frac{\chi^2_0}{\sigma_0^2}\quad, \quad \sigma_0=
\frac{f_0}{2\sqrt{1+x}}\ee
The combination of the relations (\ref{2.18a}) and (\ref{5.8})
yields a bound for the effective gauge coupling $g$. From
\be\label{5.10}
g=\frac{\sqrt7}{3}\left(\frac{1+x}{x}\right)^{1/2}\frac{\bar M_\rho}{f_0}
=5.9\left(\frac{1+x}{x}\right)^{1/2}\ee
one infers $g>5.9$. If the octet condensate dominates
(large $x$) the effective gauge coupling is $g\approx 6$.

The last field which needs a proper normalization is the $\eta'$-meson.
Including the partial Higgs effect (\ref{5.8a}), the
kinetic term for the pseudoscalar singlet has the
standard normalization for
\bear\label{5.10a}
\eta'&=&\left(\frac{2}{3}(\sigma_0^2+\frac{4}{9}\chi_0^2)-
2\Delta^2_\theta\right)^{1/2}\theta\nonumber\\
&=&\sqrt{\frac{2}{3}}(\sigma^2_0+\frac{4}{9}\chi_0^2)^{1/2}
\kappa_\theta\theta\ear
(In the leading approximation (\ref{2.6}) one should, of
course, use $\kappa_\theta=1$.) Inserting
(\ref{5.10a}) into  (\ref{4.4}) and expanding $\cos \theta$
in second order in $\theta$ one finds the mass of the
$\eta'$-meson
\be\label{5.12}
M^2_{\eta'}=\left(\frac{3}{2}\nu\sigma_0^2-\frac{1}{3}\nu'\chi_0^2
\right)(\sigma_0^2
+\frac{4}{9}\chi^2_0)^{-1}\kappa^{-2}_\theta
\sigma_0=\frac{3f_0}{4\kappa^2_\theta\sqrt{1+x}}\frac{(7\nu-
8x\nu')}{(7+16x)}.\ee
Its interactions can be extracted by inserting
\be\label{5.12a}
\theta=\frac{\sqrt6}{\kappa_\theta f_0}\left(\frac{1+x}
{1+\frac{16}{7}x}\right)
^{1/2}\ \eta'\ =\ \frac{1}{H_{\eta'}}\eta'\ee

In the gauge (\ref{5.3a}) the nonlinear Lagrangian (\ref{4.4})
can now be written in terms of the normalized physical fields
$N, \tilde V_\mu, \Pi$ and $\eta'$ as
\be\label{5.15}
{\cal L}={\cal L}_N+{\cal L}_U+{\cal L}_V\ee
with
\bear\label{5.16}
{\cal L}_N&=&i\ {\rm Tr}\{\bar N_8\gamma^\mu(\partial_\mu-i\gamma^5
\tilde a_\mu+\frac{i}{6H_{\eta'}}\gamma^5\partial_\mu\eta')N_8\}
+M_8 {\rm Tr}\{\bar N_8\gamma^5N_8\}\nonumber\\
&&+i\bar N_1\gamma^\mu(\partial_\mu+\frac{i}{6H_{\eta'}}
\gamma^5\partial_\mu\eta')N_1+M_1\bar N_1\gamma^5N_1\nonumber\\
&&+\frac{1}{\sqrt3}\ {\rm Tr}\{\bar N_8\gamma^\mu\gamma^5\tilde a_\mu\}N_1+
\frac{1}{\sqrt3}\bar N_1\gamma^\mu\gamma^5\ {\rm Tr}\{\tilde a_\mu N_8\}\ear
Here we use the octet and singlet fields
\bear\label{5.17}
&&N_1=\frac{1}{\sqrt3}N,\ N_8=N-\frac{1}{3}{\rm Tr} N=N-\frac{1}{\sqrt3}N_1
\ ,\
Tr N_8=0,\nonumber\\
&&\tilde a_\mu=a_\mu-\frac{1}{3}\ {\rm Tr}\  a_\mu=\frac{1}{2f}\lambda_z
\partial_\mu \Pi^z+O(\Pi^3)\ ,\ {\rm Tr}\ \tilde a_\mu=0\nonumber\\
&&a_\mu=
\tilde a_\mu-\frac{1}{6}\partial_\mu\theta=\tilde a_\mu-\frac{1}{6H_{\eta'}}
\eta'\ear
For the mesonic part one has
\be\label{5.18}
{\cal L}_U=\frac{f^2}{4}\ {\rm Tr}\{\partial^\mu\tilde U^\dagger
\partial_\mu\tilde U\}
+\frac{1}{2}\partial^\mu\eta'\partial_\mu\eta'-M^2_{\eta'}H^2_{\eta'}
\cos(\eta'/H_{\eta'})\ee
with
\be\label{5.21}
\tilde U=\tilde\xi^2=
\exp\left(i\frac{\Pi^z\lambda_z}{f}\right),\ \det \tilde U=1\ee
and
\bear\label{5.22}
{\cal L}_V&=&\frac{1}{2}\ {\rm Tr}\{\tilde V^{\mu\nu}\tilde V_{\mu\nu}\}
+\bar M^2_\rho\ {\rm Tr}\{\tilde V^\mu\tilde V_\mu\}+\frac{2}{g}\bar M^2_\rho
{\rm Tr}\{\tilde V^\mu\tilde v_\mu\}+\frac{1}{g^2}\bar M^2_\rho
{\rm Tr}\{\tilde v
^\mu\tilde v_\mu\}\nonumber\\
&&-g \ {\rm Tr}\{\bar N_8\gamma^\mu N_8\tilde V_\mu\}-\frac{g}{\sqrt3}(
{\rm Tr}\{\bar N_8\tilde V_\mu\}\gamma^\mu N_1+\bar N_1\gamma^\mu {\rm  
Tr}\{\tilde V_\mu
N_8\})\nonumber\\
&&-{\rm Tr}\{\bar N_8\gamma^\mu\tilde v_\mu N_8\}-\frac{1}{\sqrt3}
({\rm Tr}\{\bar N_8\tilde v_\mu\}\gamma^\mu N_1+\bar N_1\gamma^\mu
{\rm Tr}\{\tilde v_\mu N_8\})\ear
In the effective action (\ref{5.15})
we have replaced the free parameters by $M_8, M_1$,\\
$\bar M_\rho, M_{\eta'}
, f$ and $g$. We note that the number of parameters is reduced
to six since only one particular combination of $\nu$ and
$\nu'$ appears in $M_{\eta'}$.
We recall that we know the order of magnitude of $g$ by
eq. (\ref{5.10}), which relates also $g$ to $x$. The proportionality
constant $H_{\eta'}$ appearing in the couplings of
$\eta'$ is then known by eq. (\ref{5.12a}). The interactions between
physical baryons and mesons can be extracted by inserting the
explicit representations
\bear\label{5.23}
&&\Pi=\frac{1}{2}\Pi^z\lambda_z=\frac{1}{2}\left(
\begin{array}{ccc}
\pi^0+\frac{1}{\sqrt3}\eta\ ,&\ \sqrt2 \pi^+\ ,&\ \sqrt 2K^+\\
\sqrt2\pi^-\ ,&\ -\pi^0+\frac{1}{\sqrt3}\eta\ &\ \sqrt2 K^0\\
\sqrt2 K^-\ ,& \ \sqrt2\bar K^0\ ,&\ -\frac{2}{\sqrt3}\eta\end{array}
\right)\nonumber\\
&&\nonumber\\
&&\quad \tilde V_\mu=\frac{1}{2}\tilde V_\mu^z\lambda_z=\frac{1}{2}
\left(\begin{array}{ccc}
\rho^0_\mu+\frac{1}{\sqrt3}V^8_\mu\ ,&\ \sqrt2 \rho^+_\mu\ ,&\ \sqrt2K^{*+}_\mu\\
\sqrt2\rho^-_\mu\ ,&\ -\rho^0_\mu+\frac{1}{\sqrt3}V^8_\mu\ &\ \sqrt2 K^{*0}_\mu\\
\sqrt2 K^{*-}_\mu\ ,& \ \sqrt2\bar K^{*0}_\mu,&-\frac{2}{\sqrt3}V^8_\mu \end{array}
\right)\nonumber\\
&&\nonumber\\
&&N_8=\left(
\begin{array}{ccc}
\frac{1}{\sqrt2}\Sigma^0+\frac{1}{\sqrt6}\Lambda^0\ ,&\ \Sigma^+\ ,&\ p\\
\Sigma^-\ ,&\ -\frac{1}{\sqrt2}\Sigma^0+\frac{1}{\sqrt6}\Lambda^0\ ,& \ n\\
\Xi^-\ ,&\ \Xi^0\ ,&\ -\frac{2}{\sqrt6}\Lambda^0 \end{array}\right)
\nonumber\\
&&\nonumber\\
&&\bar N_8=\left(
\begin{array}{ccc}
\frac{1}{\sqrt2}\bar\Sigma^0+\frac{1}{\sqrt6}\bar\Lambda^0\ .&
\ \bar\Sigma^-\ ,&\ \bar\Xi^-\\
\bar \Sigma^+\ ,&\ -\frac{1}{\sqrt2}\bar\Sigma^0+\frac{1}{\sqrt6}
\bar\Lambda^0\ ,&\ \bar\Xi^0\\
\bar p\ ,&\ \bar n\ ,& \ -\frac{2}{\sqrt6}\bar\Lambda^0\end{array}
\right)
\ear
It is obvious that the effective action (\ref{5.15}) predicts a multitude
of different interactions between the low mass hadrons. The electromagnetic
interactions are incorporated by covariant derivatives. We will discuss
a few of these interactions in sects. 7-9 in order to check
the validity of the hypothesis of scalar Vervollst\"andigung.

Finally, the addition of the explicit chiral
symmetry breaking by the current
quark masses (\ref{2.8})
\be\label{5.28}
{\cal L}_j=-\frac{1}{2}Z_\phi^{-1/2}a_q\sigma_0\ {\rm Tr}\{\bar m(U+U^\dagger)\}\ee
leads to mass terms for the pseudoscalars \cite{Mass} and contributes
to their interactions.

\section{Baryon number}
\setcounter{equation}{0}

The correct assignment of the baryon number $B$ is the central
question for quark-baryon duality. Quarks carry $B=1/3$, whereas
baryons have $B=1$. How can this be reconciled with a dual picture
where one field describes quarks as well as baryons in
the appropriate momentum ranges? In principle, gluon-meson duality
(or the Higgs picture of QCD) is consistent with two alternatives.
In the first scenario the nonlinear fermion fields $N$ could transform
as octets with integer charge and $B=1/3$. In this case separate
baryon fields would be needed, with all quantum numbers identical
to the quark fields except for the baryon number. As in the
nonrelativistic quark model or the parton model a baryon field may
then be thought of as the composite of three quark fields. An important
puzzle would remain, however, in this scenario: Why should the
color singlet integer charged fermion octet field $N$ not
be associated to physical particles? We will argue here in favor
of the second alternative, namely that the nonlinear field $N$
indeed carries $B=1$ (and not $B=1/3$). It can therefore be
associated directly with the baryons. Possible relations of
this picture of quark-baryon duality with the nonrelativistic quark model
will be discussed in this section and sect. 11. The crucial
ingredient for the viability of this second scenario are the nontrivial
transformation properties of the various nonlinear fields.

The determination of the baryon number of the nonlinear fields
needs some care. To get familiar with the problem, we first discuss
the analogous problem for
the electric charge. The equivalence between the nonlinear language
and the Higgs picture discussed in sect. 3 requires the nonlinear
fields in $v$ to carry electric charge. Indeed, the transformation
laws
\be\label{6.1}
\delta_{em}\psi=i\beta\tilde Q \psi\ ,\ \delta_{em}\tilde B_\mu=\frac{1}
{\tilde e}\partial_\mu\beta\ ,\ \delta_{em}W_{L,R}=i\beta[\tilde Q
,W_{L,R}]\ ,\ \delta_{em} v=i\beta\tilde Q v\ee
guarantee that the quarks carry fractional electric charge
$\tilde Q=diag(\frac{2}{3},-\frac{1}{3},-\frac{1}{3})$
and the gluons are neutral, whereas the pions in $\xi$, the baryons
and the vector mesons are integer charged
\be\label{6.2}
\delta_{em}\xi=i\beta[\tilde Q,\xi]\ ,\ \delta_{em}N=i\beta[\tilde Q,N]\ ,\
\delta_{em}V_\mu=i\beta[\tilde Q,V_\mu]+\frac{1}{g}\partial_\mu\beta
\tilde Q\ee
Despite the fact that $v$ does not appear in the effective action
(\ref{4.5}), its nontrivial electric charge plays a crucial role in
the transition from the linear quark-gluon to the nonlinear baryon-meson
description.

The situation is similar for the baryon number, which we denote
by $\tilde B$ on the quark level, $\tilde B(\psi)=1/3$. The
field $v^\dagger$ transforms as a color anti-triplet. By triality
it should therefore carry baryon number $\tilde B=2/3$. The baryon
number, electric charge and color transformations of $v^\dagger$
coincide all with the ones for a diquark field. As a simple consequence,
the baryons $N$ carry $\tilde B=1$ as it should be, consistent
with triality. (Triality requires that the $SU(3)_C$ representations
in the classes $(1,8,...), (3, \bar 6, ...), (\bar 3,6,...)$ carry
$\tilde B=0,\frac{1}{3},\frac{2}{3}\ mod\  1$.) We note that $\chi$
is bilinear in $v$ and $v^\dagger$ and therefore carries $\tilde B=0$
as appropriate for a quark-antiquark composite. Besides the above
consistency considerations a direct determination of the baryon number
for $v$ becomes easy if some scalar fields are linear in $v$. In
sect. 11 we will see how the association of $v^\dagger$ with a
nonlinear diquark field arises naturally in a language where linear
diquark fields are introduced.

Let us next give the equivalent description of baryon number in
the Higgs picture. In a language with gauge-fixed
reparametrization invariance the condition $v=1$ preserves
a combination of $U(1)_P$ and baryon number transformation,
$2B_P(v)+\tilde B(v)=0$. Furthermore, $W_L$
and $W_R$ are neutral with respect to the abelian charge $B_P
+B_V$  (cf. eq. (\ref{ZA9}). After spontaneous symmetry
breaking the unbroken baryon number-type charge is therefore
\be\label{6.3}
B=\tilde B+2B_P+2B_V\ee
Both $\psi$ and $N$ have the same charge $B=1$. The Higgs
picture implies that the ``physical baryon number'' $B$ after
spontaneous symmetry breaking is three times\footnote{Note
that the normalization of $B$ is fixed by the requirement
that a hypothetical particle carrying baryon number without
strong interactions (and therefore $B_V=B_P=0$) should have
$B=\tilde B$.} the quark baryon number $\tilde B$. In table 2
we have listed the various abelian charges for the linear and
nonlinear fields, including the charge $B_N$ (cf. eq. (\ref{ZA9})).
We note the linear relation $\tilde B=B_P+B_V+B_N$.

\begin{center}
\begin{tabular}{c|cccccc}
& $\tilde B$&$B_P$&$\tilde B+2B_P$&$B_V$&$B$&$B_N$\\
\hline\\
$\psi$&$\frac{1}{3}$&0&$\frac{1}{3}$&$\frac{1}{3}$&1&0\\
$v^\dagger$&$\frac{2}{3}$&$-\frac{1}{3}$&0&0&0&1\\
$N$&1&0&1&0&1&1\\
$W_{L,R}$&0&$-\frac{1}{3}$&$-\frac{2}{3}$&$\frac{1}{3}$&0&
0\\
$U,S,X$& 0&0&0&0&0&0\\
$A_\mu,V_\mu$&0&0&0&0&0&0\\
\hline\\
\end{tabular}

Table 2: Abelian charges of linear and nonlinear fields
\end{center}
\medskip

In conclusion, the equivalence between the Higgs picture and the
nonlinear language with preserved local symmetries shows many
similarities between baryon number and electric charge. One
difference remains, however, in the Higgs picture. Whereas
some ``linear'' scalar expectation values $\chi$ carry
nonzero electric charge $\tilde Q$ (cf. eq. (\ref{3.3}) they
have all $\tilde B=0$. The ``spontaneous breaking'' of $\tilde B$
occurs only on the level of the nonlinear fields $v$ and
$W_{L,R}$ by the gauge fixing $v=1, \ W_L^\dagger=W_R$.

Let us next turn to the normalization of the fermion fields $\psi$ and
$N$ and the associated relative wave function renormalization $Z_\psi$.
In principle, we may normalize the fields according to some conserved
charge or to some standard convention for the kinetic term.
For the quark field $\psi$ we choose a normalization such that the
quark number operator $N_q$ has a standard form
\be\label{6.4}
N_q=-i\int d^3x\ {\rm Tr}\bar\psi \gamma^0\psi\ee
More precisely, the normalization specifies the way how a baryon-chemical
potential $\mu_B$ is introduced for situations with nonvanishing
baryon density. Since quarks carry $\tilde B=1/3$, we take
\be\label{6.5}
{\cal L}^{(\mu)}=\frac{i}{3}\mu_B\ {\rm Tr}\ \bar\psi\gamma^0\psi\ee
On the other hand, we have normalized the field $N$ such that its
kinetic term has a standard form\footnote{This should hold at least
in the vicinity of on-shell momenta.}
(i.e. $Z_N=1$). Since the baryons are ``physical fields'', we
assume that this also corresponds to a standard normalization of
the baryon number operator
\be\label{6.6}
N_B=-i\int d^3x\ {\rm Tr}\ \bar N\gamma^0N\ee
such that
\be\label{6.7}
{\cal L}^{(\mu)}=i\mu_B\ {\rm Tr}\ \bar N\gamma^0N\ee
The consistency of eqs. (\ref{6.5}) and (\ref{6.7}) with
the definition of the nonlinear field $N$ (\ref{4.1}) requires
\be\label{6.8}
Z_\psi=\frac{1}{3}\ee
This reflects in our language the fact that baryons carry
three times the baryon number of quarks. We emphasize that
the choice $Z_\psi=1/3$ is not a pure convention but rather a necessity if
$\psi$ and $N$ are normalized according to (\ref{6.5}), (\ref{6.7}).
Quark-baryon duality has then the interesting consequence
of a nontrivial wave function renormalization $Z_\psi$ in the
``quark number normalization'', since the kinetic terms for $\psi$ and
$N$ are directly related.

The wave function renormalization $Z_\psi$ is related to the full
quantum theory  in the low momentum limit, $Z_\psi\equiv Z_\psi(k\to0)=1/3$.
In contrast, we expect that for large $k$ a perturbative picture is valid,
with $Z_\psi(k\gg 2$ GeV) $\approx1$. This leads to a simple
speculation how quark-baryon duality can be reconciled with the
linear quark-meson model \cite{GM}, \cite{6}, \cite{8}
the nonrelativistic quark
model and the parton model. The quark model descriptions are valid
in the range of large enough $k$ (or large momenta) where
$Z_\psi(k)\approx 1$.
We speculate that around some scale $k_B$ the wave function
renormalization $Z_\psi(k)$ drops rapidly to its nonperturbative
value $Z_\psi(k\to0)=1/3$. This rapid drop is related to the
binding of three quarks to a baryon in the nonrelativistic quark
model. We therefore associate $k_B$ to a characteristic scale
where this binding takes place. The fact that $Z_\psi$ drops precisely
to the value 1/3 reflects the fact that three quarks are needed
for a baryon in the nonrelativistic quark model.

We may further speculate that the ``binding effect'' essentially
only affects the  quark wave function renormalization
$Z_\psi$ and other wave function  renormalizations related
to it by Ward identities. This has the
interesting consequence that the drop in $Z_\psi$ at $k_B$
results in a corresponding increase of those renormalized couplings which
involve negative powers of $Z_\psi$. As an example, one may take
the Yukawa couplings $h, \tilde h$ in the effective action (\ref{2.6}).
If the ``unrenormalized couplings''
$Z_\psi h$ and $Z_\psi\tilde h$ do not undergo a major change near
$k_B$ the Yukawa couplings will increase rapidly by a factor three
around this scale.
Neglecting the running of $Z_\psi h$ one obtains $h(k \stackrel
{\scriptstyle>}{\sim} k_B)=\frac{1}{3}h(k=0)$ and similar for $\tilde h$.
If also the expectation values $\sigma_0(k)$ and
$\chi_0(k)$ do not vary much in this range, the
effective fermion masses for $k \stackrel {\scriptstyle >}{\sim} k_B$
are one third the baryon masses. This is characteristic for models with
consituent quarks. A particular version of such models,
the linear quark meson model, has given a rather satisfactory description
of chiral symmetry breaking
for two flavors of light quarks\cite{6}. The relevant range for its dynamics
is 700 MeV $>k>k_B$, and one may wonder how it is related precisely
to our
scenario for $k=0$.

On the other hand, it has been pointed out \cite{7} that
fluctuations with momenta in the range $0<k<k_B$ play an important role
for the chiral properties at large baryon density. The transition
from effective nucleon degrees of freedom to quark
degrees of freedom is directly linked to the phase transition from a
hadron gas to nuclear matter at vanishing or low tempe\-ra\-ture. A change
of $Z_\psi$ by a factor of three would precisely account for the
enhancement of the contribution from baryon fluctuations as compared to
quark fluctuations to the dependence of the effective potential on the baryon  
chemical potential  \cite{7}.

Another interesting quantity is the running renormalized gauge coupling
$g(k)$. If we push perturbation theory to its limit we obtain
\be\label{6.9}
g{\rm (850\ MeV)} =\sqrt{4\pi\alpha_s{\rm (850\ MeV)}}
=2.26\ee
where the numerical value corresponds to the two-loop value in the
$\overline{MS}$-scheme\footnote{We use $\Lambda_{\rm QCD}$= 250 MeV.
The variation of $g$ in this momentum region
is still moderate, with $g(1 {\rm GeV})=2.12,\ g(700 \ {\rm MeV})
=2.48$.}. Neglecting the running of $Z_\psi g$ for scales below
the mass of the vector mesons and accounting for a drop
of $Z_\psi$ by a factor of three at $k_B$, this yields the
``perturbative estimates''
\bear\label{6.9a}
g^{(pt)}=6.8&,&\chi_0^{(pt)}=125\ {\rm MeV}\nonumber\\
x^{(pt)}=3.1&,&\sigma_0^{(pt)}=31\ {\rm MeV}\ear
With such an estimate the free parameters of our simple model
would be fixed in terms of $f$ and $\overline M_\rho$.

In order to test the consistency of
these ideas, it is instructive to couple
a hypothetical massless ``baryophoton'' $\tilde C_\mu$ to
the quarks and baryons. This would be needed if the abelian symmetry
corresponding to baryon number is gauged\footnote{In absence of
weak interactions baryon number is free of anomalies
and permits therefore a consistent local gauge symmetry.}.
According to the ``quark number'' normalization (\ref{6.4}),
the coupling of the hypothetical ``baryophoton'' is
\be\label{6.10}
{\LL}^{(C)}=\frac{1}{3}e_B\ {\rm Tr}\ \bar\psi\gamma^\mu\psi
\tilde C_\mu\ee
with $e_B$ some arbitrarily small gauge coupling
which may depend on $k$. With
$Z_\psi=1/3$ for $k=0$ this corresponds to a covariant derivative
$\partial_\mu-ie_B\tilde C_\mu$ for
the quark field $\psi$. Inserting the relation (\ref{4.1}) between
$\psi$ and $N$, the coupling (\ref{6.10}) becomes
\be\label{6.11}
{\LL}^{(C)}=\frac{1}{3Z_\psi}e_B\ {\rm Tr}
\ \bar N\gamma^\mu N\tilde C_\mu\ee
For $3Z_\psi=1$ this is indeed the expected coupling of the
``baryophoton'' to a baryon (cf. eq. (\ref{6.6})).

In conclusion, quark-baryon duality may well be compatible with
a picture where baryons are composed of three quarks. At least we
have found no obvious contradiction. The quantum numbers
match. A crucial ingredient for dynamical considerations
seems to be the value $Z_\psi=1/3$ for $k\to 0$
and $Z_\psi\approx1$ for $k\gg k_B$. A more detailed dynamical
understanding why and how the binding of three quarks to
a baryon in the nonrelativistic quark model is related
to the drop of $Z_\psi$ by a factor 1/3 in the language of
quark-baryon duality would be of great value. For the present
paper we take this as a working hypothesis and explore
phenomenological consequences of quark-baryon duality.
A discussion of linear diquark fields in sect. 11 will shed
a little more light on this issue.

\section{Electromagnetic interactions}
\setcounter{equation}{0}

As a first probe of our picture we may use the electromagnetic
interactions of mesons and baryons. The electromagnetic interactions
of the mesons are all contained in the covariant kinetic term for the scalars
\bear\label{E1}
(D^\mu\gamma_{ij,ab})^*D_\mu\gamma_{ij,ab}&=&
\chi^2_0\ {\rm Tr} \{(\hat v^\mu+g\tilde V^\mu-\tilde e
\tilde B^\mu\tilde Q)(\hat v_\mu+g\tilde V_\mu-\tilde e\tilde B_\mu\tilde
Q)\}\nonumber\\
&&+\frac{f^2}{4}\ {\rm Tr}\{(D^\mu U)^\dagger D_\mu U\}+\frac{1}{12}
\chi^2_0\ \partial^\mu\theta \partial_\mu\theta\ear
Here $D_\mu U=\partial_\mu U-i\tilde e\tilde B_\mu[\tilde Q,U]$ and the
covariant vector current reads
(with $\tilde D_\mu\xi=\partial_\mu\xi-i\tilde e\tilde B_\mu[\tilde Q,\xi])$
\bear\label{E2}
\hat v_\mu&=&-\frac{i}{2}(W_L^\dagger\tilde D_\mu W_L+W_R^\dagger
\tilde D_\mu W_R)\nonumber\\
&=&-\frac{i}{2}(\xi^\dagger\tilde D_\mu\xi+\xi\tilde D_\mu\xi^\dagger)\nonumber\\
&=&v_\mu-\frac{\tilde e}{2}\tilde B_\mu(\xi^\dagger\tilde Q\xi+\xi
\tilde Q\xi^
\dagger-2\tilde Q)\ear
Observing that both $\hat v_\mu$ and $g\tilde V_\mu-\tilde e\tilde
B_\mu\tilde Q$ transform homogeneously with respect to the
electromagnetic gauge transformations
\be\label{E2a}
\delta_{em}\hat v_\mu=i\beta[\tilde Q,\hat v_\mu],\ \delta_{em}
(g\tilde V_\mu-\tilde e\tilde B_\mu\tilde Q)
=i\beta[Q,(g\tilde V_\mu-\tilde e\tilde B_\mu\tilde Q)]\ee
the gauge invariance of (\ref{E1}) can be easily
checked. The particular combination
of vector currents in (\ref{E1}) is dictated by the combination of  
electromagnetic gauge invariance and local reparametrization symmetry.
Only this combination transforms homogeneously with respect
to both $U(3)_P$ local reparametrizations
and electromagnetic $U(1)$ gauge transformations. In particular, one
has
\be\label{E2b}
\delta(\hat v_\mu-\tilde e\tilde B_\mu\tilde Q)=i[\Theta_P,(\hat v_\mu-\tilde  
e\tilde B_\mu\tilde Q)]+i\tilde e \tilde B_\mu[\tilde Q,(\hat v_\mu-\tilde e
\tilde B_\mu\tilde Q)]-\partial_\mu \Theta_P-\partial_\mu
\beta\tilde Q\ee

The interactions (\ref{E1}) coincide with those of a ``hidden local
chiral symmetry'' \cite{Bando} if one replaces $\hat v_\mu$ by $v_\mu$.
The difference between our result and the ``hidden symmetry''
approach is due to the different electromagnetic transformation properties,
cf. eqs. (\ref{6.1}), (\ref{6.2}). If one restricts the discussion
to $\rho$-mesons and pions, $\tilde V_\mu=\frac{1}{2}\vec\rho_{V\mu}\vec\tau,  
\Pi=\frac{1}{2}\vec\pi\vec\tau$,
and neglects the difference between $\tilde e$ and $e$, one finds
for the term involving the vector mesons and currents
\bear\label{E3}
{\cal L}_{VV}&=&af^2_\pi\ {\rm Tr}(\hat v_\mu+\frac{1}{2}g_\rho\vec\rho_{V\mu}
\vec\tau-\frac{1}{2}e\tilde B_\mu\tau_3)^2\nonumber\\
&=&af^2_\pi\ {\rm Tr}\left[
\frac{1}{4f_\pi^2}((\vec\pi\times \partial_\mu\vec\pi)\vec\tau)+\frac{1}{2}g_\rho
(\vec\rho_{V\mu}\vec\tau)-\frac{1}{2}e\tilde B_\mu\tau_3
\right.\nonumber\\
&&\left.-\frac{e}{4f^2_\pi}\tilde  
B_\mu(\pi_3(\vec\pi\vec\tau)-(\vec\pi\vec\pi)\tau_3)
\right]^2+...
\ear
For the second equality
in eq. (\ref{E3}) we have only retained terms quadratic
in $\pi$ in an expansion of $\hat v_\mu$ and we have replaced $f$
by $f_\pi$.
The first expression in eq. (\ref{E3})
is actually more general than the result of the particular effective
action (\ref{2.6}) for which one has
\bear\label{E4}
g_\rho&=&g\nonumber\\
a&=&\frac{\chi_0^2}{f_\pi^2}=\frac{9}
{7\kappa_f^2}\frac{x}{1+x}\frac{f^2}{f^2_\pi}\approx 2.4\frac{x}{1+x}
\ear
In consequence of the symmetries, additional interactions will
change the values of a $a$ and $g_\rho$ without affecting the
structure of the invariant.
Such additional interactions will also generate new invariants
involving higher derivatives. We give an example in appendix A.

The contributions in eq (\ref{E3}) with canonical dimension
$\leq4$ can be written in the form
\bear\label{E5}
{\cal L}_{VV}&=&\frac{1}{2}M_\rho^2\vec\rho^\mu_V\vec\rho_{V\mu}-eg_
{\rho\gamma}\rho^\mu_{V3}\tilde B_\mu+\frac{1}{2}m_B^2\tilde B^\mu\tilde  
B_\mu\nonumber\\
&&+g_{\rho\pi\pi}\vec\rho_V^\mu(\vec\pi\times \partial_\mu\vec\pi)
+g^{(V)}_{\gamma\pi\pi}\tilde B^\mu
(\vec\pi\times \partial_\mu\vec\pi)_3\nonumber\\
&&+g_{\rho\gamma\pi\pi}\tilde B_\mu[(\vec\pi\vec\pi)\rho^\mu
_{V3}-(\vec\rho^\mu_V\vec\pi)\pi_3]\nonumber\\
&&-\frac{1}{2}ae^2\tilde B_\mu\tilde B^\mu[(\vec\pi\vec\pi-\pi^2_3]+...\ear
with
\bear\label{E6}
M^2_\rho&=&ag^2_\rho f_\pi^2\ ,\ g_{\rho\gamma}=ag_\rho f_\pi^2\ ,\
g_{\rho\pi\pi}=\frac{1}{2}ag_\rho,\nonumber\\
m_B^2&=&ae^2f_\pi^2\ ,\ g^{(V)}_{\gamma\pi\pi}
=-\frac{1}{2}ae\ ,\ g_{\rho\gamma\pi\pi}=\frac{1}{2}aeg_\rho\ear
We recover the very successful KSFR relation \cite{KSFR}, \cite{Bando}
\be\label{E7}
g_{\rho\gamma}=2f_\pi^2g_{\rho\pi\pi}\ee
which relates the decay $\rho\to 2\pi$ (with $g_{\rho\pi\pi}
\approx 6$, see next section) to the eletromagnetic properties
of the $\rho$-meson, in particular the decay $\rho_0\to e^+e^-$ (with
$\Gamma(\rho_0\to e^+e^-)=6.62$ keV and $g_{\rho\gamma}=0.12$ GeV$^2$).

The relation
\be\label{E8}
M_\rho^2=\frac{4}{a}g^2_{\rho\pi\pi}f^2_\pi\ee
requires $a=2.1$, implying
\be\label{7.10a}
\chi_0=135\ {\rm MeV}\ee
One may use eq. (\ref{E4}) for an estimate of the relative
octet contribution to the pion decay constant and eq. (\ref{2.18a})
or (\ref{5.12}) for an estimate of $g$
\be\label{7.10b}
x=7\quad, \quad \sigma_0=22\ {\rm MeV}\quad,\quad g=6.3\ee
It is striking how close these values are to the
``perturbative values'' (\ref{6.9a}).
This should, however, not be taken too literally in view of
possible substantial $SU(3)_V$-violating effects
from the nonzero strange quark mass. Also higher order operators
may affect the relation (\ref{E4}). Furthermore, one should
include the corrections from additional invariants (see appendix A).
A phenomenological
discussion including the properties of the axial-vector mesons
\cite{HSLB} favors $a\approx 1.64$. Relation (\ref{E4}) implies then
\be\label{7.10c}
x=2.16\quad,\quad \chi_0=119\ {\rm MeV}\quad, \quad \sigma_0=36\ {\rm
MeV}\quad, \quad g=7.1\ee
again close to the ``perturbative values'' (\ref{6.9a}). We point out
that the minimal
effective action (\ref{2.6}) with $\kappa_f=1$ and $f_\pi=f$ leads to a bound
$a<9/7$. We will discuss this issue and modifications of the relation
(\ref{E4}) in the next section and appendix A. The additional
invariants encountered in appendix A only affect the cubic and higher
vertices, but not the mass terms. In particular, the relation
\be\label{7.10d}
g_{\rho\gamma}=g\chi^2_0=\frac{\bar M^2_\rho}{g}\ee
will only be modified by $SU(3)$-violating effects. It can be used
for an independent estimate of $g$, yielding
\be\label{7.10e}
g=6\ee
close to (\ref{7.10b}) and the saturation of the bound implied by
eq. (\ref{5.10}).

The electromagnetic $\gamma\pi\pi$ and $\gamma\gamma\pi\pi$ vertices
also receive contributions from
\bear\label{E9}
\frac{1}{4}f^2_\pi{\rm Tr}\{D^\mu U^\dagger D_\mu U\}&=&\frac{1}{2}
\partial^\mu\vec\pi\partial_\mu\vec\pi\\
&&+e\tilde B^\mu(\vec\pi\times \partial_\mu\vec\pi)_3+\frac{1}{2}
e^2\tilde B_\mu\tilde B^\mu(\vec\pi\vec\pi-\pi^2_3)+...\nonumber\ear
One sees that for $a=2$ the two contributions (\ref{E5}) and (\ref{E9}) to
the direct $\gamma\pi\pi$ vertex cancel. The electromagnetic interactions
of the pions are then dominated by $\rho$-exchange (vector dominance),
in agreement with observation. Otherwise stated, our model leads
for the direct $\gamma\pi\pi$-coupling  to the realistic relation
\be\label{E10}
g_{\gamma\pi\pi}
=e\left(1-\frac{2g^2_{\rho\pi\pi}f^2_\pi}{M_\rho^2}\right)\ee
The vertex $\sim g_{\rho\gamma\pi\pi}$ contributes\footnote{This
vertex is absent in ref. \cite{Bando}.} to rare
decays like $\rho_0\to\pi^+\pi^-\gamma$, with
\be\label{E11}
g_{\rho\gamma\pi\pi}=eg_{\rho\pi\pi}\ee

We conclude that the electromagnetic interactions of the pseudoscalars
as well as the vector mesons can be considered as a successful test
of our simple model. The appearance of a local nonlinear reparametrization
symmetry is a direct consequence of the ``spontaneous breaking''
of color. This symmetry, combined with the simple effective action
(\ref{2.6}), has led to the KSFR relation (\ref{E7}) and to vector
dominance (\ref{E10}). At this stage the relations of the above
discussion should be taken with a 20-30 percent uncertainty. In
particular, the $SU(3)$-violation due to the nonzero strange quark
mass needs to be dealt with more carefully. Nevertheless, the
$\rho\to 2\pi$ decay and the electromagnetic decays are all
consistent and have allowed us a first determination of the size
of the octet condensate $\chi_0$, which we may take in the range between the
estimates (\ref{7.10a}) and (\ref{7.10c}), i.e.
119 MeV $< \chi_0<$ 135 MeV. The octet condensate is larger than the
singlet condensate and dominates the pseudoscalar decay constant $f$.

Finally, the connection of the above
discussion to the Higgs picture developed in sect. 3 is easily
established if one realizes that the  field
\bear\label{7.7}
G_\mu&=&-\sqrt3\ Tr\ \{\tilde Q\tilde V_\mu\}
=\sqrt3\ Tr\ \{\tilde Q(vA_\mu^T v^\dagger+\frac{i}{g}
\partial_\mu vv^\dagger)\}\nonumber\\
&=&-\left(\frac{\sqrt3}{2}\rho^0_\mu+\frac{1}{2}V^8_\mu\right)
\ear
is the nonlinear correspondence of $\tilde G_\mu$ in
sect. 3. It transforms inhomogeneously
\be\label{7.8}
\delta_{em} G_\mu=-\frac{2}{\sqrt3 g}\partial_\mu\beta=-\frac{1}{\tilde
g}\partial_\mu\beta\ee
such that the linear combinations (\ref{3.7}) $R_\mu=
\cos\theta_{em} G_\mu+\sin\theta_{em}\tilde B_\mu,B_\mu=\cos
\theta_{em}\tilde B_\mu-\sin\theta_{em}G_\mu$ have the
transformation properties of a heavy neutral boson and a photon
\be\label{7.9}
\delta_{em}B_\mu=\frac{1}{e}\partial_\mu\beta\ ,\ \delta_{em}
R_\mu=0\ee
Inserting in ${\LL}_V$ (\ref{5.22}) $\tilde V_\mu=
-\frac{\sqrt3}{2}G_\mu\tilde Q,\tilde v_\mu=0$ and adding
terms from covariant derivatives (\ref{3.1}) involving
$\tilde B_\mu$ (cf. (\ref{E3})), we recover ${\LL}^{(2)}_{em}$
(\ref{3.6}) with $\tilde G_\mu$ replaced by $G_\mu$. The
(extended) electromagnetic interactions of the baryon
octet\footnote{We omit here terms involving the baryon singlet
$N_1$ and take $\Pi=0$.} read
\bear\label{7.10}
{\LL}_{VN,0}&=&eB_\mu\ Tr\ \{\bar N_8\gamma^\mu[\tilde Q,N_8]\}\nonumber\\
&&+\tilde g\cos\theta_{em}R_\mu\ Tr\ \{\bar N_8\gamma^\mu
(N_8\tilde Q+tg ^2\theta_{em}\tilde QN_8)\}
\ear
One finds the standard coupling between the photon $B_\mu$
and the baryons according
to their charge. More generally, we conclude from
\bear\label{7.11}
&&\tilde e\tilde B_\mu=e(B_\mu+tg\theta_{em}R_\mu)\nonumber\\
&&\tilde g G_\mu=\tilde g\cos\theta_{em}R_\mu-eB_\mu\ear
that  the charged leptons have a small direct coupling to
a linear combination of the $\rho^0, \omega$ and $\phi$ vector
mesons corresponding to the term $e\  tg\theta_{em}R_\mu$ in
$\tilde e\tilde B_\mu$. Similarly, the photon has a
hadronic coupling from the term $-eB_\mu$ in $\tilde g G_\mu$.
These mixing effects are governed by the coupling $g_{\gamma\rho}$
(\ref{E7}).

\section{Vector mesons}
\setcounter{equation}{0}
The vector mesons acquire a mass through the Higgs mechanism.
They are also unstable due to the decay into two pseudoscalar
mesons. Their interactions are contained in ${\cal L}_V$ (\ref{5.22}).
In particular, the
cubic vertex between one vector meson and two pseudoscalars is
\bear\label{7.1}
{\cal L}_{V\pi\pi}&=&2g\chi_0^2\ {\rm Tr}\{\tilde V^\mu\tilde v_\mu\}
=-\frac{ig\chi^2_0}{f^2}\ {\rm Tr}\{[\Pi,\partial_\mu\Pi]\tilde V^\mu\}
\nonumber\\
&=&-2i g_{\rho\pi\pi}\ Tr\{[\Pi,\partial_\mu\Pi]\tilde V^\mu\}\ear
with
\be\label{7.2}
g_{\rho\pi\pi}=\frac{g}{2}\frac{\chi^2_0}{f^2}=\frac{M^2_\rho}{2gf^2}\ee
Considering only the effective action (\ref{2.6}), one has $\kappa_f=1$ and
infers
\be\label{7.2a}
g_{\rho\pi\pi}=
\frac{9x}{14(1+x)}g\approx 4.6\left(\frac{x}{1+x}\right)^{1/2}.\ee
If we restrict these interactions to the $\rho$-mesons
$\tilde V_\mu=\frac{1}{2}\vec\rho_{V\mu}\vec\tau$ and
pions $\Pi=\frac{1}{2}\vec\pi\vec\tau, v_\mu=\frac{1}{4f^2}(
\vec\pi\times\partial_\mu\vec\pi)\vec\tau$, we obtain
the familiar form
\be\label{7.2b}
{\cal L}_{\rho\pi\pi}=\frac{g\chi_0^2}{2f^2}\epsilon_{ijk}\pi^i\partial_\mu
\pi^j\rho_V^{k\mu}
=g_{\rho\pi\pi}(\vec\pi\times\partial_\mu\vec\pi)\vec\rho
_V^\mu\ee
It is straightforward to compute the decay rate $\rho\to 2\pi$ as
\be\label{7.3}
\Gamma(\rho\to\pi\pi)=\frac{g^2_{\rho\pi\pi}}{48\pi}
\frac{(M_\rho^2-4M^2_\pi)^{3/2}}{M_\rho^2}\simeq
150\ {\rm MeV} \ee
and one infers the phenomenological value
$g_{\rho\pi\pi}\simeq 6.0$. The discrepancy with eq. (\ref{7.2a})
reflects the difference between the realistic
value $a\approx 2$ and a value of $a$ which results from (\ref{E4})
for $\kappa_f=1,\ f_\pi=f$. The estimate (\ref{7.2a}) receives, however,
important corrections.  Omitted $SU(3)_V$-violating effects from the
strange quark mass result in corrections $\sim$ 30 \%. Furthermore,
since eq. (\ref{E4}) is no symmetry relation, it is subject
to modifications from the inclusion of additional fields. (We have already
included in eq. (\ref{E4}) the presumably most important
modification in the form of $\kappa_f<1$.)

The effective coupling between $\rho$-mesons
and baryons (\ref{5.22}) obeys
\be\label{7.6}
{\cal L}_{\bar NN_\rho}=-\frac{g}{2}\ {\rm Tr}\ \{\bar N_8\gamma^\mu
N_8\vec\tau\}\vec\rho_{V\mu}\ee
where we omit from now on electromagnetic effects.
This implies that the effective action (\ref{2.6}) does not contain
a direct coupling of the $\rho$-mesons to protons and neutrons.
The $\rho$-mesons only couple to strange baryons.
In the approximation of the effective action (\ref{2.6})
possible contributions to the nucleon-nucleon interactions in the
isospin triplet vector channel could only arise through two-pion
interactions $\sim {\rm Tr}\ \{\bar N_8\gamma^\mu\tilde v_\mu N_8\}$.
For the computation of the effective nucleon-nucleon interactions
one should solve the field equations for $\Pi$ and $\tilde V_\mu$
as functionals of the baryon fields $N_8$ in bilinear order $\sim
\bar N_8N_8$. The solution has to be reinserted into the effective
action. As a result of this procedure one finds nucleon-nucleon
interactions in the pseudoscalar channel and in the isospin-singlet
vector channel mediated by the exchange of $V_{8\mu}$ but not
in the isospin-triplet vector channel. (The-$\sigma$ exchange
term in the scalar channel is also contained in our model once
the non-Goldstone scalar excitations in $\phi$ and $\chi$ are
included.) This absence of isospin-triplet vector channel nucleon-nucleon
interactions seems not to be consistent with observation \cite{9}.

In summary, three shortcomings indicate that the effective action
(\ref{2.6}) gives only an insufficient picture of the hadronic
interactions in the vector channel: (i) the absence of a physical
$SU(3)$-singlet vector state (the ninth vector meson),
(ii) the inaccurate estimate of the parameter $a$ in eq. (\ref{E4}) for
$\kappa_f=1$, (iii) the absence of nucleon-nucleon interactions in the
isospin-triplet vector channel. In addition, no axial-vector mesons
are present. These shortcomings can be overcome once we include the effective
four-quark interactions in the color singlet vector and axial-vector
channel. The successful  relations (\ref{E7}), (\ref{E10})
and (\ref{E11}) can be maintained, whereas  the parameter $a$
and the relation  between $M_\rho, g_{\rho\pi\pi}$
and $f_\pi$ will be modified. Mixing effects with the
divergence of the axial vector induce the correction (\ref{5.8a}) and
therefore $\kappa_f<1$.

The effects from vector and axial-vector four-quark interactions
are discussed in detail in appendix A. Besides the ``partial Higgs
effect'' (\ref{5.8a}) they account for the missing nucleon-nucleon
interactions in the isospin triplet channel. They also contain
the missing ninth vector meson, i.e. the $SU(3)_C$-singlet state,
as well as the axial-vector mesons. Unfortunately, the additional effective
interactions are parametrized by new unknown couplings. We will
retain here only the ``partial Higgs effect'', the singlet vector
meson $S^\mu$ with mass $\mu_V^2$ and the effective vector channel
four-quark interaction. We omit the physics of axial-vector mesons.
The relevant interactions discussed in appendix A lead then to
an addition to the effective action (\ref{2.6}) of the form
$(S_{\mu\nu}=\partial_\mu S_\nu-\partial_\nu S_\mu)$
\bear\label{8.7}
{\cal L}_\rho&=&\frac{1}{4}S^{\mu\nu}S_{\mu\nu}+\frac{\mu_V^2}{2}S^\mu S_\mu
+\frac{1}{\sqrt6}\tilde c_{\rho\bar qq}S^\mu{\rm Tr}\bar N\gamma_\mu N
\\
&&+\Delta{\cal L}_{kin}^{(P)}+y_N{\rm Tr}\{\bar N\gamma_\mu\gamma^5\tilde
{\hat a}^\mu N\}-\tau_V\ {\rm Tr}\{\bar N\gamma^\mu\lambda_z N\}
\ {\rm Tr}\{\bar N\gamma_\mu\lambda_z N\}\nonumber\ear
where the partial Higgs effects results in $\Delta{\cal L}_{kin}^{(P)}$ as
given by eq. (\ref{5.8a}) and the term $\sim y_N$ with
\be\label{8.8}
\tilde{\hat a}_\mu=-\frac{i}{2}(\xi^\dagger D_\mu\xi-\xi D_\mu\xi^\dagger
)-\frac{1}{3}{\rm Tr} a_\mu.\ee
The couplings
$\tilde c_{\rho\bar qq}, y_N$ and $\tau_V$
may be determined from phenomenology and we assume $\mu_V^2
\approx \bar M^2_\rho$. There is actually
no symmetry argument why $\mu^2_V$ should equal $\bar M^2_\rho=g^2\chi_0^2$.
In particular, $\mu^2_V$ is not related to chiral and color symmetry breaking.
We speculate that a partial fixed point in the renormalization flow
of the ratio $\mu_V^2/(g^2\chi_0^2)$ could lead to an understanding
of the puzzle of the phenomenologically required approximate
``ninefold degeneracy'' of the light vector meson masses.

All other effective interactions discussed in appendix A will be
considered as subleading and neglected. Some of the couplings can be
estimated from observations like the decay reates of axial-vector mesons
into vector mesons and pseudoscalars. This may later be used for an
estimate of the typical size of the neglected subleading terms.

\section{Interactions of pseudoscalar mesons}
\setcounter{equation}{0}
The cubic interactions between the pseudoscalar and the baryon
octets are usually parametrized by
\be\label{99.1}
{\cal L}_N^{(p)}=F\ {\rm Tr}\{\bar N_8\gamma^\mu \gamma^5
[\tilde a_\mu,N_8]\}+D\ {\rm Tr}\{\bar N_8\gamma^\mu\gamma^5\{\tilde a_\mu,
N_8\}\}\ee
Experimental values are $F=0.459\pm0.008$ and
$D=0.798\pm0.008$. The effective action (\ref{2.6}) leads to
(cf. eq. (\ref{5.16})) $F=D=0.5$. This can
be considered as a good achievement since the reparametrization
symmetry does not restrict the coupling constants $F$ and $D$. (This
contrasts with the coupling of the vector current $\tilde v_\mu$.)
In our context the relation $F=D=0.5$ is directly
connected to the origin of these couplings from the quark kinetic term
and therefore to quark-baryon duality. The term $\sim y_N$ from
the partial Higgs effect (\ref{8.7}) provides for a correction
$F+D=1+y_N$ and one infers
$y_N\approx 0.26$. Contributions to $D-F\approx0.34$ have to be generated
from other higher-order invariants, as, for example, a momentum-dependent
Yukawa coupling involving $D_\mu\chi$. We note that $D-F$ contributes to the
$\eta$-nucleon coupling but not to the interaction between protons,
neutrons and pions. The latter can be written in a more conventional
form with the nucleon doublet ${\cal N}^T=(p,n)$ and $\tilde a_\mu$ restricted
to a $2\times 2$ matrix (by omitting the last line and column)
\bear\label{9.1a}
{\cal L}^{(\pi)}_{\cal N}&=&g_A\bar{\cal N}\gamma^\mu\gamma^5
\tilde a_\mu{\cal N}\nonumber\\
g_A&=&F+D\ear
The inclusion of weak interactions will replace the derivative
in the definition (\ref{5.4}) of $\tilde a_\mu$ by a covariant
derivative involving a coupling to the $W$-boson (see next section).
The constant $g_A$ will therefore appear in the $\beta$-decay rate of the
neutron.

For a discussion of the self-interactions of the
pseudoscalar mesons we first
expand eq. (\ref{5.18}) in powers of $\Pi=\frac{1}{2}\Pi^z\lambda_z$
\bear\label{99.2}
{\cal L}^{(\pi)}&=&\frac{f^2}{4}\ Tr\{\partial^\mu\exp(-\frac{2i}{f}
\Pi)\partial_\mu\exp(\frac{2i}{f}\Pi)\}\\
&=&Tr\{\partial^\mu\Pi\partial_\mu\Pi\}+\frac{1}{f^2}\ Tr\{
\partial^\mu\Pi^2\partial_\mu\Pi^2-\frac{4}{3}\partial^\mu
\Pi\partial_\mu\Pi^3\}+...\nonumber\ear
Similarly, the current quark mass term (\ref{5.28}) contributes
\bear\label{9.3M}
{\cal L}_j&=&{\rm Tr}\left\{ M^2_{(p)}\Pi^2\right\}-{\rm Tr}
\left\{\frac{M^2_{(p)}}{3f^2}\Pi^4\right\}+...\nonumber\\
M^2_{(p)}&=&2Z^{-1/2}_\phi a_q\sigma_0f^{-2}\bar m\ear
This yields the low-momentum four-pion interactions.
Further effective
interactions arise from the exchange of
vector mesons according to eq. (\ref{5.22}).
They are obtained by substituting for
the vector mesons the solution of the field equation in
presence of pseudoscalars
\be\label{99.3}
\tilde V_\mu=-\frac{\bar M^2_\rho}{g}G^{(\rho)\nu}_\mu\tilde v_\nu+...\ee
with the vector meson propagator $G^{(\rho)}$ obeying
\be\label{99.4}
G^{(\rho)\ \nu}_{\ \mu}[(\bar M_\rho^2-\partial^2)\delta^\sigma_\nu+
\partial_\nu\partial^\sigma]=\delta^\sigma_\mu\ee
One obtains
\be\label{99.5}
{\cal L}_V=\chi^2_0\ {\rm Tr}\{\tilde v^\mu(\delta^\sigma_\mu-\bar M_\rho^2
G^{(\rho)\ \sigma}_{\ \mu})\tilde v_\sigma\}+...\ee
and we note that the term $\chi^2_0\ {\rm Tr}\{\tilde v^\mu\tilde v_\mu\}$
in the expression (\ref{5.22}) is cancelled by the lowest order
of a derivative expansion of $G^{(\rho)}$. It should be remarked
that ${\rm Tr}\{\tilde v^\mu\tilde v_\mu\}$ is invariant
under the chiral transformations. It involves only two derivatives and
cannot be reduced to the term ${\rm Tr}\{\partial^\mu U^\dagger\partial_\mu  
U\}$ which appears in chiral perturbation theory. It is, however,
not consistent with the local reparametrization symmetry.

The next order in a derivative expansion of eq. (\ref{99.5})
reads
\be\label{99.6}
{\cal L}_V=\frac{1}{2g^2}{\rm Tr}\{\tilde v^{\mu\nu}\tilde v_{\mu\nu}\}+...\ee
where
\be\label{99.7}
\tilde v_{\mu\nu}=\partial_\mu\tilde v_\nu-\partial_\nu\tilde v_\mu+
i[\tilde v_\mu,\tilde v_\nu]\ee
has been completed to a covariant ``field strength'' such that
(\ref{99.6}) is invariant under the local reparametrization
symmetry. With
\be\label{99.8}
\tilde v_{\mu\nu}=-\frac{i}{4}\xi^\dagger(\partial_\mu U
\partial_\nu U^\dagger
-\partial_\nu U\partial_\mu U^\dagger)\xi\ee
one obtains
\bear\label{99.9}
{\cal L}_V&=&\frac{1}{16g^2}\ {\rm Tr}\{\partial_\mu U^\dagger
\partial^\mu U\partial_\nu U^\dagger \partial^\nu U-
\partial_\mu U^\dagger\partial_\nu U\partial^\mu U^\dagger \partial^\nu
U\}\nonumber\\
&=&\frac{1}{32g^2}[6\ {\rm Tr}\{\partial_\mu U^\dagger\partial^\mu  
U\partial_\nu U^\dagger \partial^\nu U\}
-({\rm Tr}\{\partial_\mu U^\dagger\partial^\mu U\})^2\nonumber\\
&&-2{\rm Tr}\{\partial_\mu U^\dagger\partial_\nu U\}\ {\rm Tr}\{
\partial^\mu U^\dagger\partial^\nu U\}]\ear
One can infer the contribution of ${\cal L}_V$ to the parameters
$L_i$ appearing in next to leading order in chiral perturbation
theory \cite{GL}
\be\label{99.10}
L^{(V)}_1= \frac{1}{32g^2}\quad,\quad L_2^{(V)}=\frac{1}{16g^2}\quad,\quad  
L_3^{(V)}=-
\frac{3}{16g^2}\ee
We observe that with the hypothesis of scalar Vervollst\"andigung
(\ref{2.6}) the constants depend only on the effective gauge
coupling $g$.
For $g=6$ the values $L_1^{(V)}=0.87\cdot10^{-3},\ L_2^{(V)}=1.74\cdot
10^{-3},\
L_3^{(V)}=-5.2\cdot10^{-3}$ compare
well with the values  \cite{BEG} extracted from observation
$L_1=(0.7\pm0.3)\cdot10^{-3},\
 L_2=(1.7\pm0.7)\cdot 10^{-3},\ L_3=-(4.4\pm2.5)\cdot10^{-3}$.
This is consistent with the hypothesis that these constants are
dominated\footnote{Further contributions arise from the exchange of scalars
and have been estimated
\cite{JWPT} as $L_2^{(S)}=0,\ L_3^{(S)}=1.3\cdot10^{-3}$.} by vector-meson  
exchange \cite{Ecker}. The momentum
dependence of the effective four-pion vertex extracted from eq.
(\ref{99.5}) describes the fact the $\pi-\pi$ scattering at
intermediate energies is dominated by the $\rho$-resonance.
We conclude that the hypothesis of scalar
Vervollst\"andigung (\ref{2.6}) gives a
very satisfactory picture of the pion interactions.

\section{Weak interactions}
\setcounter{equation}{0}

Kaons decay into two pions by weak interactions. It is an old
puzzle how the strong enhancement of the $\Delta I=1/2$ decays of
$K^0_S$ as compared to the $\Delta I=3/2$ decay of $K^\pm$ should be
explained in QCD. The semileptonic weak decays of kaons
and pions are directly related to the corresponding decay constants
$f_K, f_\pi$. These characteristic properties of the weak decays
are described by our model once the weak interactions are
incorporated. New parameters specify the strength
of effective vertices which are one-particle
irreducible with respect to cutting a $W^\pm$-boson line. Following
the philosophy underlying the effective action
(\ref{2.6}) we only include ``renormalizable'' interactions
with dimension $\leq4$. These invariants only contribute to
the $\Delta I=1/2$ decays. Comparison with the $K^0_S$-decays
determines the relevant parameter -- ultimately it should be
computed from QCD. The contribution of (one-particle reducible)
$W$-boson exchange diagrams to the $\Delta I=3/2$ processes
is substantially smaller. The corresponding effective vertex
for $\phi$ and $\chi$ involves two more powers of these fields
as compared to the leading invariant. Together  with higher
1PI-vertices it should be considered as subleading. Scalar
Ver\-voll\-st\"an\-digung leads to a qualitative understanding
of the $\Delta I=1/2$ enhancement.

\noindent {\bf a) Strangeness-violating local interactions}

The exchange of the heavy $W$ and $Z$ bosons in tree and loop
diagrams produces multi-fermion interactions which may
violate strangeness. On the momentum scale of interest here these
interactions can be taken as local.
After bosonisation, they result in two extensions of the
effective action (\ref{2.6}). All diagrams which decay into pieces
by cutting a $W$ or $Z$ propagator can be represented by the
couplings of a heavy boson
to the fields present in the effective action (\ref{2.6}).
Those are retricted by the weak
gauge symmetry. Keeping only the lowest dimension terms, we
implement these couplings by appropriate covariant derivatives.
The semileptonic decays are completely described by them. On the
other hand, the diagrams which are one-particle irreducible
with respect to cutting a $W$ or $Z$-boson line cannot be
represented in this way. They result in additional local
interactions which have to obey the appropriate symmetries.

First of all, the inclusion of weak interactions in our framework
requires the effective action to be invariant under local
$SU(2)_L\times U(1)_Y$ symmetry. On the other hand, the global
flavor symmetry  $SU(3)_L\times SU(3)_R$ is
reduced to $SU(2)_R\times U(1)_R$. Here the global $SU(2)_R$ transformations  
act between right-handed
down and strange quarks (which carry identical hypercharge)
and $U(1)_R$ consists of phase rotations of the right-handed quarks
with generator $\tilde Q$.
In lowest order the
strangeness-violating interactions obey, however, the larger $SU(3)_R$
flavor symmetry. (The latter is  broken by the couplings of the $Z$-boson
and the photon.)
Also $C$ and $P$ need not be conserved
separately any more, whereas $CP$ is an exact symmetry
if effects from the third generation quarks are neglected. Of
course, local $SU(2)_L$ symmetry requires the charmed quark.
In this section we use an effective action where the heavy charmed
quark has been integrated out.

Local $SU(2)_L\times U(1)_Y$ gauge symmetry is easily implemented
by replacing all derivatives in the effective action (\ref{2.6}) by
appropriate covariant derivatives. Besides this we have to include
the effects from additional local interactions which become possible
due to the reduced flavor symmetry. They arise from effective
short distance four-fermion interactions mediated by the exchange
of $W^\pm$ or $Z^0$ in loops (i.e. ``penguin diagrams'') and therefore
are suppressed by
inverse powers of the
squared $W$-boson mass $M^2_W$. As an example, a new local mass term
for $\phi$ reads for a vanishing Cabibbo angle
\be\label{9.A}
{\LL}_{\tilde\mu}=\frac{g_W^2\tilde\mu^4_W}{4M_W^2}\ {\rm Tr} \  
\{\phi^\dagger\phi\lambda_8\}\ee
Here the weak gauge coupling $g_W$ and $M_W$ are related to
the Fermi constant by $G_F=g_W^2/(4\sqrt2 M^2_W)$. The value of
the parameter $\tilde\mu_W(k)$ can be computed for large $k$
by translating (\ref{9.A}) into an effective four-quark
interaction according to sect. 2 and comparing with the perturbatively
computed value. For the present purpose we treat $\tilde\mu_W\equiv\tilde
\mu_W(k=0)$ and similar constants as free parameters. The mass term
(\ref{9.A}) violates $C$ and $P$ and preserves $CP$. For a nonzero
Cabibbo angle the field $\phi$ in (\ref{9.A}) has to be replaced
by the ``weak interaction eigenstate'' $\hat\phi$, which is related
to the basis of mass eigenstates $\phi$ by
\be\label{9.B}
\hat\phi=\phi R_\theta=\phi\left(\begin{array}{ccc}
1&0&0\\
0&c_\theta&-s_\theta\\
0&s_\theta&c_\theta\end{array}\right)\ee
Here $s_\theta$ and $c_\theta$ are the sine and cosine of the Cabibbo
angle. The mass term (\ref{9.A}) becomes
\bear\label{9.C}
{\LL}_{\tilde\mu}&=&\frac{g_W^2\tilde\mu_W^4}{4M_W^2}\ {\rm Tr}\{
\phi^\dagger\phi\lambda_W\}\nonumber\\
\lambda_W&=&R_\theta\lambda_8 R^T_\theta=(1-\frac{3s_\theta^2}{2})
\lambda_8+\sqrt3c_\theta s_\theta\lambda_6+\frac{\sqrt3}{2}
s^2_\theta\lambda_3\ear
and we underline the strangeness-violating contribution
$\sim\lambda_6$. Inserting the nonlinear pseudoscalars (\ref{4.3})
this term vanishes, however, and does not contribute to strangeness
violation in the sector of the Goldstone bosons $\Pi$.

In leading order $M^{-2}_W$ the terms involving two derivatives
of $\phi$ or $\chi$ are given by
\be\label{9.E}
{\LL}_\mu=\frac{g_W^2\mu^2_W}{4M_W^2}[{\rm Tr}\ \{D^\mu\phi^\dagger
D_\mu\phi\lambda_W\}
+\delta_W(D^\mu\chi)^*_{ij,ab}(D_\mu\chi)_{ij,ac}(\lambda_W)_{cb}]
\ee
The effective action (\ref{9.E}) involves two additional real parameters
$\mu^2_W$ and $\delta_W$. It is invariant under the local symmetry
$SU(3)_C\times SU(2)_L\times U(1)_Y$, global $SU(3)_R$ flavor symmetry
and $CP$, whereas it violates $P, C$ and strangeness. In lowest order
in $M_W^{-2}$ and neglecting electromagnetic effects we may replace
$D_\mu\phi$ by $\partial_\mu\phi$ and use (\ref{2.6a}) for $D_\mu\chi$.
Furthermore, we will be interested only in strangeness-violating effects and
replace $\lambda_W$ by $\sqrt3 c_\theta s_\theta\lambda_6$. Concentrating
on the pseudoscalar interactions we insert (\ref{5.4}) and
obtain
\bear\label{9.F}
{\LL}_{\mu U}&=&\frac{1}{2}A_Wg_8f^2[{\rm Tr}\ \{ \partial^\mu U^\dagger
\partial_\mu U\lambda_6\}
+4g_\chi\ {\rm Tr}\{\partial^\mu\xi^\dagger\partial_\mu\xi
\lambda_6\}\nonumber\\
&&+\frac{4i}{3}g_\chi\partial^\mu\theta\ {\rm Tr}\ \{\xi^\dagger\partial_\mu
\xi\lambda_6\}]\ear
with
\bear\label{9.G}
A_W&=&\frac{c_\theta s_\theta g_W^2f^2}{4M_W^2}=3.98\cdot10^{-8}
\nonumber\\
g_8&=&\frac{\sqrt3}{2}\frac{\mu^2_W}{f^2}\frac{1-2\delta_Wx/7}{1+x}
\nonumber\\
g_\chi&=&\frac{9\delta_Wx}{14-4\delta_Wx}\ear
replacing $\mu_W$ and $\delta_W$. Adding the interactions with gluons
leads to the replacements
\bear\label{16.6AA}
i\xi^\dagger\partial_\mu\xi&\longrightarrow& i(v_\mu+g\tilde V_\mu+a_\mu)\nonumber\\
\partial^\mu\xi^\dagger\partial_\mu\xi&\longrightarrow&
(v^\mu+g\tilde V^\mu+a_\mu)(v_\mu+g\tilde V_\mu+a_\mu)\ear
This makes the local reparametrization invariance of ${\cal L}_{\mu U}$
manifest.
We show in appendix B that only
derivative terms contribute to strangeness-violating interactions
for the pseudoscalars. One infers that ${\cal L}_\mu$ (\ref{9.E})
is the only relevant bosonic term with dimension $\leq 4$.
The invariant of dimension four for strangeness-violating fermionic
interactions can be found in appendix B. This should dominate the
hyperon decays.

\medskip\noindent
\noindent{\bf b) Vector-meson-W-boson mixing}

We also have to include the effects of local $SU(2)_L\times
U(1)_Y$ symmetry in the derivative terms of eq. (\ref{2.6}). The
photon coupling has already been discussed in sect. 3.
The coupling of the weak gauge bosons $W_\mu^\pm,$ to the
quarks is implemented by inserting in the effective action (\ref{2.6})
the appropriate covariant derivatives
\bear\label{9.1}
&&\partial_\mu\psi_{L,R}\to\partial_\mu\psi_{L,R}+D_\mu^{(L,R)}
\psi_{L,R}\nonumber\\
&&D_\mu^{(L)}=-\frac{i}{2}g_W\vec{\tilde W}_\mu\vec\tau_W=-i\frac{g_W}{\sqrt2}\left(
\begin{array}{ccc}
0_\mu\ ,&\ c_\theta\tilde W_\mu^+\ ,&\ s_\theta \tilde W_\mu^+\\
c_\theta\tilde W^-_\mu\ ,&\ 0\ ,&\ 0\\
s_\theta\tilde W^-_\mu\ ,&\ 0\ ,& \ 0\end{array}\right)\\
\nonumber
&&D_\mu^{(R)}=0\ear
We have omitted in the explicit form of $D_\mu^{(L,R)}$ the couplings of the
$Z$-bosons since they do not contribute to strangeness-violating
interactions $(\tau_{3W}=\tau_3)$. They can easily be inserted in the formalism
below.
The weak coupling to the quark
bilinears $\gamma_{ij}\sim\psi_{Ri}\bar\psi_{Lj}$ follows
correspondingly\footnote{Covariant derivatives have
also to be used in the additional interactions
(\ref{8.7}),(\ref{8.8}).} by replacing in eq. (\ref{2.6})
\be\label{9.2}
D_\mu\gamma_{ij}\to D_\mu\gamma_{ij}+D^{(R)}_\mu
\gamma_{ij}-\gamma_{ij}D_\mu^{(L)}\ee
Finally, we add the mass terms for the gauge bosons\footnote{The
kinetic terms for the weak gauge bosons and the Higgs sector
are not important here.}
\be\label{9.3}
{\cal L}_W=M_W^2\tilde W^{+\mu}\tilde W^-_\mu
+\frac{1}{2}M^2_Z\tilde Z^\mu\tilde Z_\mu\ee
Eliminating the fields $W_\mu^\pm, Z_\mu$ by solving their field
equations as functionals of $\psi, \phi$ and $\chi$ yields
effective four Fermi interactions\footnote{In the
full standard model this also involves the
leptons.} relevant for weak interactions
at low momenta as well as weak mesonic interactions.
Furthermore, spontaneous color symmetry breaking by the octet
expectation value $<\chi>$ (\ref{2.17}) leads to a mixing between
$W,Z$-bosons and gluons similar to the photon mixing discussed
in sect. 3. The mixing angle is tiny due to the large masses
$M_W, M_Z$. It needs, however,
to be included for strangeness-violating processes.

In the language of the nonlinear meson fields
the weak interactions add to the effective Lagrangian a term
${\LL}_W={\LL}_{WN}+{\LL}_{WM}+{\LL}_{WU}+{\LL}_{\mu U}$
\bear
{\LL}_{WN}&= &i\ {\rm Tr}\ \{\bar N_L\gamma^\mu\xi D_\mu^{(L)}\xi^\dagger N_L+
\bar N_R\gamma^\mu \xi^\dagger D_\mu^{(R)}\xi N_R\}\label{9.4}\\
{\LL}_{WM}&=&M^2_W\tilde W^{+\mu}\tilde W^-_\mu+\frac{1}{2}M^2_Z
\tilde Z^\mu\tilde Z_\mu\nonumber\\
&&-ig\chi^2_0\ {\rm Tr}\ \{(\xi^\dagger D_\mu^{(R)}\xi+\xi D_\mu^{(L)}
\xi^\dagger)\tilde V^\mu\}\nonumber\\
&&-(\sigma_0^2+\frac{5}{9}\chi_0^2)\ {\rm Tr}\ \{D^{(L)\mu}D^{(L)}_\mu+D^{(R)
\mu}D^{(R)}_\mu\}\label{9.5}\\
&&+(2\sigma^2_0+\frac{1}{9}\chi_0^2)\ Tr\{ U^\dagger D^{(R)\mu}
U D^{(L)}_\mu\}+\frac{1}{3}\chi^2_0\ Tr D_\mu^{(R)}\ {\rm Tr}\ D_\mu^{(L)}
\nonumber\\
{\LL}_{WU}&=& (2\sigma_0^2-\frac{1}{9}\chi_0^2)\ {\rm Tr}\ \{U^\dagger \partial^\mu
UD_\mu^{(L)}+U\partial^\mu U^\dagger D_\mu^{(R)}\}\nonumber\\
&&+i\chi^2_0\ {\rm Tr}\ \{(\tilde v^\mu+\tilde a^\mu)D_\mu^{(L)}+
(\tilde v^\mu-\tilde a^\mu)D_\mu^{(R)}\}
\nonumber\\
&&-\frac{i}{3}\frac{\chi_0^2}{H_{\eta'}}\partial^\mu \eta'\ {\rm Tr}\  
\{D_\mu^{(L)}-D_\mu^{(R)}\}\label{9.6}
\ear
with ${\LL}_{\mu U}$ given above. These are the leading interactions
for the strangeness-violating pseudoscalar meson decays. For other
quantities like the $K^0_L-K^0_S$-mass difference they have to
be supplemented by interactions discussed in appendix C.

For a computation of the mixing between $W^\pm_\mu, \rho^\pm_\mu$ and
$K^{*\pm}_\mu$ we can put $\xi=1$ and neglect the terms
$\sim (D_\mu^{(L,R)})^2$
in ${\LL}_{WM}$. Inserting
\be\label{9.7}
\tilde V_\mu=\frac{1}{\sqrt2}\left(\begin{array}{ccc}
0\ ,&\ \tilde \rho_\mu^+\ ,&\ \tilde K^{*+}_\mu\\
\tilde\rho_\mu\ ,&\ 0\ ,&\ 0\\
\tilde K^{*-}_\mu\,&\ 0\ ,&\ 0\end{array}\right)\ee
and combining with the vector-meson mass term one obtains the squared
mass matrix (with $b=(\tilde\rho^-, \tilde K^{*-}, \tilde W^-), \ {\LL}
_M=b^\dagger {\cal M}^2_b b)$
\bear\label{9.8}
{\cal M}^2_b&=&\left(\begin{array}{ccc}
M^2_\rho\ ,&\ 0\ ,&\ Bc_\theta\\
0\ ,&\ M^2_{K^*}\ ,&\ Bs_\theta\\
Bc_\theta\ ,&\ Bs_\theta\ ,&\ M_W^2\end{array}\right)\nonumber\\
B&=&-\frac{1}{2}g_Wg\chi^2_0\ear
In a good approximation the fields $(\rho, K^*, W)$ for the physical
spin-one bosons, i.e. the mass eigenstates, obey
\bear\label{9.9}
\tilde\rho_\mu&=&\rho_\mu-s_\rho W_\mu\ ,\ \tilde K^*_\mu=K^*_\mu
-s_KW_\mu,\nonumber\\
\tilde W_\mu&=&W_\mu+s_\rho\rho_\mu+s_KK^*_\mu\ear
with
\be\label{9.10}
s_\rho=\frac{1}{2}g_Wg\frac{\chi_0^2}{M^2_W}c_\theta\ ,
\ s_K=\frac{1}{2}g_Wg\frac{\chi_0^2}{M_W^2}s_\theta\ \ee
The mixing angles $s_\rho, s_K$ between $\tilde\rho,\tilde K^*$ and
$\tilde W$ are suppressed by $B/M^2_W$ and therefore indeed tiny.
The mass eigenvalues get only negligible corrections. Also the slight
modification of the couplings of the physical $W$-boson
(the mass eigenstate with mass $\approx M_W$) is of no importance.
There remains, however, one relevant effect: The couplings of the
physical $\rho^\pm$ and $K^{*\pm}$ mesons acquire from the mixing
a small strangeness-violating correction. Indeed,
expressed in terms of the
``physical fields'' one finds the contribution of charged spin-one
bosons to $D_\mu^{(L)}$ (cf. eq. (\ref{9.1})
\be\label{9.11}
D_\mu^{(L)}=-i\frac{g_W}{2\sqrt2}[W^+_\mu+\frac{g_Wg}{2}
\frac{\chi_0^2}{M_W^2}(c_\theta
\rho^+_\mu+s_\theta K^{*+}_\mu)]
[c_\theta(\lambda_1+i\lambda_2)+s_\theta(\lambda_4+i\lambda_5)]
+{\rm h.c.}\ee
Inserting this term in ${\LL}_{MN}$, one sees that the
$\rho$ meson couples not only to baryon-bilinears with zero strangeness
but also has a small contribution of
a coupling to $\bar p\Sigma$. It therefore contributes
to strangeness violating weak decays. Similarly, the exchange
of $\rho_\mu$ and $K^*_\mu$
induce additional strangeness violating interactions for the
pseudoscalar mesons. We note that the mixing effects of the
neutral weak boson $Z_\mu$ do not lead to violations of quantum
numbers and can therefore be neglected.  In summary,
in leading order in $G_F$ the only relevant effect of the
mixing between gluons and weak gauge bosons is the expression
(\ref{9.11}) for $D_\mu^{(L)}$. From
${\LL}_{WM}$ we need then only to retain the piece $M_W^2W^{+\mu}
W^-_\mu+\frac{1}{2}M^2_ZZ^\mu Z_\mu$.

\medskip\noindent
\noindent {\bf c) Leptonic meson decays}

Let us next turn to the leptonic decays of the pseudoscalar
pions and kaons. The relevant vertex for the coupling of
$W^\pm$ to the charged
leptons $l$ and neutrinos $\nu_l$ is given by
\bear\label{9.12}
{\LL}_l&=&\sum_l\frac{g_W}{2\sqrt2}\{\bar l\gamma^\mu
(1+\gamma^5)\nu_l\tilde
W^-_\mu+\bar\nu_l\gamma^\mu(1+\gamma^5)l\tilde W^+_\mu\}\\
&=&\sum_l\bar\nu_l\gamma^\mu (1+\gamma^5)l\left\{
\frac{g_W}{2\sqrt2}W^+_\mu
+\frac{g_W^2g}{4\sqrt2}\frac{\chi_0^2}{M_W^2}
(c_\theta\rho^+_\mu+s_\theta K^{*+}_\mu)\right\}+\ {\rm h.c.}
\nonumber\ear
The semileptonic pion decay can therefore in principle proceed via
intermediate $W, \rho$ or $K^*$ exchange. In leading order in an
expansion in $G_F$, however, only the $W$ exchange contributes. The relevant
bilinear between $\Pi$ and $W$ obtains from
\be\label{9.13}
{\LL}_{W\Pi}=\frac{i}{2}f\partial^\mu\Pi^z\ Tr\{\lambda_zD_\mu^{(L)}\}\ee
After eliminating $W$ by the field equation, one finds the effective
cubic vertex
\be\label{9.14}
{\LL}_{\pi \mu\nu}=\frac{g_W^2c_\theta f}{4\sqrt2 M^2_W}
\bar\mu\gamma^\rho(1+\gamma^5)\nu_\mu\partial_\rho\pi^-\ee
Comparison with the pion decay rate yields $f=f_\pi=92.5$ MeV. In our
approximation where $SU(3)$-violating vacuum expectation values
are neglected we also have $f_K=f$. We have therefore taken in
eq. (\ref{5.8}) an average value. A detailed discussion of
semileptonic decays including $SU(3)$-violating expectation values
due to the nonvanishing mass of the strange quark can be found
in ref. \cite{8}.

\noindent
{\bf d) Nonleptonic kaon decays}

For the non-leptonic weak decays of the kaons
four effects need to be considered
in linear order in $G_F$. The most
important coupling arises from the one particle
irreducible coupling (\ref{9.F}). This is the only
contribution in leading order. The second concerns
the  effective strangeness violating
cubic vertices induced by the exchange of $\tilde W^\pm$.
A third contribution reflects the weak mixing between the
pseudoscalars $\pi^\pm, K^\pm$ and the scalars $a^\pm, K^{*\pm}$.
This last effect needs information about the effective potential
$U(\gamma)$ and we will not discuss it here in a quantitative way.
Finally, one expects ``nonleading'' strangeness violating vertices
of the type (\ref{9.E}) involving higher powers of $\phi,\chi$ or
higher derivatives. A combination of these effects should explain the
observed values
\bear\label{9.26}
\Gamma(K^0_S\to\pi^+\pi^-)&=&5.061\cdot 10^{-12}\ {\rm MeV}\nonumber\\
\Gamma(K^0_S\to\pi^0\pi^0)&=&2.315\cdot 10^{-12}\ {\rm MeV}\nonumber\\
\Gamma(K^-\to\pi^-\pi^0)&=&1.126\cdot 10^{-14}\ {\rm MeV}\ear
which reflect the strong enhancement of the $\Delta I=1/2$ decays.

The cubic coupling for the pseudscalar octet arising from (\ref{9.F})
needs some care. At first sight the terms
$\sim g_\chi$ seem to contribute. These terms generate, however,
also parity-violating mixings between the
vector mesons $\tilde v_\mu$ and the pseudoscalars $a_\mu\sim\partial_\mu\Pi$  
(cf. eq. (\ref{16.6AA})). Subsequent ``decay'' of intermediate vector bosons
into two Goldstone bosons leads to contributions to the effective
cubic vertex. This issue is most easily dealt with by ``integrating out''
the vector mesons, similar to sect. 9. In lowest order in the derivatives
one has (cf. eq. (\ref{99.5})) $v_\mu+g\tilde V_\mu=0$. As a consequence
one finds $\partial^\mu\xi^\dagger\partial_\mu\xi\to a^\mu a_\mu$
and concludes that the terms $\sim g_\chi$ do not contribute to
the strangeness-violating cubic vertices. The effective cubic
interaction therefore
reads $(\partial^2=\partial^\mu\partial_\mu)$
\bear\label{9.14a}
{\LL}^{(3)}_{\mu U}&=&\frac{2i}{f}A_W g_8
{\rm Tr}\ \{[\partial^\mu\Pi\partial_\mu\Pi,\Pi]\lambda_6\}\\
&=&\frac{1}{2f}A_Wg_8\{K^0_S[6\partial^\mu\pi^+
\partial_\mu\pi^-+3\partial^\mu\pi^0\partial_\mu\pi^0\nonumber\\
&&+\pi^+\partial^2\pi^-+\pi^-\partial^2\pi^++\pi^0\partial^2\pi^0]
+iK^0_L[\pi^-\partial^2\pi^+-\pi^+\partial^2\pi^-]\nonumber\\
&&-i\pi^0[K^+\partial^2\pi^--K^-\partial^2\pi^+]
+i\partial^2\pi^0[K^+\pi^--K^-\pi^+]\}\nonumber\ear
Here we have only retained terms involving one kaon and two pions in
the second expression and we use
\be\label{10.22a}
K^0_S=\frac{i}{\sqrt2}(K^0-\bar K^0)=\Pi^7\quad,\quad
K^0_L=\frac{1}{\sqrt2}(K^0+\bar K^0)=\Pi^6\ee
Evaluated on mass-shell
$(\partial^2\to M^2)$, this vertex reduces to
\bear\label{9.14b}
{\LL}^{(3)}_{\mu U}&=&-\frac{1}{2f}A_Wg_8
\{(3M^2_K-4M^2_\pi)K^0_S\pi^+\pi^-\\
&&+(\frac{3}{2}M^2_K-2M_\pi^2)K^0_S\pi^0\pi^0+i(M^2_{\pi^+}-M^2_{\pi^0}
)(K^-\pi^+\pi^0-K^+\pi^-\pi^0)\}\nonumber\ear
Up to tiny isospin-violating corrections from the difference between
charged an neutral pion masses it only contributes to the decay
$K^0_S\to2\pi$. With
\bear\label{10.23a}
&&\Gamma^{(3)}(K^0_S\to 2\pi^0)=\Gamma^{(3)}(K^0_S\to\pi^+\pi^-)=\\
&&\qquad-\frac{A_W}{2f}g_8(3M^2_K-4M^2_\pi)
=-1.26\cdot10^{-4}
g_8\ {\rm MeV}\nonumber\ear
one obtains the partial width for $K^0_S\to 2\pi^0$
\be\label{9.14c}
\Gamma(K^0_S\to 2\pi^0)=2.66\cdot 10^{-13}g^2_8
\ {\rm MeV}\ee
Comparison with the experimental value requires
\be\label{9.14d}
g_8=2.95\ee
This corresponds to a very reasonable value\footnote{For
$|\delta_Wx|\ll7/2$ the contribution $\sim\delta_W$
is small. We have used $x=4,\ \delta_W=0$ for the
numerical value. Negative $\delta_W$
lead to smaller values of $\mu_W$.} $\mu_W\approx$ 440 MeV. From eq. (\ref{10.23a})
one also infers that the decay rate $\Gamma(K^0_S\to\pi^+\pi^-)$
is twice the one for the decay into neutral pions, in accordance
with observation (see below
eq. (\ref{10.36b}) for a correction
to this relation). The vertex (\ref{9.F}) does not contribute to the
charged kaon decay. The latter will only be induced by $\tilde W$-exchange,
pseudoscalar-scalar mixing and nonleading 1PI vertices.
We will see below that these effects
are suppressed compared to the vertex (\ref{9.14b}). If a computation of the
effective coupling $g_8$ from the standard model couplings yields
indeed the value (\ref{9.14d}) the dominance of the $\Delta I=1/2$
kaon decays can be naturally understood.

In order to compute the contribution of $\tilde W$-exchange to the
effective
strangeness-violating cubic vertex we need the strangeness-violating
two-point functions (\ref{9.13}) ($\Pi=\frac{1}{2}\Pi^z\lambda_z$)
\be\label{9.15}
{\LL}_{W\Pi}=\frac{g_Wf}{\sqrt2}\ {\rm Tr}\{\partial^\mu\Pi\hat\lambda_+\}
(W^+_\mu
+s_\rho \rho^+_\mu+s_KK^{*+}_\mu)\ +\ {\rm h.c.}
\ee
and the cubic vertex
\bear\label{9.16}
{\LL}_{W\Pi\Pi}&=&\frac{1+5x/14}{1+x}\ {\rm Tr}\ \{[\Pi,\partial^\mu\Pi]
D_\mu^{(L)}\}\\
&=&-\frac{ig_W}{\sqrt2}\frac{1+5x/14}{1+x}\ {\rm Tr}\ \{[\Pi,\partial^\mu\Pi]
\hat\lambda_+\}
W^+_\mu\ +\ {\rm h.c.}\ +\ ...\nonumber\ear
with
\be\label{9.17}
\hat\lambda_+=\frac{1}{2}c_\theta(\lambda_1+i\lambda_2)+\frac{1}{2}
s_\theta(\lambda_4+i\lambda_5)\ee
Consider first the contribution
from $W$-exchange. Inserting the solution of the field equation
\be\label{9.18}
W^+_\mu=-\frac{g_W}{\sqrt2 M^2_W}\ {\rm Tr}\ \{(f\partial_\mu\Pi-
i\frac{1+5x/14}{1+x}[\Pi,\partial_\mu\Pi])\hat\lambda^\dagger_+\}
\ee
into (\ref{9.15}), (\ref{9.16}), (\ref{9.5}), one finds
\bear\label{9.19}
{\LL}^{(W)}_{\Delta  
S}&=&i\frac{g_W^2f}{2M^2_W}\frac{1+5x/14}{1+x}\hat{\LL}_{\Delta S},
\nonumber\\
\hat{\LL}_{\Delta S}&=& {\rm Tr}\ \{\partial^\mu\Pi\hat\lambda_+\}\
{\rm Tr}\ \{ [\Pi,\partial_\mu\Pi]\hat\lambda_+^\dagger\}+
{\rm Tr}\ \{\partial^\mu\Pi\hat\lambda_+
^\dagger\}\ Tr\{ [\Pi,\partial_\mu\Pi]\hat\lambda_+\}\nonumber\\
&=&-\frac{1}{4}c_\theta s_\theta\{2K^+\partial^\mu\pi^-\partial_\mu
\pi^0+2K^+\pi^-\partial^\mu\partial_\mu\pi^0-K^+\partial^\mu
\partial_\mu\pi^-\pi^0\nonumber\\
&&+2\sqrt2K^0\partial^\mu \pi^-\partial_\mu\pi^++\sqrt2 K^0
\partial^\mu\partial_\mu\pi^-\pi^+\}-c.c.\ear
In the last expression we again have only listed the vertices
involving one kaon and two pions. Another contribution arises
from the exchange of $\rho$ and $K^*$ mesons due to the
strangeness-violating mixing (\ref{9.15}). It is computed similarly by inserting 
\bear\label{9.20}
&&\tilde V_\mu^{\Delta S}=-\frac{g_Wf}{4}[s_\rho(M^2_\rho-\partial^2)^{-1}
(\lambda_1+i\lambda_2)+s_K(M^2_{K^*}-\partial^2)^{-1}(\lambda_4
+i\lambda_5)]\cdot\nonumber\\
&& \qquad\qquad {\rm Tr}\ \{\partial^\mu\Pi\hat\lambda_+^\dagger\}\ + \
{\rm c.c.}\ear
into eqs. (\ref{7.1}), (\ref{9.15}). One finds
\bear\label{9.21}
{\LL}^{\rho,K^*}_{\Delta S}&=&i\frac{g_W^2g^2\chi_0^4}{4 fM^2_W}\ {\rm Tr}\  
\{[\Pi,\partial_\mu\Pi][c_\theta(M^2_\rho-\partial^2)^{-1}
\frac{\lambda_1+i\lambda_2}{2}\nonumber\\
&&+s_\theta(M^2_{K^*}-\partial^2)^{-1}\frac
{\lambda_4+i\lambda_5}{2}]\}{\rm Tr}\ \{\partial^\mu\Pi\hat\lambda^\dagger_+\}
\ +\ {\rm c.c}\ear
We observe that in the approximation $M_\rho^2-\partial^2=
M^2_{K^*}-\partial^2=\bar M^2_\rho$ the cubic vertex (\ref{9.21})
is proportional to (\ref{9.19}), with a relative factor
$\frac{9x}{14+5x}$. In this limit the total strangeness-violating
contribution to the cubic vertex from $W, \rho$ and $K^*$
exchange
\be\label{9.22}
{\LL}^{\tilde W}_{\Delta S}={\LL}^W_{\Delta S}+{\LL}^{\rho,K^*}_{
\Delta S}=i\frac{g_W^2f}{2M_W^2}\hat{\LL}_{\Delta S}\ee
does not depend\footnote{Alternatively, the effective interactions
${\cal L}_{\Delta S}^{\tilde W}$ can be evaluated by first
integrating out the vector bosons $\tilde V_\mu$. The leading
term follows from inserting convariant derivatives in ${\cal L}_U$ (\ref{5.18})
and then integrating out the $\tilde W$-bosons.}
on $x$. From eq. (\ref{9.22}) we can extract the
contributions to the on-shell cubic vertices
\bear\label{9.23}
i\Gamma^{(3)}_{(\tilde W)}(K^+\to\pi^+\pi^0)&=&-\frac{1}{2}
\Gamma^{(3)}_{(\tilde W)}(K^0_S\to\pi^+\pi^-)
=\frac{g_W^2f}{8M_W^2}c_\theta s_\theta(M_K^2-M^2_\pi)\nonumber\\
&=&4.2\cdot10^{-5}\ {\rm MeV}\nonumber\\
\Gamma^{(3)}_{(\tilde W)}(K^0_S\to\pi^0\pi^0)&=&0\ear

We finally have to inlcude the contributions from scalar pseudoscalar mixing
and from nonleading 1PI vertices. As discussed in appendix B, the interesting
contribution can be parametrized as
\be\label{10.34a}
\Delta{\cal L}_{\Delta S}=-i\frac{g_W^2f}{2M_W^2}
z_W\hat{\cal L}_{\Delta S}\ee
with $z_W$ of order one. In consequence, this leads to the
multiplication of the $W$-exchange contribution (\ref{9.22}) by
a factor $(1-z_W)$.

The partial decay width for $K\to\pi\pi$ is given by
\bear\label{9.24}
\Gamma(K\to\pi\pi)&=&F_{K\pi\pi}|\Gamma^{(3)}(K\to\pi\pi)|^2
\nonumber\\
F_{K\pi\pi}&=&\frac{1}{8\pi}\frac{|\vec p_\pi|}{M_{K^2}}=3.347\cdot
10^{-5}\ {\rm MeV}^{-1}\ear
with an additional factor $1/2$ for two identical $\pi^0$. One finds
\be\label{9.25}
\Gamma(K^-\to\pi^-\pi^0)=5.92\cdot 10^{-14}(1-z_W)^2\ {\rm MeV}\ee
For the reasonable value $z_W=0.56$ this is compatible with observation.
The same parameter $z_W$ also determines the ratio (for $\mu_W^2>0$)
\be\label{10.36b}
\frac{\Gamma(K^0_S\to\pi^+\pi^-)}{\Gamma(K^0_S\to2\pi^0)}
=2\left(1+\frac{0.84(1-z_W)}{3.71}\right)^2=2.2\ee
This is in perfect agreement with the observed value 2.19, reflecting
the fact that the kaon decays are governed by only two independent
amplitudes.

The overall scales appearing in the kaon decays can be visualized
easily from the effective vertex which obtains from ``integrating
out'' the one-particle reducible $W$-boson exchange
\be\label{10.41AA}
{\cal L}^{\tilde W}_{\Delta S}=\frac{g_W^2}{8M^2_W}{\rm Tr}
\{(\phi^\dagger\partial_\mu\phi-\partial_\mu\phi^\dagger\phi)
\vec\tau_W\}{\rm  
Tr}\{(\phi^\dagger\partial^\mu\phi-\partial^\mu\phi^\dagger\phi)\vec\tau_W
\}+...\ee
where the dots stand for terms involving $\chi$. Comparison with
the leading invariant (\ref{9.E}) shows that essentially $\mu^2_W$ is
replaced by two powers of $\sigma_0$ or $\chi_0$. The resulting
relative suppression factor $f^2/\mu^2_W$ accounts for most of the
relative suppression of the $\Delta I=3/2$ amplitude. A smaller additional
suppression arises from the partial cancellation
$1-z_W=0.44$. We observe that the suppression is not visible
on the level of chiral perturbation theory. The invariant
relevant for the $\Delta I=3/2$ decays
\be\label{10.42AA}
{\cal L}_{\Delta I=3/2}=(1-z_W){\cal L}^{\tilde W}_{\Delta S}
=\frac{(1-z_W)g_W^2f^4}{32 M_W^2}{\rm Tr}\{U^\dagger
\partial^\mu U\vec\tau_W\}{\rm Tr}\{U^\dagger\partial_\mu U^\dagger
\vec\tau_W\}
\ee
contains two derivatives just as the one (\ref{9.F})
relevant for the $\Delta I=1/2$ decays.

In summary, the $\Delta I=1/2$ enhancement of the hadronic kaon
decays can be qualitatively explained by two ingredients. Since
for the pseudoscalar interactions only derivative terms are relevant,
the leading dimensionless operator consistent with the symmetries
is quadratic in $\phi$ or $\chi$. On the quark level it corresponds
to diagrams involving two left-handed quarks. These penguin-type diagrams
contribute only to $\Delta I=1/2$ decays. On the other hand,
the diagrams contributing to $\Delta I=3/2$ decays involve at least
four left-handed quarks. On the mesonic level they correspond
to effective operators involving at least four powers of $\phi$ or
$\chi$. Such higher dimension operators are suppressed according
to the hypothesis of scalar Vervoll\-st\"an\-di\-gung. Operators of this
type can arise from one particle reducible $W$-boson exchange
or from box-type diagrams where the $W$-boson exchange is
supplemented by gluon exchange. An estimate of the direct $W$-exchange
does not involve unknown parameters and shows indeed a small $\Delta
I=3/2$ amplitude. The box-type 1PI diagrams have the same structure
as the $W$-exchange contribution. Their negative relative sign
is related to the sign of the appropriate anomalous dimension
of the relevant four-quark operator. The corresponding factor
$(1-z_W)$ leads to a partial cancellation of the $\Delta I=3/2$ amplitude.
This further reduces the decay width for the hadronic decays of the charged
kaons. The small parameter appearing in the relative suppression
of the $\Delta i=3/2$ amplitude is the size of chiral symmetry
breaking $\sim f$ divided by a typical hadronic scale $\sim \mu_W$.

\bigskip

\section{Diquark condensates}
\setcounter{equation}{0}

At high baryon density one expects quark pairs to condensate
\cite{4}. In particular, the color-flavor locking \cite{CF} at high
density ressembles in several aspects the spontaneous color
symmetry breaking in the vacuum discussed in the preceeding
sections. In this section we reformulate the diquark condensation
in terms of effectice bosonic diquark fields $\Delta\sim\psi\psi$.
In contrast to the color octet $\chi\sim\bar\psi\psi$ the diquark
fields carry nonzero baryon number. If diquark condensation occurs
at high density, the global symmetry corresponding to baryon number is
spontaneously broken. There is therefore an order parameter which
distinguishes the high density phase with $<\Delta>\not=0$ from
the low density phase with $<\Delta>=0$. This implies the existence
of a true phase transition.

In addition to the  fields discussed previously we consider
in this section scalar
fields with the transformation properties of quark-quark pairs. In our  
notation, these additional scalar fields
are represented by complex $3\times 3$ matrices $\Delta_L, \Delta_R$ obeying
\be\label{10.1}
\delta\Delta_L=-i\Theta^T_C\Delta_L-i\Delta_L\Theta_L\quad,
\quad \delta\Delta_R=-i\Theta^T_C\Delta_R-i\Delta_R\Theta_R\ee
The ``quark pairs'' $\Delta_{L(R)}\sim\psi_{L(R)}\psi_{L(R)}$
belong to a color antitriplet and carry baryon number 2/3. The discrete  
symmetries act
as
$P:\ \Delta_L\leftrightarrow \Delta_R,
\quad C:\ \Delta_L\leftrightarrow \Delta_R^*$.
For an appropriate scaling of the fields, the most general effective
Lagrangian consistent with these symmetries and containing operators
up to dimension four is
\bear\label{10.2}
{\cal L}_\Delta &=&\ {\rm  
Tr}\{(\partial^\mu\Delta^\dagger_L-ig\Delta^\dagger_LA^*_\mu)(\partial
_\mu\Delta_L+ig A^T_\mu\Delta_L)\nonumber\\
&&\qquad +(\partial^\mu\Delta^\dagger_R-ig\Delta^\dagger_R A^*_\mu)
(\partial_\mu\Delta_R+ig A^T_\mu\Delta_R\}\nonumber\\
&&+U_\Delta(\phi,\chi,\Delta_L,\Delta_R)\nonumber\\
&&+[h_\Delta\ {\rm Tr}\{\Delta_L^\dagger\tilde\delta_L+\Delta
^\dagger_R\tilde\delta_R\}+c.c]\ear
Here the quark-quark bilinears $(\tilde\delta_L)_{ia},
(\tilde\delta_R)_{ia}$ are
defined as
\be\label{11.3}
(\tilde\delta_{L,R})_{ia}=(\psi_{L,R})_{bj\beta}c^{\beta\gamma}
(\psi_{L,R})_{ck\gamma}\epsilon_{ijk}\epsilon_{abc}\ee
They transform under all continuous symmetries precisely like
(bo\-so\-nic) antiquarks
$(\psi^\dagger_L)_{ia},(\psi^\dagger_R)_{ia}$. (Here $\beta,
\gamma=1,2$ are (Weyl-)spinor indices and  
$c^{\beta\gamma}=\epsilon^{\beta\gamma}$ is the appropriate
charge conjugation matrix. Our conventions \cite{Conv}
are chosen such that parity
transforms $\tilde\delta_L\leftrightarrow\tilde\delta_R$.)
One observes that $\Delta_L$ transforms as $\delta_L$ and
similarly for $\Delta_R$ and $\delta_R$. The
addition to the effective potential
\bear\label{11.4}
U_\Delta&=&V_\Delta(\Delta_L,\Delta_R)+\gamma_\phi\ Tr\ (\phi^\dagger
\Delta^\dagger_R
\Delta_L+\phi\Delta^\dagger_L\Delta_R)\nonumber\\
&&+\gamma_\chi[(\Delta_R)_{ia}\chi_{ij,ab}(\Delta^\dagger_L)_{bj}+(\Delta_L)
_{ia}\chi^*_{jiba}(\Delta_R^\dagger)_{bj}]\nonumber\\
&&+\epsilon[(\Delta_L\phi^\dagger)_{ia}\chi_{ij,ab}(\Delta_L^\dagger)
_{bj}+(\Delta_R\phi)_{ia}\chi^*_{jiba}(\Delta^\dagger_R)_{bj}\nonumber\\
&&+(\Delta_L)_{ia}\chi^*_{jiba}(\phi\Delta^\dagger_L)_{bj}+(\Delta_R)
_{ia}\chi_{ijab}(\phi^\dagger\Delta^\dagger_R)_{bj}]\nonumber\\
&&+...\ear
involves real parameters $\gamma_\phi,\gamma_\chi$, and
$\epsilon$. The dots stand for terms involving $\Delta,\Delta^\dagger$
and two powers of $\phi$ or two powers of $\chi$. The additional
potential $U_\Delta$
conserves the axial $U(1)$ symmetry. Electromagnetic and
weak interactions are implemented by inserting the appropriate
covariant derivatives.

In order
to visualize the physical content in this regime, it is convenient
to use  nonlinear coordinates in field space, given by
hermitean matrices $D_L,D_R$,
\be\label{11.5}
\Delta_L=v^\dagger D_LW_L^\dagger,\quad \Delta_R=v^\dagger D_R W
_R^\dagger \ee
Both $D_L$ and $D_R$ are color singlets with the
transformation properties
\be\label{11.6}
\delta D_L=i[\Theta_P,D_L],\quad \delta D_R=i[\Theta_P,D_R]\ee
With respect to the local reparametrization
symmetry and the physical global $SU(3)$-symmetry
$D_L$ and $D_R$ transform as singlets plus octets.
A possible expectation value of the singlets (proportional
to the unit matrix)
\be\label{11.7}
<D_L>=<D_R>=\delta_0\ee
also preserves parity and charge conjugation. For $\delta_0\not=0$
baryon number is, however, spontaneously broken. From
\be\label{11.8}
\delta_{em}\Delta_{L,R}=-i\beta\Delta_{L,R}\tilde Q\quad,
\quad\delta_{em}
D_{L,R}=i\beta[\tilde Q, D_{L,R}]\ee
one infers that the electromagnetic
$U(1)$-symmetry is modified in the Higgs
picture (similar to the octet condensate)
whereas the electric charge of $\delta_0$ is zero.

The diquarks $\Delta_{L,R}$ carry baryon number $\tilde B=2/3$.
From
\be\label{11.9}
\Delta_L^\dagger\Delta_L=W_LD^2_LW_L^\dagger\ ,\
\Delta_R^\dagger\Delta_R=W_RD_R^2W_R^\dagger\ee
and the fact that $W_{L,R}$ have $\tilde B=0$,
one concludes that $D_L$ and $D_R$ cannot carry nonzero $\tilde B$.
The definition of the nonlinear fields (\ref{11.5}) implies then
directly that $v^\dagger$ must have $\tilde B=2/3$
in accordance with the discussion in sect. 6. In fact, up
to reparametrization transformations the decomposition of
$\phi,\Delta_L$ and $\Delta_R$ into the nonlinear fields is
unique, except of special points in field space where some
fields are not invertible or degenerate. For $D_L=D_R=\delta$,
$\delta\not=0$
the association of $v^\dagger=\Delta_LW_L/\delta$ with a
diquark field is particularly apparent.

For small $\Delta_L=\Delta_R=\delta=\delta^*$
one may expand $V_\Delta=m^2_\Delta\ {\rm Tr}\{\Delta_L^\dagger
\Delta_L+\Delta_R^\dagger\Delta_R\}$ such that $U_\Delta=\frac{1}{2}
M^2_\delta\delta^2+...$ with
\be\label{11.11}
M^2_\delta=12 m^2_\Delta+12\gamma_\phi\sigma_0+\frac{32}{\sqrt{6}}
\gamma_\chi\chi_0(1+2\epsilon\sigma_0)+...\ee
Typically, the masses and couplings will depend on the baryon
chemical potential $\mu_B$. Baryon number conservation in the
vacuum requires $M^2_\delta>0$ for $\mu_B=0$. In the spirit
of the discussion in sect. 2 the couplings also depend on
the renormalization scale $k$. In particular, for large $k$
(and $\mu_B=0$)
perturbative QCD should lead to a positive $m_\Delta$ and one can neglect
$\sigma_0$ and $\chi_0$. In presence of quarks the field equation for
$\Delta_{L,R}$ becomes then (for $q^2\ll m_\Delta^2$)
\be\label{11.12}
\Delta_{L,R}=-\frac{h_\Delta}{m_\Delta^2}\tilde\delta_{L,R}\ee
This shows the connection between the composite fields $\Delta_{L,R}$
and quark-quark bilinears.

For $k=0$ and large baryon density (large $\mu_B$) one expects
the minimum of $U_\Delta$ to occur for $\delta_0\not=0$. The
particle content in the sectors of pseudoscalars, vector mesons
and fermions may again be obtained from a non-linear representation by
setting $D_L=D_R=\delta_0$ in eq. (\ref{11.5}). As
before, the field $v$ disappears from the effective action and only
color singlets for possible physical excitations remain. Inserting
eq. (\ref{11.5}) into eq. (\ref{10.2}) leads to expressions that
are similar to sects. 4,5 in many respects. This is not surprising
-- except for the spontaneous breaking of baryon symmetry the symmetries
are the same.

At this place we note that for $\delta_0\not=0$ there are
terms in the effective potential  linear in $\chi$. They arise
from the cubic term $\sim \gamma_\chi$
or the quartic term $\sim\epsilon$ in eq. (\ref{11.4}).
As a consequence, no solution $\chi_0=0$ is possible for $\delta_0\not=0$
-- the diquark condensation necessarily induces a color octet
$<\bar\psi\psi>$-condensate! (In the limit where the nonlinear
part of the effective potential for $\chi_0$ is dominated
by a quadratic term $\sim\frac{1}{2}\mu^2_\chi
(\sigma_0,\delta_0)\chi_0^2$, the expectation value is given
by $\chi_0=-16\gamma_\chi(1+2\epsilon\sigma_0)\delta^2_0/
(\sqrt6\mu_\chi^2)$.)

In our scenario the octet condensate $\chi_0$ is therefore
different from zero both for small and large baryon density.
The only qualitative change at high density is the spontaneous
breaking of baryon number by $\delta_0\not=0$. A second- or first-order
phase transition line separates the low density phase with $\delta_0
=0$ from the ``quark matter'' phase at high density. Due to the
spontaneously broken abelian baryon number
symmetry quark matter is a superfluid \cite{CF}. If the phase
transition is of second order it belongs to the universality class of the
$O(2)$-Heisenberg model. It is plausible that the transition line
to quark matter could bifurcate from the line of first order
transitions between a nucleon gas and nuclear matter (nucleon
liquid). (This latter transition line ends in an endpoint
with critical Ising behavior.) If so, there is no distinction
between nuclear matter and quark matter at low enough temperature.
For $T=0$ both nuclear and quark matter would be superfluids
with spontaneously broken baryon number. They should be identified
thus leading to a further manifestation of quark-baryon duality.

\section{Heavy quarks}
\setcounter{equation}{0}
In this short section we extend our description to
the heavy quarks carrying quantum numbers of charm, beauty and
topness. These quarks are neutral under the global $SU(3)_L\times SU(3)_R$
symmetry. With respect to the physical $SU(3)_V$ symmetry they therefore
transform as antitriplets. The physical electric charge has a
contribution from $Q_c$ similar to table 1. In consequence, the
charmed fermions carry electric charge 0, 1, 1. In complete analogy
with the discussion in sect. 6 they also carry integer baryon
number. We identify them with the charmed baryon antitriplet
$\Xi^0_c(2472\ {\rm MeV}, dsc)$, $\Xi^+_c$ (2466 MeV, $usc)$ and $\Lambda_c^+$  
(2284 MeV, $udc)$ where we have given the
mass and the standard quark content in brackets. In the nonlinear
language the charmed baryons consist of the $c$-quark and the
nonlinear diquark field $v^\dagger$ which carries the quantum numbers of
two light quarks. Similarly, the fermions with beauty carry electric
charge -1,0,0. They are identified with
$(dsb),\ (usb)$ and $\Lambda^0_b$ (5641 MeV, $udb)$. The quantum
numbers of the fermions with topness are the same as the charmed
baryons. The weak interactions follow from the standard assignment
into doublets with the appropriate Kobayashi-Maskawa matrix.

The description of the charmed mesons needs the introduction
of scalar fields for the bilinears $q\bar c, \bar qc, \bar cc$ with $q$
representing the three light quarks $(u,d,s)$. As for the light
quark-antiquark pairs these scalars transform as singlets or
octets with respect to the color symmetry $SU(3)_C$. From the transformation
properties with respect to $SU(3)_L\times SU(3)_R\times SU(3)_C$
we can read the representations of the physical $SU(3)_V$-group
\bear\label{X12.1}
(q\bar c):&&(3,1,8+1)\longrightarrow 3+3+\bar 6+15\nonumber\\
&&(1,3,8+1)\longrightarrow 3+3+\bar 6+15\nonumber\\
&&\nonumber\\
(c\bar c):&& (1,1,8+1)\longrightarrow 8+1\ear
Only the singlet $(c\bar c)$ can acquire a vacuum expectation
value consistent with $SU(3)_V$-symmetry. We also note that
a direct Yukawa coupling of the charmed quark to $\phi$ or $\chi$
is not allowed by the chiral $SU(3)_L\times SU(3)_R$ symmetry.
The relation between current quark mass and baryon mass (or
constituent quark mass) may therefore differ between charmed quarks
and up, down, or strange quarks. In summary, the charmed particles
do not constitute an obstacle for the picture of spontaneous
breaking of color.
The quantum numbers of the physical fields agree with those of
observed particles. The same holds for beauty and top.

\section{Conclusion}
\setcounter{equation}{0}

We conclude that the ``spontaneous breaking'' of color
is compatible with observation. The simple effective action
(\ref{2.6}) gives
a realistic approximate description of the masses of all low-lying
mesons and baryons and of their interactions. Gluon-meson duality
is associated to the well-known Higgs phenomenon with colored composite
scalar fields,  corresponding to quark-antiquark pairs.

The most important characteristics of our scenario can be summarized
in the following points.

(i) The Higgs mechanism generates a mass for the gluons. In the
limit of equal (current) masses for the three light
quarks the masses of all gluons are equal. This leads to a
simple picture of confinement: If one places color charges
at different positions in the vacuum, the gauge fields cannot
vanish. Due to the nonzero mass of the gauge fields, the energetically
most favorable configurations are color-magnetic and color-electric
flux tubes. This is in complete analogy with the Meissner effect
in superconductors,  with the additional ingredient that the Yang-Mills
self-interactions link color-electric and color-magnetic fields.
The string tension of the flux tubes provides for a simple
mechanism for the confinement of color charges. More precisely,
these statements hold only as long as string breaking due to
quark-antiquark pairs is energetically suppressed. One may attempt
a computation of the string tension appearing in the heavy
quark potential by solving the field equations derived from the
effective action (\ref{2.6}) in presence of static color charges.

(ii) The physical fermion fields are massive baryons with integer
electric charge. The low mass baryons form an octet with respect
to the $SU(3)$-symmetry group of the ``eightfold way'' In a gauge-invariant
language the baryons are quarks with a dressing of nonlinear fields.
In a gauge-fixed version quarks and baryons can be described
by the same field. This is quark-baryon duality. The
main contribution to the mass of these baryons arises from chiral
symmetry breaking through quark-antiquark condensates in the color
singlet and octet channels. These considerations extend to the heavy
quarks $c, b,t$, except that the
mass is now dominated by the current quark mass.
The lightest charmed baryons (and $t$-baryons)
belong to a $SU(3)$-antitriplet with electric charge 0,1,1.
Correspondingly, the
$SU(3)$-antitriplet of the lightest $b$-baryons carries electric
charge -1,0,0.

(iii) Our scenario shares the properties of chiral symmetry breaking with
many other approaches to long-distance strong interactions. In
particular, chiral perturbation theory is recovered in the low energy
limit. This guarantees the observed mass pattern for pions, kaons
and the $\eta$-meson and the structure of their (low momentum)
interactions.

(iv) Spontaneous color symmetry breaking generates a nonlinear
local $SU(3)_P$-reparametrization symmetry. The gauge bosons of
this symmetry originate from the gluons. They form the octet of
low masses physical vector mesons.  (The singlet discussed in
sect. 8 can be associated with the gauge boson of the abelian
$U(1)_P$-symmetry and the ninth vector meson.) The symmetry relations
following from $SU(3)_P$ symmetry appear in the electromagnetic
and strong interactions of the vector mesons. They are compatible
with observation and provide for a successful test of our scenario.

(v) The $\Delta I=1/2$ rule for the weak hadronic kaon decays
arises naturally once weak interactions are incorporated into our
picture. It is a consequence of the properties of the lowest
dimension operators which are consistent with the symmetries.

Beyond the important general symmetry relations arising from the
nonlinear local reparametrization symmetry the simple effective
action (\ref{2.6}) leads to particular predictions. They are related
to the assumption that the effective action can be described in leading
order by effective couplings with positive or zero mass dimension. This
should hold once composite scalar fields are introduced for
quark-antiquark bilinears and counted according to the canonical
dimension for scalar fields. Expressed in other words we assume
that the dominant nonperturbative effects in QCD result in the
formation of scalar and pseudoscalar bound states and in large
effective couplings. This assumption
of scalar Vervoll\-st\"andi\-gung can be tested by comparing
the predictions of the effective action (\ref{2.6})
with observation. For this purpose we fix the parameters
$\chi_0, \sigma_0$ and $g$ by $M_\rho,f_\pi$ and $\Gamma(\rho
\to e^+e^-$). More precisely, we use here eqs. (5.16), (5.17)
(7.14) with $\overline M_\rho$= 850 MeV, $f_0$= 128 MeV, $g_{\rho
\gamma}=0.12\ {\rm GeV}^2$. (With $g=6$ we may take the limit
$x\to\infty$ for many expressions, keeping in mind that
the precise value of $x$ is needed mainly for a determination
of $\sigma_0$.) In addition, the Yukawa couplings $h, \tilde h$
are fixed by the baryon masses $M_8, M_1$ (cf. eq. (4.10))
whereas the strength of the chiral anomaly $\nu$
and $\nu'$ determines  $M_{\eta'}^2$ by eq. (5.19). (The precise ratio
$\nu'/\nu$ is not relevant here.) We concentrate on the following
predictions of scalar Vervollst\"andigung
which are not dictated purely by symmetry considerations:

(1) The pion nucleon couplings are found as $F=D=0.5$ to be
compared  with the observed values $F=0.459\pm0.008,\ D=0.798\pm 0.008$.

(2) The decay width $\Gamma(\rho\to2\pi)=115$ MeV turns out to be
somewhat lower than the observed value of 150 MeV. (Note that
we use here directly eq. (\ref{7.2}) and $\kappa_f$ equals one in
leading order, such that $g_{\rho\pi\pi}=4.6$.)

(3) The direct coupling of the photon to pions is suppressed
$g_{\gamma\pi\pi}/e=0.04$ (by virtue of eq. (\ref{E10}) with
$M_\rho\to \overline M_\rho,\ f_\pi\to f_0$).
This phenomenon of vector dominance describes well the
observations.

(4) The effective next-to leading order couplings
$L_1, L_2, L_3$ of chiral perturbation theory come out
compatible with observation. In particular, scalar
Vervollst\"andigung predicts $L_2=1.7\cdot 10^{-3},
L_3=-(3.9-5.2)\cdot10^{-3}$.

(5) The inclusion of weak interactions
leads to a suppression of the $\Delta I=3/2$ hadronic kaon
decays compared to the ones with $\Delta I=1/2$. A realistic
picture of the three hadronic kaon two-body decays is obtained. This
involves, however, two additional parameters.

For a first approximation we consider the effective action (\ref{2.6})
as very satisfactory. Corrections arise from two sources: (a)
The nonvanishing quark masses, in particular the strange quark
mass, break the physical global
$SU(3)$-symmetry. (b) Higher order invariants are certainly
present in the full effective action. One important source
relevant for the vector mesons is discussed in appendix A.
Some of the corrections are known phenomenologically as, for example,
the partial Higgs effect which reduces the average meson decay
constant from
$f_0$ to the mean value $f=(f_\pi+2f_K)/3$=106 MeV.
Together with the further lowering of $f_\pi$ as compared
to $f$ by $SU(3)_V$ violation this leads to a realistic
value of $\Gamma(\rho\to2\pi)$. The neglected
higher-order invariants can be used to improve the agreement with
observation. Typically, these corrections are below 30 \%.

Once the singlet vector meson $S_\mu$ is added (cf. eq. (\ref{8.7}))
the effective action (\ref{2.6}) may be viewed as the
lowest order of a twofold systematic
expansion. First, higher (``non-renormalizable'')
invariants involving additional powers of $\phi$ or $\chi$
lead to corrections $\sim f^2/\mu_s^2\approx 0.1-0.3$ with
$\mu_s$ a typical strong interaction scale. We have encountered
this factor in hadronic weak decays $(\mu_s \hat=\mu_W)$, the
partial Higgs effects (\ref{5.8a}), (\ref{V27}) $(\mu_s\hat=
\mu_\rho/e_{\rho\pi\pi})$ or the coupling between nucleons
and the axial vector pion current $g_A-1\sim y_N$ (\ref{9.1a}),
(\ref{12.54a}) $(\mu_s\hat= \mu_\rho/
\sqrt{c_{\rho\bar qq}e_{\rho\pi\pi}}
)$. Second, the expansion in the number of derivatives involves some
characteristic squared momentum $q^2/M^2_s$ as a small parameter.
Typically $M_s$ may be of the size of the mass of particles
not included in our description, i. e. $M_s\approx 1$ GeV. A typical operator
is $(Z_\psi/M_s)\bar\psi_i[\gamma_\mu,\gamma_\nu]G^{\mu\nu}_{ij}\psi_j$.
This is responsible for the anomalous magnetic moment of the
nucleons via the mixing between the photon and the gluon $\tilde G_\mu$
(see sects. 3,7). We observe a typical relation $M_s/\mu_s=c_s$ with
$c_s$ around three a dimensionless coupling constant of the
type $e_{\rho\pi\pi}$. Furthermore, the expansion
in $SU(3)$ violating effects originating from the
strange quark mass is governed \cite{8} by $(f_K-f_\pi)/f\approx 0.2$.
This last expansion may remain within the hypothesis of scalar
Ver\-voll\-st\"andi\-gung.

The parameters of the model being fixed additional predictions become
possible. An example are the rare electromagnetic decays of mesons.
Particularly interesting is the strangeness splitting of masses
of the various $SU(3)_V$-muliplets. This involves, however,
some additional parameters
of the effective scalar potential (\ref{2.7}). On a phenomenological
basis the $SU(3)_V$-violation will apear in the form of different
expectation values of $\phi$ and $\chi$ in the strange and
nonstrange directions.

The present framework opens new perspectives for a calculation
of the properties of hadronic matter at high temperature and density.
Especially for high density the issue of baryons vs. quarks plays
a crucial role due to the different Fermi surfaces \cite{7}.
Quark-baryon duality allows for a simple approach to this problem
by using only one field.

Quark-baryon duality has profound implications. The nonrelativistic
quark model where baryons are bound states of three quarks is
now supplemented by the view of baryons as ``dressed quarks''. In
a high energy scattering process ``physical quarks'' will come out,
but they will come out with the quantum numbers and masses of
baryons. In this sense, quarks are not confined particles, despite
the fact that color charges remain exactly confined and cannot
appear connected with free particles.
When leaving the interaction region, the physical
particles (asymptotic states) acquire always
a dressing by (nonlinear) fields that makes them color neutral
and integer charged. (In precise formulations of
quantum field theory this also happens to electrons --
the physical particles being neutral with respect to the
electromagnetic gauge symmetry.) This view may have important
consequences for our picture of the parton model, both for structure
functions at small $Q^2$ and for fragmentation.

Let us finally address this perhaps most important open
issue in our approach,
i.e. its
connection to the parton model at high $Q^2$ and to the
nonrelativistic quark model which describes rather well the
higher bound states. In short, one has to understand the correspondence
between quarks dressed by a cloud of diquarks and bound states of
three quarks. The effective action (\ref{2.6}) is thought to be
valid at low momenta. Going to higher momenta, the couplings become
typically momentum-dependent.
This also holds for the definition of the nonlinear fields, e.g. $Z_\psi\to  
Z_\psi(-D^\mu D_\mu)$ in eq. (\ref{4.1}). Along the lines of the
discussion in sect. 2 the effective action at high momentum (or, more
precisely, large off-shell momentum) should be given by perturbative
QCD, with $g$ the perturbative gauge coupling and $Z_\psi(-D^2)\approx 1$.
The scalar part of the effective action
should express appropriate 1PI contributions in the
perturbative quark-gluon language. For high momentum scattering
the dominant effect of spontaneous color symmetry breaking is then
the mass term for the gluons which provides for an infrared cutoff.
``Infrared safe'' quantities which can be  reliably described by
perturbative QCD are not very sensitive to this mass.

In a momentum basis the nonlinear field decomposition (\ref{4.1})
reads for $Z_\psi(-D^2)=1$
(with $\int_p=\int\frac{d^4p}{(2\pi)^4}$)
\be\label{13.1}
\psi_L(\tilde p)=\int_p\int_{p'}W_L(\tilde p-p-p')N_L(p)v(p')\ee
This may be inserted, for example, into the electromagnetic vertex
\be\label{13.2}
\sim\tilde e{\rm Tr}\{\bar\psi(\tilde p+q)\gamma^\mu\tilde Q\psi(\tilde p)\}
\tilde B_\mu(q).\ee
The scattering of an off-shell photon with momentum
$q$ and an off-shell quark with momentum $\tilde p$ to an off-shell
quark with momentum $\tilde p+q$ describes  the annihilation
of an on-shell proton $N$ with momentum $p$ and
simultaneous production of the fields contained in
the ``spectator jet'' $Y(p-\tilde p)$. Here the spectators consist of
the particles described by the nonlinear fields $Y\sim W^\dagger\circ  
v^\dagger$, with $\circ$ denoting the appropriate algebra for the
particular combination relevant for the proton. The ``parton''
$\psi(\tilde p)$ carries a (longitudinal) fraction $x=
-q^2/(2(pq))\approx(\vec{\tilde p}\vec p)/(\vec p^2)$
of the spacelike momentum of the proton. Similarly, the produced
quark $\psi(\tilde p+q)$ ``fragments'' into the hadrons $X$ contained in
its nonlinear decomposition (\ref{13.1}). In a process with high
off-shell parton momenta final states involving many on-shell hadrons
become kinematically accessible . On a hadronic level, this
corresponds to inelastic multihadron production $N+\gamma^*\to X+Y$.
If the parton structure functions can be suitably implemented
in this setting\footnote{It is conceivable that the nonlinear
constraints $v^\dagger v=1,\ W^\dagger W=1$ are not appropriate
at high off-shell momenta. Fluctuations away from the
nonlinear constraints may have to be included.}
the conservation of global quantum numbers should lead to the
appropriate sum rules. A deviation of $Z_\psi$ from one modifies
the composition of $X$ and $Y$ without effecting the general
setting.

Instead of the momentum dependence of the effective couplings
one may also investigate the related (but not identical!)
dependence of the effective action $\Gamma_k$ on an appropriate
infrared cutoff $k$ (see sect. 2). The use of nonperturbative
flow equations \cite{5} for $\Gamma_k$ may hopefully permit to connect
perturbative QCD with the effective action $\Gamma_0$ (\ref{2.6}).

These last ideas are only sketches for further developments. The
phenomenological success of the simple low momentum effective
action proposed in this paper should motivate a serious
investigation in these directions. We hope that quark-baryon duality
will permit further insights into the connection between the parton
model and the QCD-vacuum.

\section*{Acknowledgement}
This work was supported by the TMR-network FMRX-CT97-0122 and
by the Deutsche Forschungsgemeinschaft We 1056/3-2.

\section*{Appendix A: Four-quark interactions in the color singlet vector
channel}
\renewcommand{\theequation}{A.\arabic{equation}}
\setcounter{equation}{0}
In this appendix we discuss the effects of four-quark interactions in
the color singlet vector and axial vector channels. As for the scalar and
pseudoscalar channels the effective action is described in a partially
bosonized form. For this purpose we introduce fields for the composite
quark-antiquark bilinears $\rho^\mu_L, \rho^\mu_R$ which correspond
to $\bar\psi_i\gamma^\mu(1\pm\gamma^5)\psi_i$.

Let us add to the effective action (\ref{2.6}) a piece accounting for the
four-quark interactions in the vector channel. We express it in terms of
effective interactions\footnote{We follow here closely appendix B
of ref. \cite{8}. The normalization of the
matrix-valued fields $\rho_{L,R}^\mu$
used here differs by a a factor of two from \cite{8}.} for
left-handed and right-handed vector fields $\rho_L^\mu,\ \rho^\mu_R$.
\be\label{EQ1}
{\LL}_\rho={\LL}_{\rho\bar qq}+{\LL}_{\rho^2}+{\LL}_{\phi^2\rho}
+{\LL}_{\phi^2\rho^2}+{\LL}_{\chi^2\rho}+{\LL}_{\chi^2\rho^2}+{\LL}_
{\rho B}\ee

\bear\label{EQ2}
{\cal L}_{\rho\bar qq}&=&\sqrt2 Z_\psi c_{\rho\bar q q}{\rm  
Tr}\{\bar\psi_L\gamma_\mu\tilde\rho_L^\mu\psi_L+\bar\psi_R\gamma_\mu
\tilde\rho_R^\mu\psi_R\}\nonumber\\
&&+\frac{\sqrt2}{3}Z_\psi\tilde c_{\rho\bar qq}({\rm Tr}\{{\bar\psi}_L\gamma
_\mu\psi_L\}{\rm Tr}\rho^\mu_L+{\rm Tr}\{\bar\psi_R\gamma_\mu
\psi_R\}{\rm Tr}\rho^\mu_R)\ear
\bear\label{EQ3}
{\LL}_{\rho^2}&=&
\Bigl[\frac{1}{2}{\rm Tr}\{(D_\mu\rho_{L\nu}-D_\nu\rho_{L\mu})
(D^\mu\rho^\nu_L-D^\nu\rho^\mu_L)\\
&&+\frac{1}{\alpha_\rho}{\Tr}\{(D_\mu\tilde\rho_L^\mu)(D_\nu\tilde  
\rho_L^\nu)\}+\frac{1}{3\alpha_\rho'}
(\partial_\mu{\Tr}\rho_L^\mu)^2+\mu^2_\rho{\Tr}\{\tilde\rho_{L\mu}
\tilde\rho^\mu_L\}+(L\leftrightarrow R)\Bigr]\nonumber\\
&&+\frac{1}{6}\mu^2_A{\Tr}\{\rho_L^\mu-\rho^\mu_R\}{\Tr}
\{\rho_{L\mu}-\rho_{R\mu}\}+\frac{1}{6}\mu^2_V{\Tr}
\{\rho^\mu_{L\mu}+\rho^\mu_R\}{\Tr}\{\rho_{L\mu}+\rho_{R\mu}\}\nonumber\ear
\bear\label{EQ4}
{\LL}_{\phi^2\rho}&=&i\sqrt2  
c_{\rho\pi\pi}{\Tr}\{(D_\mu\phi^\dagger\phi-\phi^\dagger  
D_\mu\phi)\tilde\rho_L^\mu+(D_\mu\phi\phi^\dagger-\phi  
D_\mu\phi^\dagger)\tilde\rho_R^\mu\}\nonumber\\
&&+i\frac{\sqrt2}{3}\tilde c_{\rho\pi\pi}{\Tr}
\{D_\mu\phi^\dagger\phi-\phi^\dagger D_\mu\phi\}({\Tr}\rho_L^\mu-
{\Tr}\rho_R^\mu)\ear
\bear\label{EQ5}
{\cal L}\phi^2\rho^2&=&2(c^2_{\rho\pi\pi}+f_1)({\Tr}\{\phi^\dagger\phi
\tilde\rho_L^\mu\tilde\rho_{L\mu}\}+{\Tr}
\{\phi\phi^\dagger\tilde\rho_R^\mu\tilde\rho_{R\mu}\})\nonumber\\
&&
-4(c^2_{\rho\pi\pi}+f_2){\Tr}\{\phi^\dagger\tilde\rho_R^\mu\phi\tilde
\rho_{L\mu}\}-\frac{4}{3}\tilde f_2({\Tr}\{\phi^\dagger\phi\}
-3\sigma^2_0){\Tr}\rho^\mu_R{\Tr}\rho_{L\mu}\nonumber\\
&&+4f_3({\Tr}\{\phi^\dagger\phi\}-3\sigma_0^2){\Tr}\{\tilde\rho^\mu
_L\tilde\rho_{L\mu}+\tilde\rho_R^\mu\tilde\rho_{R\mu}\}\nonumber\\
&&+\frac{4}{3}\tilde f_3({\rm Tr}\{\phi^\dagger\phi\}
-3\sigma_0^2)({\Tr}\rho_L^\mu{\Tr}\rho_{L\mu}+{\Tr}\rho_R^\mu{\Tr}\rho_{R\mu})
\ear
\bear\label{EQ6}
{\LL}_{\chi^2\rho}&=&i\sqrt2 e_{\rho\pi\pi}\{[(D_\mu\chi)^*_{ijab}\chi_{ijac}
-\chi^*_{ijab}(D_\mu\chi)_{ijac}](\tilde\rho_L^\mu)_{cb}\\
&&
+[(D_\mu\chi)_{ijab}\chi^*_{ijcb}-\chi_{ijab}(D_\mu\chi)^*
_{ijcb}](\tilde\rho_R^\mu)_{ca}\}\nonumber\\
&&+i\frac{\sqrt2}{3}\tilde e_{\rho\pi\pi}[(D_\mu\chi)^*_{ijab}\chi_{ijab}-
\chi^*_{ijab}(D_\mu\chi)_{ijab}]({\Tr}\rho^\mu_L-{\Tr}\rho_R^\mu)
\nonumber\ear
\bear\label{EQ7}
{\LL}_{\chi^2\rho^2}&=&e_1[\chi^*_{ijab}\chi_{ijac}(\tilde\rho_L^\mu)_{cd}
(\tilde\rho_{L\mu})_{db}+\chi_{ijab}\chi^*_{ijcb}(\tilde\rho^\mu_R)
_{cd}(\tilde\rho_{R\mu})_{da}]\nonumber\\
&&+e_2\chi_{ijab}^*(\tilde\rho_R^\mu)_{ac}\chi_{ijcd}(\tilde\rho_{L\mu})
_{db}\nonumber\\
&&+\frac{1}{3}\tilde e_2(\chi^*_{ijab}\chi_{ijab}-\frac{4}{3}
\chi_0^2){\Tr}\rho_R^\mu{\Tr}\rho_{L\mu}\nonumber\\
&&+e_3(\chi^*_{ijab}\chi_{ijab}-\frac{4}{3}\chi_0^2)
{\Tr}\{\tilde\rho_L^\mu\tilde\rho_{L\mu}+\tilde \rho_R^\mu\tilde\rho_{R\mu}\}
\nonumber\\
&&+\frac{1}{3}\tilde e_3(\chi^*_{ijab}\chi_{ijab}-\frac{4}{3}
\chi_0^2)({\Tr}\rho_L^\mu{\Tr}\rho_{L\mu}+{\Tr}\rho_R^\mu{\Tr}\rho_{R\mu}
)\ear
\be\label{EQ8}
{\LL}_{\rho B}=\frac{1}{6}e_{\rho\gamma\gamma}{\Tr}\{\partial_\mu
\rho_{L\nu}-\partial_\nu\rho_{L\mu}+\partial_\mu
\rho_{R\nu}-\partial_\nu\rho_{R\mu}\}\tilde B^{\mu\nu}\ee
Here we have defined
\be\label{V2}
\tilde\rho_{L,R}^\mu=\rho^\mu_{L,R}-\frac{1}{3}{\Tr}\rho_{L,R}^\mu=
\frac{1}{2}\tilde\rho^{z\mu}_{L,R}\lambda_z\ee
and the electromagnetic covariant derivatives read
\be\label{V3}
D_\mu\rho^\nu_{L,R}=\partial_\mu\rho^\nu_{L,R}-i\tilde e B_\mu
[\tilde Q,\rho_{L,R}^\nu]\ee
Up to omitted terms involving three or four vector fields
$\rho_{L,R}^\mu$ this is the most general effective action
with terms up to dimension four and consistent with chiral,
color, electromagnetic and discrete symmetries. The fields
$\rho_{L,R}^\mu$ are color-neutral and transform with respect to
chiral $SU(3)_L\times SU(3)_R$ as octets $\tilde\rho_{L,R}^\mu$
and singlets ${\Tr}\rho^\mu_{L,R}$
\be\label{V4}
\delta\rho_{L\mu}=i[\Theta_L,\rho_L^\mu],\
\delta\rho^\mu_R=i[\Theta_R,\rho^\mu_R]\ee
whereas the discrete transformations read
\bear\label{V5}
&&P:\quad \rho_L^\mu\leftrightarrow \rho^\mu_R\nonumber\\
&&C:\quad \rho_L^\mu\to -(\rho_R^\mu)^T,\quad\rho_R^\mu\to-
(\rho_L^\mu)^T\ear

As discussed in sect. 2, the effective four-quark interactions
in the vector channel (as well as other multiquark interactions)
can be recovered from (\ref{EQ1}) by solving the field equations
for $\rho_L^\mu$ and $\rho_R^\mu$ as functionals of $\psi,\phi$ and
$\chi$. It is obvious that the reinsertion of this solution into the
effective action produces an addition to the effective action (\ref{2.6})
for $\psi, \phi,\chi$ and $A_\mu$ which now also contains terms of
dimensions larger than four. In this sense the piece (\ref{EQ1})
can be interpreted as a suggestion for the most important higher
order operators not yet contained in eq. (\ref{2.6}). In particular, the  
induced effective four-quark interaction for low momenta
reads
\bear\label{12.25a}
\Delta{\cal L}_\rho&=&-Z^2_\psi[\tau_V\ {\rm Tr}
\{\bar\psi\gamma^\mu\lambda_z
\psi\}\ {\rm Tr}\{\bar\psi\gamma_\mu\lambda_z\psi\}\nonumber\\
&&-\tau_A\ {\rm Tr}\{\bar\psi\gamma^\mu\gamma^5\lambda_z
\psi\}\ {\rm Tr}\{\bar\psi\gamma_\mu\gamma^5\lambda_z\psi\}]
\nonumber\\
&&-\frac{Z^2_\psi\tilde c^2_{\rho\bar qq}}{12\mu_V^2}\ {\rm  
Tr}\{\bar\psi\gamma^\mu\psi\}\ {\rm Tr}\{\bar\psi\gamma_\mu\psi\}\nonumber\\
&&-\frac{Z_\psi^2\tilde c^2_{\rho\bar qq}}{12\mu_A^2}\ {\rm  
Tr}\{\bar\psi\gamma^\mu\gamma^5\psi\}\ {\rm Tr}\{\bar\psi\gamma_\mu
\gamma^5\psi\}\ear
with
\be\label{12.25b}
\tau_V=\frac{c^2_{\rho\bar qq}}{8M^2_V}\quad ,\quad
\tau_A=\frac{c^2_{\rho\bar qq}}{8M^2_A}\ee
The first term contributes to the isospin triplet vector channel
for the nucleon-nucleon interactions (with $M_V$ and $M_A$ given below).

The effective interactions in (\ref{EQ1}) involve many new parameters.
We concentrate here only on the most important modifications
as compared to the effective action (\ref{2.6}). First of all, the
fields $\rho^\mu_L$ and $\rho^\mu_R$ contain vector $(r_V^\mu)$ and
axial vector $(\rho_A^\mu)$ degrees of freedom
\be\label{V6}
r^\mu_V=\frac{1}{\sqrt2}(\rho_R^\mu+\rho^\mu_L),\quad
\rho^\mu_A=\frac{1}{\sqrt2}(\rho^\mu_R-\rho^\mu_L)\ee
The squared mass for the axial vector octet obtains as
\be\label{V7}
M_A^2=\mu^2_\rho+2\sigma_0^2(2c^2_{\rho\pi\pi}+f_1+f_2)
+(\frac{2}{9}e_1+\frac{1}{36}e_2)\chi^2_0\ee
whereas the axial vector singlet mass term is given by $\mu_A^2$. The
vector field $r_V^\mu$ provides now for the missing singlet vector meson
\be\label{V8}
S_\mu=\sqrt{\frac{2}{3}}{\Tr}\  r^\mu_V\ee
with squared mass $\mu^2_V$. We note that the axial vectors cannot
mix with the vectors because they have different parity. Also the
singlets cannot mix with the octets in the limit of an unbroken
physical $SU(3)$ symmetry. On the other hand, the octet in $r^\mu_V$
has the same transformation properties with respect to the physical
global symmetries as the gluons. One therefore expects a mixing
with the fields in $\tilde V^\mu$.

For an investigation of the interactions with the pseudoscalar mesons
we introduce again nonlinear fields
\be\label{V9}
T^\mu_L=W_L^\dagger\rho^\mu_LW_L,\quad T^\mu_R=W_R^\dagger\rho_R^\mu W_R\ee
and we replace eq. (\ref{V6}) by
\be\label{10}
r^\mu_V=\frac{1}{\sqrt2}(T^\mu_R+T^\mu_L),\quad\rho^\mu_A
=\frac{1}{\sqrt2}(T_R^\mu-T^\mu_L)\ee
These fields transform homogeneously as octets and singlets under
the local reparametrization symmetry
\be\label{V11}
\delta T^\mu_{L,R}=i[\Theta_P,T_{L,R}^\mu]\ee
and are neutral with respect to $SU(3))_L\times SU(3)_R\times
SU(3)_C$. They also transform homogeneously with respect to the
electromagnetic gauge symmetry
\be\label{V12}
\delta_{em}r^\mu_V=i\beta[\tilde Q,r^\mu_V],\quad
\delta_{em}\rho_A^\mu=i\beta[\tilde Q,\rho_A^\mu]\ee
If we define
\bear\label{V12a}
\hat D_\mu r_{V\nu}&=&\partial_\mu r_{V\nu}-ie\tilde B_\mu[\tilde
Q,r_{V\nu}]+i[\hat v_\mu,r_{V\nu}]+i[\hat a_\mu,\rho_{A\nu}]\nonumber\\
\hat D_\mu\rho_{A\nu}&=&\partial_\mu\rho_{A\nu}-ie\tilde B_\mu
[\tilde Q,\rho_{A\nu}]+i[\hat v_\mu,\rho_{A\nu}]+i[\hat a_\mu,
r_{V\nu}]\ear
\be\label{V12b}
\hat a_\mu=-\frac{i}{2}(W_R^\dagger D_\mu W_R-W_L^\dagger D_\mu W_L)\ee
the terms quadratic in $\rho$ read
\bear\label{V12c}
{\LL}_{\rho^2}&=&\frac{1}{2}{\Tr}\{(\hat D_\mu r_{V\nu}-\hat D_\nu r_{V\mu})
(\hat D^\mu r_V^\nu-\hat D^\nu r_V^\mu)\}\nonumber\\
&&+\frac{1}{2}{\Tr}\{(\hat D_\nu \rho_{A\nu}-\hat D_\nu \rho_{A\mu})
(\hat D^\mu \rho_A^\nu-\hat D^\nu \rho_A^\mu)\}\nonumber\\
&&+\frac{1}{\alpha_\rho}{\Tr}\{(\hat D_\mu\tilde r_V^\mu)^2+(\hat
D_\mu\tilde\rho_A^\mu)^2\}\nonumber\\
&&+\frac{1}{2\alpha_\rho'}[(\partial^\mu S_{A\mu})^2+(\partial^\mu  
S_\mu)^2]\nonumber\\
&&+\mu^2_\rho({\Tr}\{\tilde r_V^\mu\tilde r_{V\mu}\}+
{\Tr}\{\tilde\rho_A^\mu\tilde\rho_{A\mu}\})\nonumber\\
&&+\frac{1}{2}\mu^2_A S^\mu_A S_{A\mu}+\frac{1}{2}\mu^2_VS^\mu S_\mu
\ear
with
\be\label{V12d}
\tilde r_V^\mu=r^\mu_V-\frac{1}{3}{\Tr}\ r_V^\mu,\ \tilde \rho_A^\mu
=\rho_A^\mu-\frac{1}{3}{\Tr}\rho_A^\mu,
\ S^\mu_A=\sqrt{\frac{2}{3}}{\Tr}\rho^\mu_A\ee

The existence of an additional homogeneous vector field $r_V^\mu$ allows
us to construct new invariants and leads to mixing. Indeed, the
invariant for the vector currents in (\ref{E1}) is now enlarged
by a piece
\bear\label{V13}
{\LL}_{VV}&=&\chi^2_0{\Tr}\{(\hat v_\mu+g\tilde V_\mu-e\tilde B_\mu
\tilde Q)(\hat v^\mu+g\tilde V^\mu-e\tilde B^\mu\tilde Q)\}
\nonumber\\
&&+M^2_V{\Tr}\{\tilde r_{V\mu}\tilde r_V^\mu\}
+2\nu_\rho{\Tr}\{(\hat v_\mu+g\tilde V_\mu-e\tilde B_\mu \tilde Q)
{\tilde r}_V^\mu\}\ear
with
\be\label{V14}
M^2_V=\mu^2_\rho+2\sigma_0^2(f_1-f_2)+
(\frac{2}{9}e_1-\frac{1}{36}e_2)\chi_0^2\ee
The mixing term $\sim \nu_\rho$ arises from the covariant
derivative in the cubic $\chi-\chi-\rho$ term $(\tilde{\hat a}_\mu
=\hat a_\mu-\frac{1}{3}{\Tr}\hat a_\mu)$
\be\label{V15}
{\LL}_{\chi^2\rho}=-2e_{\rho\pi\pi}\chi_0^2[{\Tr}\{(\hat v_\mu
+g\tilde V_\mu-e\tilde B_\mu\tilde Q)\tilde r_V^\mu\}
+\frac{7}{9}{\Tr}\{\tilde{\hat a}_\mu\tilde\rho_A^\mu\}]\ee
\be\label{V16}
\nu_\rho=-e_{\rho\pi\pi}\chi_0^2\ee
The low mass vector meson octet $\tilde\rho_V^\mu$
and a heavy state $\tilde H_V^\mu$
\bear\label{V17}
\tilde\rho_V^\mu&=&\cos\tau\ \tilde V^\mu+\sin \tau\ \tilde r_V
^\mu,\nonumber\\
\tilde H_V^\mu&=&\cos\tau\ \tilde r^\mu_V-\sin \tau\ \tilde V_\mu\ear
are the mass eigenstates with mass eigenvalues
\bear\label{V18}
M^2_\rho&=&\frac{1}{2}(g^2\chi_0^2+M^2_V-\sqrt{
(M_V^2-g^2\chi_0^2)^2+4g^2\nu^2_{V\rho}}\nonumber\\
M^2_H&=&\frac{1}{2}(g^2\chi_0^2+M^2_V+\sqrt{
(M_V^2-g^2\chi_0^2)^2+4g^2\nu^2_{V\rho}}\ear
and mixing angle
\be\label{12.41a}
tg(2\tau)=-\frac{2g\nu_\rho}{M_V^2-g^2\chi^2_0}\ee

For the interactions it is most convenient to eliminate
$\tilde r_V^\mu$ by its field equation. In the local approximation,
where the kinetic term for $\tilde r_V^\mu$ is small compared
to the mass term $(q^2\ll M_V^2)$, one finds
\be\label{V19}
\tilde r_V^\mu=-\frac{\nu_\rho}{M_V^2}(\hat v^\mu+g\tilde V^\mu-e\tilde  
B^\mu\tilde Q)\ee
and therefore
\be\label{V20}
{\LL}_{VV}=(\chi_0^2-\frac{\nu^2_\rho}{M^2_V}){\Tr}\{(\hat v_\mu+g\tilde V_\mu
-e\tilde B_\mu\tilde Q)(\hat v^\mu+g\tilde V^\mu-e\tilde B^\mu \tilde Q)\}\ee
This is the same structure as (\ref{E1}), leading for the
$\rho-\pi$-interaction (\ref{E3}) to
\be\label{V21}
a=\frac{1}{f_\pi^2}(\chi_0^2-\frac{\nu_\rho^2}{M_{V^2}})\ee
hereby lowering the value of $a$ as compared to (\ref{E4}).
The kinetic term for $\tilde V^\mu$ also obtains a contribution
from the insertion of (\ref{V19}) such that the $\rho$-mesons
with a standard normalization correspond to
\be\label{V22}
\tilde\rho_V^\mu=\left(1+\frac{g^2\nu^2_\rho}{M_V^4}
\right)^{-1/2}\tilde V^\mu\ee
After a replacement of $\tilde V$ by $\tilde \rho_V$ this
modifies (\ref{E4}) according to
\be\label{V23}
g_\rho=g\left(1+\frac{g^2\nu^2_\rho}{M_V^4}\right)
^{1/2}\ee
As announced before the structure of the interactions between
vector mesons, pseudoscalar mesons and photons is dictated
by the local $U(3)_P\times U(1)_{em}$ symmetries. The low
momentum limit is not modified by the inclusion of additional
interactions or degrees of freedom. In the following we work in the
approximation
\be\label{V24}
\nu^2_\rho\ll\chi_0^2M^2_V,\quad e^2_{\rho\pi\pi}
\ll\frac{M_V^2}{\chi^2_0}\ee
such that the modifications of $a$ and $g_\rho$ can be neglected.

Let us next turn to the contribution to the pseudoscalar kinetic term
(\ref{5.8a}) from the interaction $\sim{\Tr}\{\tilde\rho_A^\mu\tilde
{\hat a}_\mu\}$. In addition to (\ref{V15}) such a term is also
induced by
\be\label{V25}
{\LL}_{\phi^2\rho}=-8c_{\rho\pi\pi}\sigma^2_0{\Tr}\{\tilde
\rho_A^\mu\tilde{\hat a}_\mu\}\ee
With the electromagnetically covariant axial vector current
\bear\label{V26}
\hat a_\mu&=&a_\mu+\frac{e}{2}\tilde B_\mu(W_L^\dagger QW_L-W_R^\dagger
QW_R)\nonumber\\
&=&-\frac{i}{2}W_R^\dagger D_\mu UW_L=\frac{i}{2}W_L
^\dagger D_\mu U^\dagger W_R\nonumber\\
\tilde{\hat a}_\mu&=&\hat a_\mu-\frac{1}{3}{\Tr}\hat a_\mu,\quad
{\Tr}\hat a_\mu={\Tr} a_\mu=-\frac{1}{2}\partial_\mu\theta\ear
the term linear in $\tilde\rho_A^\mu$ is also linear in
the pseudoscalar fields. It corresponds to a mixing between the
pseudoscalar in $U$ and the pseudoscalar $\partial_\mu\rho_A^\mu$ (both are
$0^{-+}$ states). This so-called ``partial Higgs effect''
has been extensively discussed in \cite{8}. After elimination
of the field $\rho^\mu_A$ by virtue of the field equation it
leads to an additional negative contribution to the
pion decay constant $f$. Indeed, for
\bear\label{V27}
{\LL}_{\rho A}&=&M_A^2{\Tr}\{\tilde\rho_A^\mu\tilde\rho_{A\mu}\}
+2\alpha{\Tr}\{\tilde\rho_A^\mu\hat a_\mu\}
-c_{\rho\bar qq}\ {\rm Tr}\{\bar N\gamma_\mu\gamma^5\tilde\rho^\mu_AN\}
+...\nonumber\\
\alpha&=&-4\sigma_0^2(c_{\rho\pi\pi}+xe_{\rho\pi\pi})
\ear
the insertion of the solution of the field equation
\be\label{V28}
\tilde\rho_A^\mu=-\frac{\alpha}{M^2_A}\tilde{\hat a}^\mu+
\frac{c_{\rho\bar qq}}{4M_A^2}\ {\rm Tr}\{\bar  
N\gamma^\mu\gamma^5\lambda_zN\}\lambda_z+...\ee
yields the bosonic contribution
\be\label{V29}
{\LL}^{(1)}_{\rho A}=-\frac{\alpha^2}{M^2_A}{\Tr}\{\tilde{\hat a}^\mu
\tilde{\hat a}_\mu\}+...=-
\frac{\alpha^2}{4M^2_A}{\Tr}\{D^\mu\tilde U^\dagger D_\mu\tilde U\}
+...\ee
This constitutes the contribution $\Delta_f^2$ mentioned in sect. 5
\be\label{V30}
\frac{\Delta^2_f}{f^2}=\frac{\alpha^2}{M^2_Af^2}=
\frac{(c_{\rho\pi\pi}+xe_{\rho\pi\pi})^2}{(1+x)^2}\frac{f^2}
{\kappa^4_fM^2_A}\approx 0.45\ee
Similarly, the contribution quadratic in the baryon fields reads
\bear\label{12.54a}
{\cal L}^{(2)}_{\rho A}&=&y_N\ {\rm Tr}\{\bar N\gamma_\mu\gamma^5\tilde {\hat  
a}^\mu N\}
\nonumber\\
y_N&=&\frac{\alpha c_{\rho\bar qq}}{M_A^2}\approx 0.45\frac{c_{\rho\bar  
qq}f^2}{\alpha}\approx
-0.31\frac{c_{\rho\bar qq}(1+x)}{c_{\rho\pi\pi}+xe_{\rho\pi\pi}}\ear
There are similar contributions from the elimination of the singlet
axial vector Tr $a^\mu$.

We finally note that the insertion\footnote{The contributions
from inserting corrections $\sim r_V\hat a$ to
(\ref{V28}) into (\ref{V27}) vanish.} of eq. (\ref{V28})
into (\ref{V12c}) introduces a term (after partial integration)
\bear\label{AppA1}
{\cal L}_{\rho 2}&=&-2i\frac{\alpha}{M_A^2}{\rm Tr}
\{(D^\mu r_V^\nu-D^\nu r^\mu_V)[\hat a_\mu,\hat a_\nu]\}
+...\nonumber\\
&=&2i\frac{\alpha}{M_A^2f^2}{\rm Tr}\{\partial^2r^\nu_{VT}
[\Pi,\partial_\nu\Pi]\}+...\ear
with $r^\nu_{VT}$ the transversal part of $r^\nu_V(\partial^2
=\partial_\mu\partial^\mu)$
\be\label{AppA2}
r^\nu_{VT}=r^\nu_V-\frac{\partial^\nu\partial^\rho}{\partial^2}r^\rho_V
\ee
Using (\ref{V20}), this replaces for the $\rho\to 2\pi$
decay rate
\bear\label{AppA3}
g_{\rho\pi\pi}&\to&g_{\rho\pi\pi}+\frac{\nu_\rho\alpha g}{M_A^2f^2}
\\
&=&g_{\rho\pi\pi}+\frac{9}{7}g\frac{x}{(1+x)^2}
\frac{e_{\rho\pi\pi}(c_{\rho\pi\pi}+xe_{\rho\pi\pi})f^2}{M^2_A
\kappa_f^4}\nonumber\\
&=&g_{\rho\pi\pi}+0.45\frac{\nu_\rho}{\alpha}g=g_{\rho\pi\pi}
+0.58(1+\frac{c_{\rho\pi\pi}}{xe_{\rho\pi\pi}}
)^{-1}g\ear
where we have employed the on-mass-shell condition $(\partial^2\to M^2_V)$.
As a result, the phenomenologically required value of $g_{\rho\pi\pi}$
in eq. (\ref{E6}) could be smaller than 6, and in consequence, the value of
$a$ required by eq. (\ref{E8}) could be somewhat lower than 2.

We observe that after eliminating $r^\mu_V$ by (\ref{V19})
the term (\ref{AppA1}) leads to an invariant
\bear\label{12.58}
I_3&=&\ {\rm Tr}\{(D^\mu v^\nu_I-D^\nu v^\mu_I)[\hat a_\mu,\hat a_\nu]\}
\nonumber\\
v^\mu_I&=&\hat v^\mu+g\tilde V^\mu-e\tilde B^\mu\tilde Q\nonumber\\
D_\mu v_{I\nu}&=&\partial_\mu v_{I\nu}-ie\tilde B_\mu[\tilde Q,v_{I\nu}]
+i[\hat v_\mu,v_{I\nu}]\ear
Since $v^\mu_I$ and $\hat a^\mu$ transform homogeneously with
respect to $U(3)_P\times U(1)_{em}$ and (\ref{12.58}) involves a fully  
covariant derivative, it is
obvious that $I_3$ adds a new invariant structure to the terms contained
in eq. (\ref{E1}). This invariant influences the cubic and higher
interactions but does not contribute to the two-point functions.
Concerning eq. (\ref{E6}), it does not affect $M_\rho^2,g_{\gamma\rho}$
and $m^2_B$, whereas quantities like $g_{\rho\pi\pi}$ or $g_{\gamma\pi\pi}$
receive corrections. One concludes that the KSFR relation (\ref{E7})
is not a pure symmetry relation. Its validity also requires that the effective  
action for $\rho$-mesons is dominated by the invariant
${\cal L}_{VV}$ (\ref{E3}). The observed success of the KSFR relation
implies that corrections from $I_3$ must be small. This is the case
for $e_{\rho\pi\pi}\ll c_{\rho\pi\pi}/x$, implying that the mixing
between $\tilde r_V^\mu$ and $\tilde V^\mu$ can indeed be neglected (cf.
eq. (\ref{V16})). For processes not involving axial vectors we therefore
retain from (\ref{EQ1}) only ${\cal L}_{\rho^2}$ (for the singlet
vector meson), ${\cal L}_{\phi^2\rho}$ (for the partial Higgs effect)
and ${\cal L}_{\rho\bar qq}$ (for the effective nucleon interactions in the
vector channel (\ref{12.25a}). This leads to the additional terms
(\ref{8.7}) displayed in sect. 8.

\section*{Appendix B}
\renewcommand{\theequation}{B.\arabic{equation}}
\setcounter{equation}{0}
In this appendix we discuss invariants with $SU(3)_C\times SU(2)_L
\times U(1)_Y$ local and $SU(3)_R$ global symmetry beyond
the ones mentioned in sect. 10. We first turn to terms without
derivatives. One may
ask if local terms not involving derivatives
could generate a term involving only one strange meson as, for
example, the $CP$-even state
$K^0_L=\frac{1}{\sqrt2}(K^0+\bar K^0)$. Such terms would contribute
to the weak kaon decays.
The answer is negative, as can be seen
by inserting into the infinitesimal transformation of $\Pi$ up to linear
order
\be\label{9.D}
\delta\Pi=\frac{f}{2}(\Theta_R-\Theta_L)+i(\Theta_R\Pi-\Pi\Theta_L)
+0(\Pi^2)\ee
the appropriate global $SU(2)_R\times U(1)_R$ or local
$SU(2)_L\times U(1)_Y$ transformations. In fact, the remaining
symmetry is still powerful enough to forbid for $s_\theta=0$
any nonderivative terms involving the pseudoscalars $\pi^\pm,
\pi^0, \eta,K^0, \bar K^0$. Any nonderivative coupling must
therefore involve an even number of charged kaons of the form $(K^+
K^-)^n$. For $s_\theta
\not=0$ this is replaced by $(c_\theta K^+-s_\theta\pi^+)\cdot
(c_\theta K^--s_\theta\pi^-)^n$. On the other hand, strangeness-violating
processes always have to involve charged $W^\pm$-bosons. In lowest
order $\sim M^{-2}_W$ we can neglect the exchange of $Z^0$. The
strangeness-violating short-distance four-quark interactions
have then a global $SU(3)_R$ flavor symmetry. This forbids
also the terms mentioned above. One infers that there are no
strangeness-violating non-derivative interactions in order $M^{-2}_W$.
We conclude that we can omit all additional non-derivative interactions
for a discussion of the weak decays of pseudoscalars.

We next turn to terms involving two derivatives and we restrict the
discussion for simplicity to $\phi$. It is straightforward
to see that terms involving only one trace and only one matrix $\lambda_W$
lead to the same structure as (\ref{9.E}) since
$\phi^\dagger\phi=\sigma_0^2$ as far as the pseudoscalars are concerned.
We therefore omit such terms. The invariant
involving four powers of $\phi$ and one factor of $\lambda_W$ is
proportional ${\rm Tr}\{\phi^\dagger D^\mu\phi\lambda_W\}{\rm Tr}\{
\phi^\dagger D_\mu\phi\}$ and does not contribute to the hadronic
kaon decays into two pions.
This is obvious by expanding
for $\vartheta=0$
\be\label{B.A2}
\phi^\dagger D_\mu\phi=\sigma_0^2\tilde U^\dagger\partial_\mu\tilde  
U=\frac{2\sigma_0^2}{f}(i\partial_\mu\Pi+\frac{1}{f}[\partial_\mu
\Pi,\Pi])+...\ee
A relevant invariant reads
\be\label{B.3AA}
{\cal L}^{(4)}_{W,\phi}=-\frac{z_W^\phi g_W^2}{8M^2_W}{\rm Tr}\{
(\phi^\dagger D_\mu\phi-D_\mu\phi^\dagger\phi)
\vec\tau_W\}{\rm Tr}\{(\phi^\dagger D^\mu\phi-D^\mu\phi^\dagger
\phi)\vec\tau_W\}
\ee
with
\be\label{B.3BB}
\tau_{W1}=c_\theta\tau_1+s_\theta\lambda_4\ ,\quad
\tau_{W2}=c_\theta\tau_2+s_\theta\lambda_5\ ,\quad \tau_{W3}=
\tau_3\ee
Here the fact that $\tau_{W3}$ is not rotated is related
to the absence of flavor-changing neutral currents. (The rotation
of the type (\ref{9.B}) has to be realized by a four by four
matrix including the charm quark.) Up to a factor $-z^\phi_W$
the invariant (\ref{B.3AA}) has precisely the same structure
as the one generated from one-particle reducible $W$-exchange
by inserting the field equation for $W$ as a
functional of $\phi$. This is no surprise since the underlying
structure on the quark level corresponds to box-type diagrams where
the exchange of a $W$-boson is supplemented by gluon exchange.
A similar argument holds for operators
involving $\chi$. The effects of the
weak mixing between scalars and pseudoscalars
may also be represented by effective higher-order operators
of the type (\ref{B.3AA}) which obtain after elimination of the scalar
fields via their field equation.
We include all these effects\footnote{There are also
contributions to the $\Delta I=1/2$ amplitude which may be
absorbed by a small change in $\mu_W$.} by multiplying
the $W$-boson exchange contribution (\ref{9.22}) by
a factor $(1-z_W)$. The relative minus sign $(z_W>0)$ reflects
the negative sign of the corresponding anomalous dimension
in perturbation theory.

Strangeness-violating local interactions can also appear on the level
of the Yukawa couplings between quarks and mesons
\be\label{9.K}
{\LL}_\beta=Z_\psi\bar\psi_{Ri}\left(\frac{g^2_W\beta_W^2h}{4M_W^2}
\phi\delta_{ij}+\frac{g_W^2\tilde\beta_W^2\tilde h}{4M_W^2}
\chi_{ij}\right)\lambda_W\psi_{Lj}+h.c.\ee
The interaction (\ref{9.K}) is the only dimensionless invariant
contributing to strangeness-violating baryon interactions. According
to the hypothesis of scalar Ver\-voll\-st\"an\-digung it should
dominate the hyperon decays. It has to be supplemented by
strangeness-violating mixings of the pseudoscalars discussed
in appendix C.

We recall that we work in a basis where both the kinetic
and the mass term for the quarks conserve strangeness. This
imposes constraints on the effective strangeness-violating
couplings. As discussed in sect. 2, we can recover the multi-quark
interactions by inserting the solutions of the field equations
for $\phi$ and $\chi$ as functionals of $\bar\psi\psi$. The
terms (\ref{9.C}), (\ref{9.E}), (\ref{9.K}) result in a shift of the
solution $\phi[\psi]=\phi_0[\psi]+\delta\phi[\psi]$. For large
enough $k$ this reads approximately
\be\label{9.L}
\delta\phi_{ab}=\frac{g_W^2}{4M_W^2}(m^2_\phi-\partial^2)^{-1}
[Z_\psi\beta_W^2h\bar\psi_{L,ic}(\lambda_W)_{cb}\psi_{R,ai}
-(\tilde\mu_W^4-\mu^2_W\partial^2)(\phi_0\lambda_W)_{ab}]\ee
with
\be\label{9.M}
(\phi_0)_{ab}=Z_\psi h(m^2_\phi-\partial^2)^{-1}\bar\psi_{Lib}\psi
_{Rai}+\frac{1}{2}Z^{-1/2}_\phi j_{ab}\ee
Insertion in the effective action (including the source
term (\ref{2.8})) leads to a strangeness-conserving quark mass
term only for $\beta_W^2=\tilde\mu_W^4m_\phi^{-2}$. A similar,
but more complicated relation exists
between the strangeness violation
in the effective potential for $\phi,\chi$
and the strangeness violation in the Yukawa-type couplings.

The insertion of (\ref{9.L}) in the effective action leads to four-quark  
interactions involving two left-handed
and two right-handed quarks. Assume that $\tilde\mu_W(k=0)$
can be related to $\tilde\mu_W(k)$ at a typical transition
scale around 1 GeV by following the renormalization
group flow of the effective action. It can then be matched
to the perturbatively (RG-improved) computed value of the
corresponding four-quark interaction. Since a tree exchange
of $W$-bosons always involves at least four left-handed
fermions, it is obvious that the four-quark interaction
of interest can only be generated by a $W$-boson loop,
as appropriate for a one-particle irreducible contribution.
We will not pursue this road for a computation of the effective
couplings $\mu_W, \tilde\mu_W$ etc. in the present work.

\section*{Appendix C: $K^0_L-K^0_S$-mass difference}
\renewcommand{\theequation}{C.\arabic{equation}}
\setcounter{equation}{0}

For the discussion of the mass difference between $K^0_S$ and
$K^0_L$ one needs  the one-particle irreducible
$\Delta S=2$ contributions in order $M_W^{-4}$
\bear\label{9.I}
{\LL}_\mu^{(\Delta S=2)}&=&\frac{g_W^4\mu^2_{W,2}}{16 M^4_W}(\eta_1
\ {\rm Tr}\{\phi^\dagger D^\mu\phi\lambda_W\phi^\dagger D_\mu\phi\lambda_W
\}\nonumber\\
&&+\eta_2{\rm Tr}\{\phi^\dagger D^\mu\phi\lambda_W\}{\rm Tr}\{
\phi^\dagger D_\mu\phi\lambda_W\})+...\ear
Here the dots stand for terms involving $\chi$. This effective
interaction corresponds to an eight-quark interaction. It is
related to a vertex with four left-handed quarks
(e.g. box diagrams with exchange of two $W$-boson lines)
by supplementing
the four right-handed quarks needed to form scalars. For the
kinetic term of the pseudoscalar mesons it reduces to
\bear\label{9.J}
{\LL}_\mu^{(\Delta S=2)}&=&\frac{3}{16}A^2_W\mu^2_{W,2}\
[\eta_1{\rm Tr}\{\partial^\mu U\lambda_6
\partial_\mu U\lambda_6\}\nonumber\\
&&+\eta_2{\rm Tr}\{\partial^\mu U\lambda_6\}{\rm Tr}\{\partial_\mu
U\lambda_6\}]+...\ear
This induces a difference in the wave function renormalization
for $K^0_L$ and $K^0_S$
\bear\label{C.B1}
{\cal L}_{kin}&=&\frac{Z_L}{2}\partial^\mu K^0_L\partial_\mu K^0_L+\frac
{Z_S}{2}\partial^\mu K^0_S\partial_\mu K^0_S+...\nonumber\\
Z_L-Z_S&=&-\frac{3}{2}\frac{A_W^2\mu^2_{W,2}}{f^2}\bar\eta\nonumber\\
\bar\eta&=&\eta_1+\eta_2+...\ear
where the dots in $\bar\eta$ stand again for contributions
from invariants involving the octet $\chi$. We normalize $\mu^2_{W2}$
such that $\bar\eta=1$. Using a normalization with $Z_S=1,\ M_{K_S}=M_K$,
one has $M_{K_L}=M_KZ_L^{-1/2}$ or
\be\label{C.B2}
(M_{K_L}-M_{K_S})_d=\frac{3}{4}\frac{A_W^2\mu^2_{W2}M_K}{f^2}
=0.59\cdot 10^{-12}\ {\rm MeV}\ \left(\frac{\mu_W^2}{f^2}
\right)\ee
to be compared with the observed value
\be\label{C.B3}
M_{K_L}-M_{K_S}=3.522\cdot10^{-12}\ {\rm MeV}\ee

The index $d$ in eq. (\ref{C.B2}) indicates that this
is only the direct contribution from the 1PI effective vertices.
A further contribution arises from mixing effects of the pseudoscalar
mesons in first order in $\Delta S$. In fact, due
to weak interaction effects the fields in eq. (\ref{5.23}) have small  
off-diagonal corrections in their propagators.
Expanding ${\LL}_{\mu U}$ (\ref{9.F})
in second order in $\Pi,\theta$, one finds
\bear\label{9.27a}
{\LL}_{\mu U}^{(2)}&= &2A_W \hat g_8
[Tr\{\partial^\mu\Pi\partial_\mu\Pi \lambda_6\}-\frac{f}{3}
\frac{1+\delta_Wx}{1+5\delta_Wx/14}
\partial^\mu\theta \ Tr\{\partial_\mu \Pi \lambda_6\}
]\nonumber\\
&=&A_W\hat
g_8[\partial^\mu\tilde K^+\partial_\mu\tilde\pi^-+\partial^\mu\tilde
\pi^+\partial_\mu\tilde K^--\partial^\mu\tilde \pi^0\partial_\mu
\tilde K^0_L-\frac{1}{\sqrt3}\partial^\mu\tilde\eta\partial_\mu
\tilde K_L^0]\nonumber\\
&&-\tilde g_8\partial^\mu\tilde\eta'\partial_\mu\tilde K^0_L\nonumber\\
\hat g_8&=&\frac{14+5\delta_Wx}{14-4\delta_Wx}g_8\quad,\quad\tilde  
g_8=\frac{2f}{3H_{\eta'}}\frac{1+\delta_Wx}{1-2\delta_W
x/7}g_8\ear
Here we use a tilde for the fields appearing in eq. (\ref{5.23}) which
correspond to the basis where the quark mass term (\ref{2.8}) is
diagonal. The physical meson fields $\pi^0, K^0_{L,S},
\eta, \eta'$ and $\pi^\pm,K^\pm$ obtain after diagonalization of
the inverse propagator. The total quadratic term for the neutral
mesons reads
\bear\label{9.27b}
{\LL}^0_{(2)}&=&\frac{1}{2}(\partial^\mu\tilde\pi^0\partial_\mu\tilde
\pi_0+\partial^\mu\tilde\eta\partial_\mu\tilde\eta+\partial^\mu
\tilde\eta\,'\partial_\mu\tilde\eta\,'+\partial^\mu\tilde  
K^0_L\partial_\mu\tilde K^0_L+\partial
^\mu\tilde K^0_S\partial_\mu\tilde K^0_S)\nonumber\\
&&-A_W \hat g_8(\partial^\mu\tilde\pi^0\partial_\mu\tilde K^0_L+\frac{1}
{\sqrt3}\partial^\mu\tilde\eta\partial_\mu\tilde K^0_L)
-A_W\tilde g_8\partial^\mu\tilde\eta\,'\partial_\mu\tilde K^0_L\\
&&+\frac{1}{2}(M^2_\pi(\tilde\pi^0)^2+M_\eta^2\tilde\eta^2+M^2_{\eta'}\tilde
\eta\,^{'2}+M^2_K[(\tilde K^0_L)^2+(\tilde K^0_S)^2])\nonumber\ear
and the ``physical fields''
are given by
\bear\label{9.27c}
\tilde\pi^0&=&\pi^0+\frac{M_K^2}{M_K^2-M^2_\pi}A_W\hat g_8K^0_L\nonumber\\
\tilde\eta&=&\eta-\frac{M_K^2}{M_\eta^2-M_K^2}\frac{A_W\hat g_8}{\sqrt3}
K^0_L\nonumber\\
\tilde\eta\,'&=&\eta'-\frac{M_K^2}{M_{\eta'}^2-M_K^2}A_W\tilde g_8K^0_L
\nonumber\\
\tilde K^0_L&=&K^0_L-\frac{M_\pi^2}{M_K^2-M_\pi^2}A_W\hat g_8\pi^0
+\frac{M_\eta^2}{M_\eta^2-M_K^2}\frac{A_W\hat g_8}{\sqrt3}\eta
+\frac{M^2_{\eta'}}{M^2_{\eta'}-M^2_K}
A_W\tilde g_8\eta'\nonumber\\
\tilde K^0_S&=&K^0_S\ear
In terms of the physical fields the relevant contributions to ${\LL}
^0_{(2)}$ read
\bear\label{9.27d}
&&{\LL}^0_{(2)}=\frac{1}{2}(\partial^\mu\pi^0\partial_\mu\pi^0+\partial^\mu
\eta\partial_\mu\eta
+\partial^\mu\eta'\partial_\mu\eta'+\partial^\mu K^0_S\partial_\mu
K^0_S\\
&&\qquad\qquad+(1+A^2_WQ_Z)\partial^\mu K^0_L\partial_\mu K^0_L)\nonumber\\
&&+\frac{1}{2}(M^2_\pi(\pi^0)^2+M^2_\eta\eta^2+M^2_{\eta'}{\eta'}^2+M_K^2
(K^0_S)^2+M^2_K(1+A^2_WQ_M)(K^0_L)^2)\nonumber
\ear
where we note the difference in the kinetic and mass terms between
$K^0_S$ and $K^0_L$, with
\be\label{9.27e}
Q_Z=-\frac{M_K^2(M_K^2-2M^2_\pi)\hat g^2_8}{(M_K^2-M_\pi^2)^2}+
\frac{1}{3}\frac{M_K^2(2M^2_\eta-M_K^2)\hat g_8^2}{(M_\eta^2-M_K^2)^2}
+\frac{M_K^2(2M^2_{\eta'}-M^2_K)\tilde g_8^2}{(M^2_{\eta'}-
M^2_K)^2}\ee
\be\label{9.27f}
Q_M=\frac{M_K^2M^2_\pi \hat g^2_8}{(M_K^2-M_\pi^2)^2}+
\frac{1}{3}\frac{M_K^2M^2_\eta \hat g^2_8}{(M^2_\eta-M_K^2)^2}
+\frac{M_K^2M_{\eta'}^2\tilde g_8^2}{(M_{\eta'}^2-
M_K^2)^2}\ee

\bigskip
In consequence, the renormalized masses for $K^0_L, K^0_S$ read
without the 1PI-irreducible contribution (\ref{C.B1})
\be\label{9.31}
M^2_{K_L}=M^2_K\left(\frac{1+A^2_WQ_M}{1+A^2_WQ_Z}\right)\ , \
M^2_{K_S}=M^2_K\ee
This corresponds to a contribution of the
mixing effect to the mass difference
\bear\label{9.32}
\Delta(M_{K_L}-M_{K_S})&=&\frac{A^2_WM_K}{2}(Q_M-Q_Z)\nonumber\\
&=&\frac{A^2_WM_K}{2}\left(\frac{M^2_K\hat g_8^2}{M_K^2-M_\pi^2}-\frac{1}{3}
\frac{M_K^2\hat g_8^2}{M_\eta^2-M_K^2}-\frac{M_K^2\tilde g_8^2}{M_{\eta'}^2-M_K^2}
\right)\nonumber\\
&=&-\frac{A_W^2M_K}{2}(0.465 \hat g^2_8+\tilde g^2_8)\ear
Summing (\ref{C.B2}) and (\ref{9.32}) we obtain
\be\label{D.C1}
\frac{M_{K_L}-M_{K_S}}{(M_{K_L}-M_{K_S})_{obs}}=
0.17\left(\frac{\mu^2_{W2}}{f^2}-0.31 \hat g^2_8-\frac{2}{3}\tilde g^2_8\right)
\ee
With typical values $\hat g_8\approx 1.5$ (cf. eq. (\ref{9.14d})), and
$\tilde g_8\approx 1$ this would imply  $\mu_{W2}\approx$ 300 MeV.
This is of the order of the expected characteristic mass scale.

We complete this appendix by a discussion of
the effective mixing between $K^\pm$ and $\pi^\pm$. The
relevant contribution of $\tilde W$-exchange to the pseudoscalar kinetic
term is given by the square of the term linear in $\partial_\mu\Pi$ in eq.
(\ref{9.15})
\bear\label{9.27}
\Delta{\LL}_{kin}^{(W)}&=&-M^2_W(W_\mu^+)^{(1)}(W^{-\mu})^{(1)}
\nonumber\\
&=&-\frac{g_W^2f^2}{2M_W^2}\ {\Tr}\{\partial^\mu\Pi\hat\lambda_+
^\dagger\}
{\rm Tr}\{\partial_\mu\Pi\hat\lambda_+\}
\nonumber\\
&=&-\frac{g_W^2f^2}{4M^2_W}\partial^\mu(c_\theta\tilde\pi^++s_\theta\tilde
K^+)\partial_\mu(c_\theta\tilde\pi^-+s_\theta\tilde K^-)\ear
This has to be combined with similar terms in eq. (\ref{9.27a}).
Here we use again $\tilde K^\pm,\tilde\pi^\pm$ for the fields appearing
in eq. (\ref{5.23}) which correspond to the basis where the
quark mass term (\ref{2.8}) is diagonal. In presence of weak
interactions the fields $\tilde K^\pm,\tilde\pi^\pm$ have to be
related to the fields $K^\pm,\pi^\pm$ for the physical pseudoscalar
mesons. The latter have  diagonal propagators. If we only
consider effects linear in $G_F$, the relevant bilinear
\bear\label{9.28}
{\LL}^{K\pi}_{(2)}&=&\partial^\mu\tilde\pi^+\partial_\mu\tilde\pi^-
+\partial^\mu\tilde K^+\partial_\mu\tilde K^-+M^2_\pi
\tilde\pi^+\tilde\pi^-+M^2_K\tilde K^+\tilde K^-\nonumber\\
&&+A_W(\hat g_8-1)(\partial^\mu\tilde  
K^+\partial_\mu\tilde\pi^-+\partial^\mu\tilde\pi^+\partial_\mu\tilde
K^-)\ear
is diagonalized by
\be\label{9.29}
\tilde \pi^+=\pi^+-\frac{M_{K^2}}{M_{K^2}-M_{\pi^2}}A_W
(\hat g_8-1)K^+\ ,\ \tilde K^+=
K^++\frac{M_{\pi^2}}{M_{K^2}-M_{\pi^2}}A_W(\hat g_8-1)\pi^+\ee
such that
\be\label{9.30}
{\LL}^{K\pi}_{(2)}=\partial^\mu\pi^+\partial_\mu\pi^-+
\partial^\mu K^+\partial_\mu K^-
+M^2_\pi\pi^+\pi^-+M^2_KK^+K^-\ee
This mixing effect is not relevant for the hadronic kaon two-body decays
since there are no cubic vertices involving only pseudoscalars
in the absence of weak interactions. It contributes, however, to the
decay of $K^0_L$ into three pions. Furthermore, it leads
to an additional contribution to the
strangeness-violating vertices between baryons and pseudoscalars.

\end{document}